\documentclass[%
 aip,
 jmp,%
 amsmath,amssymb,
 reprint,%
]{revtex4-2}

\usepackage{graphicx}
\usepackage{dcolumn}
\usepackage{bm}
\usepackage{xcolor}

\def\AP#1{{\textcolor{black}{#1}}}
\def\DR#1{{\textcolor{black}{#1}}}
\def\NY#1{{\textcolor{black}{#1}}}

\def\grad{\nabla}

\begin{document}

\preprint{AIP/123-QED}

\title[Helicity SGS model\AP{s} and 
\AP{their} validation]{Helicity subgrid-scale model\AP{s} and 
\AP{their} numerical validation}

\author{N. Yokoi}
 \email{nobyokoi@iis.u-tokyo.ac.jp}
\altaffiliation[Also at ]{Nordic Institute for Theoretical Physics (NORDITA), Stockholm University and Royal Institute of Technology (KTH).}
\affiliation{ 
Institute of Industrial Science, University of Tokyo
}%

\author{P.D.~Mininni}%
\affiliation{ 
Universidad de Buenos Aires (UBA), Facultad de Ciencias Exactas y Naturales, Departamento de Física, and CONICET-UBA, Instituto de Física Interdisciplinaria y Aplicada (INFINA), and CNRS-CONICET-UBA, Institut Franco-Argentin de Dynamique des Fluides pour l’Environnement (IFADyFE), IRL2027, Ciudad Universitaria, 1428 Buenos Aires, Argentina
}%

\author{A. Pouquet}
\affiliation{%
National Center for Atmospheric Research (NCAR), Boulder, Colorado (USA)
}%

\author{D. Rosenberg}
\affiliation{%
Cooperative Institute for Research in the Atmosphere/NOAA
}%

\author{R. Marino}
\affiliation{%
\'{E}cole Centrale de Lyon, Lyon 182-8522, France
}%

\date{19 January 2026}

\begin{abstract}
Large-eddy simulations (LES) with an appropriate subgrid-scale (SGS) model provide a powerful tool for investigating real-world turbulence. The Smagorinsky model, one of the simplest and most used SGS models, often shows an over-dissipative behavior even when using dynamic procedures to adjust the model coefficient. By incorporating the structural or geometrical information of turbulence provided by helicity (velocity–vorticity correlations), the helicity SGS model is expected to alleviate these issues in the standard Smagorinsky 
\AP{framework}, in which only information of turbulence intensity is considered through the turbulent energy. The validity of 
helicity SGS model\AP{s} is investigated here with the aid of direct numerical simulations (DNSs). Using configurations with and without net rotation, and with large-scale helicity gradients sustained by a mechanical forcing, we show that to better model SGS turbulence, SGS helicity effects should be incorporated into the model together with the Smagorinsky-like eddy viscosity.
\end{abstract}
\maketitle

\section{\label{sec:I}Introduction}

Astrophysical, geophysical, plasma physics, and engineering flows are almost invariantly extremely turbulent. Because of huge spatial dimensions, high speeds, and very small diffusivities (including viscosity, thermal diffusivity, resistivity, etc.), the Reynolds numbers of such flows are huge. And due to the strong nonlinearity, the vast scales of motions are coupled with each other. The ratio of the largest energy-containing scale $L$ to the smallest viscous dissipation scale $\ell_\textrm{D}$ is expressed in terms of the Reynolds number $Re$ as $L/\ell_{\rm{D}} \sim Re^{3/4}$. The number of grid points required for three-dimensional (3D) direct numerical simulations (DNSs), $N_{\rm{G}}$, scales as a result as $N_{\rm{G}} \sim Re^{9/4}$. Even when considering the flow associated with our own walking, the Reynolds number can be as large as \NY{$Re = UL/\nu = 10^6$ ($U = 
1\ {\rm{m}}\ {\rm{s}}^{-1}$, $L = 
1\ {\rm{m}}$, $\nu = 
10^{-6}\ {\rm{m}}^2\ {\rm{s}}^{-1}$)}.
It becomes then impossible to simultaneously solve all scales in flows of astro/geophysical or engineering interest, with even larger $Re$. As a result, we have to reduce the number of degrees of freedom needed to perform numerical simulations. \AP{Thus,}  
turbulence modeling is an indispensable approach to turbulence flows.

In the Reynolds-averaged Navier--Stokes (RANS) numerical approach and its associated modeling, a statistical averaging of ﬁeld quantities is adopted. The statistically averaged fields are called the mean fields. In this formulation, the fluctuations are not directly calculated. \AP{They are} 
represented instead by a turbulence model with a small number of degrees of freedom, \AP{and with the} 
fluctuation effects in the mean-ﬁeld equations \AP{given} 
by turbulent fluxes such as the Reynolds stress, the turbulent mass flux, the turbulent heat flux, and so on. RANS models are used mostly for engineering calculations, and in some cases for some geo/astrophysical applications. Because of the statistical character of the averaging, the applicability of RANS for a fine description of unsteady turbulence is limited.

Similar to the Reynolds-averaged turbulence model, large-eddy simulations (LESs) of turbulence with subgrid-scale (SGS) models provide a powerful tool for analyzing turbulent flows at very large Reynolds numbers. In LESs, by adopting an appropriate filter (with filter width denoted by $\Delta$), the turbulent motions are divided into grid-scale (GS) and SGS components. If the filtering is operated in the space domain, the GS component represents the turbulent motions at large scales, while the SGS component corresponds to small scales. In a simulation, the large-scale or GS motions (large eddies) are explicitly solved, while the small-scale (SGS) turbulence motions are somehow modeled. The scales of motions 
in the SGS model are much smaller than their counterparts in RANS models. As a result, SGS fluctuations tend to be more homogeneous and isotropic. As a result, the SGS models in LESs are expected to be much simpler than the models in RANS simulations.

There are several approaches to SGS modeling.\cite{sag2006} One of them is the so-called functional modeling, in which the effect of the small-scale (SGS) fluctuations on the large-scale (GS) motions is represented through the energetic action associated with the forward energy cascade. The GS energy is drained 
\AP{using} a SGS tensor and given to the SGS energy. This SGS stress tensor, which represents the effect of unresolved scale (SGS) fluctuations on the resolved scale (GS) motions, is expressed \AP{specifically} 
in terms of the GS strain rate tensor. Another approach is the so-called structural modeling, in which a part of the SGS structural information is reconstructed by assuming some relation between the statistical structure of the ﬂow at different filtering levels. Because of the additional information somehow preserved from the unresolved scales, the prediction of the energy transfer is often greatly improved in structural models, without resorting to any prior knowledge of the nature of the interaction between the GS and SGS fluctuations.

Perhaps the simplest existing functional model is the Smagorinsky model.\cite{sma1963} In this model, the SGS stress is expressed by the GS strain rate tensor, $\mbox{\boldmath${\overline{s}}$} = \{ {\overline{s}^{ij}} \}$, using a SGS viscosity which is determined by $\nu_{\rm{S}} = (C_{\rm{S}} \Delta)^2 \overline{s}$ with $\overline{s} = \sqrt{{\overline{s}}^{ij} {\overline{s}}^{ij} /2}$. The model constant $C_{\rm{S}}$ is called the Smagorinsky constant, whose value can be evaluated from the Kolmogorov energy spectrum as $C_{\rm{S}} \approx 0.18$ assuming the Kolmogorov constant to be equal to $1.4$.

There are some drawbacks of the Smagorinsky model. The first one is the need to adjust its constant depending on the ﬂow considered. The optimised Smagorinsky constant $C_{\rm{S}}$ is $0.18$ for homogeneous isotropic turbulence,\cite{ant1981} $0.15$ for mixing layer turbulence,\cite{ham1987} and $0.10$ for wall turbulence\cite{dea1970} (see Table~\ref{tab:table1}). Since the SGS viscosity $\nu_{\rm{S}}$ depends on $C_{\rm{S}}^2$, if we adopt the value of $0.18$ for wall turbulence, the SGS viscosity is overestimated by more than three times [$(0.18/0.10)^2 = 3.24$] 
\AP{compared to}  the corresponding  $\nu_{\rm{S}}$ in the bulk with the optimised value of $0.10$. This is not at all good from the viewpoint of the universality of the model constant. The second drawback of the Smagorinsky model is also related to this point: The Smagorinsky model is generally over-dissipative in regions of large GS strain. As a result, \AP{it} tends to exaggerate the dissipative properties of turbulence transport. Finally, the Smagorinsky representation of the SGS stress does not automatically vanish very close to walls where the turbulence effects vanish. In other words, the Smagorinsky model in its simple form has no self-adaptivity in the near-wall region. We therefore have to implement a self-adaptive scheme, such as a dynamic procedure to determine the model coefficient, or adopt a damping function to satisfy the near wall turbulence behavior. In dynamic procedures some of these drawbacks are alleviated by adopting a sequential application of two filters at different scales, and by evaluating the effective coefficient through a minimization of errors.\NY{\cite{ger1991}}  However, even with this dynamic procedure, it is known that if the ﬂow displays strong turbulent structures, e.g., in the streamwise vorticity, results obtained with the Smagorinsky model still have strong limitations.

In this work we address the drawbacks of the lack of universality in the \AP{$C_{\rm{S}}$} constant and the over-dissipative properties of the Smagorinsky model from another physical perspective. In wall turbulence, localized and very strong vortical \AP{structures} develop mostly in the streamwise direction. They are positively and negatively aligned to the streamwise ﬂow, and show a characteristic spatio-temporal distribution.\cite{ham2006, wal2013, liu2024} Such streamwise structures are often called streaks, and are observed also in mixing-layer turbulence. As we list in Table~\ref{tab:table1}, the appearance of streamwise vorticity depends on the ﬂow. Streamwise vorticity is strongest in wall turbulence, weakest in homogeneous isotropic turbulence even in the presence of a mean flow (e.g., in wind tunnels), while mixing-layer turbulence is in-between. This dependence suggests that the need to adjust the Smagorinsky constant may be related to the strength of the streamwise vorticity in each ﬂow. In the case of wall turbulence, where the streamwise vorticity is strongest, the effective dissipation (as measured by the optimised value of $C_{\rm{S}} = 0.10$) is much smaller than in the case of homogeneous isotropic turbulence. In the latter, the streamwise vorticity is weaker, while the effective dissipation is much stronger as measured by the optimised value of $C_{\rm{S}} = 0.18$.
Since the streamwise vorticity is best measured by the local helicity
(i.e., the local velocity-vorticity correlation), the $C_{\rm{S}}$ flow dependence might be explained in terms of its dependence on the turbulent helicity. 

\begin{table}
\caption{\label{tab:table1}Smagorinsky constant $C_{\rm{S}}$ and strength of vortical structures in several flows.}
\begin{ruledtabular}
\begin{tabular}{ccc}
Flow
&Optimised $C_{\rm{S}}$
&Streamwise vorticity\\
\hline
Homogeneous isotropic 
& 0.18 
& Weak\\
Mixing-layer 
& 0.15 
& Intermediate\\
Wall 
& 0.10 
& Strong\\
\end{tabular}
\end{ruledtabular}
\end{table}

The important role of helicity in turbulence has been considered in various contexts.\cite{wol1958, mof1969, ber1984, pou2022, yok2024} 
\NY{Its relevance was first argued for magnetic-field generation in turbulent media (the so-called $\alpha$ dynamo).\cite{par1955} Following the pioneering work on the invariance of magnetic helicity $\int_V {\bf{a}} \cdot {\bf{b}}\ dV$ (where {\bf{a}} is the magnetic potential, ${\bf{b}} = \nabla \times {\bf{a}}$ is the magnetic field, and $V$ is the fluid volume) in the absence of magnetic diffusivity by Woltjer,\cite{wol1958} the topological properties of magnetic helicity have been intensively explored in magnetohydrodynamics (MHD).\cite{ber1984} It has been recognized that helical properties of turbulence play an essential role in the structuring of large-scales in MHD. With the aid of the eddy-damped quasi-normal Markovian (EDQNM) approximation, it was shown that the magnetic helicity density ${\bf{a}} \cdot {\bf{b}}$, or more directly the current helicity density ${\bf{b}} \cdot {\bf{j}}$ (where ${\bf{j}} = \nabla \times {\bf{b}}$ is the  electric current), alters the behavior of the $\alpha$ dynamo which had been originally considered to depend only on the kinetic helicity.\cite{pou1976} In hydrodynamics, it was shown by Moffatt that the kinetic helicity $\int_V {\bf{u}} \cdot \mbox{\boldmath$\omega$}\ dV$ (where ${\bf{u}}$ is the velocity and $\mbox{\boldmath$\omega$} = \nabla \times {\bf{u}}$ is the vorticity) is an invariant in the absence of dissipation.\cite{mof1969} With the aid of the EDQNM approximation, the role of kinetic helicity on the energy cascade was investigated shortly after.\cite{and1977}}

In particular, and more recently, helical fluid turbulence has been explored in the context of turbulent transport,\cite{yok1993,ang2021}  global ﬂow generation,\cite{yok2016b,ina2017} and angular-momentum transport in  stellar convection.\cite{yok2025} 
\NY{Another possible application of helicity in geophysical flows is given by tornadoes and tropical cyclones. In a tropical cyclone, in the eye and eyewall regions, strong dry descents and eyewall updrafts are ubiquitous. In the combination with swirling winds, this up and down configuration gives inhomogeneous mean-flow helicity, that naturally results in inhomogeneous turbulent helicity. In particular, tropical cyclones have strong inhomogeneous turbulent helicity near the eyewall. This, coupled with the azimuthal large-scale vorticity (caused by the up- and down-drafts) should induce a strong swirling azimuthal wind. This effect may explain the observed wind acceleration near the eyewall, and may provide a mechanism for some aspects of the genesis of cyclone and tornadoes.\cite{yok1993, yok2024, lev2020}}
In the presence of inhomogeneous turbulent helicity, the turbulent viscosity can also be decreased by the coupling of the inhomogeneous helicity with the large-scale vorticity. 

Considering these results, we therefore explore this inhomogeneous-helicity effect in the framework of SGS modeling of LESs.\cite{yok2017} 
\NY{There have been previous attempts to incorporate helicity into SGS models. For instance, in \citet{li2006}, helicity was implemented into the SGS stress in coupling with the vorticity-gradient tensor, and the models were tested {\it a priori} in steady isotropic helical turbulence. The authors found that the role of their helicity terms in the energy and helicity dissipation rates was quite small. Here we focus instead on effects associated with inhomogeneities of helicity, which couple with the absolute vorticity of the flow.}
The effect of inhomogeneous helicity enters the SGS stress expression in addition to the gradient diffusion approximation of the turbulent momentum transport as used in the Smagorinsky model. This helicity effect may therefore alleviate the fundamental problems of the \AP{original}  Smagorinsky model, \AP{and hopefully narrow the range of variation of $C_{\rm{S}}$.}

The organisation of this paper is as follows. In Sec.~\ref{sec:II}, helicity and its effects in turbulence transport are presented. This includes a general introduction to helicity, a theoretical analysis of the effect of helicity on the Reynolds stress, the presentation of turbulence modeling based on theoretical results, how to incorporate structural effects into the models, the transport equations for one-point one-time turbulence statistical quantities, and a review of previous numerical validations of the role of helicity in turbulence models. In Sec.~\ref{sec:III}, SGS models with the structure effects incorporated through the SGS helicity  are proposed (\AP{models which we call} ``helicity SGS models''\AP{, or H-SGS model}). In Sec.~\ref{sec:IV}, the numerical set-up for the validation of the helicity SGS models is presented. The numerical results comparing the model expressions and the direct numerical simulations (DNSs) are shown in Sec.~\ref{sec:V}. The properties of helicity SGS models are discussed in Sec.~\ref{sec:VI}, followed by the concluding remarks in Sec.~\ref{sec:VII}.

\section{\label{sec:II}Helicity effects}
\subsection{\label{sec:II.A}Helicity}

The kinetic helicity, previously defined as $H = \int_V {\bf{u}} \cdot \mbox{\boldmath$\omega$} dV$, as well as the energy $E = \int_V {\bf{u}}^2/2 \, dV$, are inviscid invariants of the incompressible fluid equations of motion. The local density of helicity, $h = {\bf{u}} \cdot \mbox{\boldmath$\omega$}$, represents the right- and left-handed twist of a fluid element. Hereafter, we call the helicity density $h$ just helicity. Helicity is a pseudo-scalar, which changes its sign under reflection or inversion of the coordinates. This is in marked contrast to the energy density $e = {\bf{u}}^ 2 /2$ (in the following, the energy), a positive-deﬁnite pure scalar, which does not change sign under reflection or inversion. As a reminder, from the definition of mirror-symmetry, any statistical quantity in a mirror-symmetric system satisﬁes $f({\bf{x}}) = f(-{\bf{x}})$ under inversion ${\bf{x}} \to -{\bf{x}}$. At the same time, a pseudo-scalar quantity changes its sign under inversion as $f(-{\bf{x}}) = - f({\bf{x}})$. As a consequence, a pseudo scalar in a mirror-symmetric system satisfies $f({\bf{x}}) = f(-{\bf{x}}) = - f({\bf{x}})$. Thus, a pseudo scalar vanishes, $f({\bf{x}}) = 0$, in mirror-symmetric systems. In other words, a finite pseudo-scalar such as the helicity $h$ measures the break down of mirror-symmetry in a flow.

In turbulence, whereas the turbulent energy $\langle {{\bf{u}'}^2} \rangle /2$ measures the intensity of turbulence associated with transverse and longitudinal velocity correlations (here the prime denotes fluctuations), the turbulent kinetic helicity $\langle {{\bf{u}'} \cdot \mbox{\boldmath$\omega$}'} \rangle$ represents the structural information corresponding to the cross velocity correlation. Studies on helical fluid and (Hall-)MHD turbulence were brieﬂy reviewed by \citet{pou2022}. Transport in helical fluid turbulence was described in \citet{yok2024} (see also \citet{ang2021} for laboratory measurements).

\subsection{\label{sec:II.B}Theoretical analysis}

The mean velocity ${\bf{U}}$ of an incompressible turbulent flow (possibly rotating with fixed angular velocity $\mbox{\boldmath$\omega$}_{\rm{F}}$) obeys the momentum equation
\begin{equation}
	\frac{\partial {\bf{U}}}{\partial t}
	+ ({\bf{U}} \cdot \nabla) {\bf{U}}
	= - \nabla P
	+ {\bf{U}} \times 2 \mbox{\boldmath$\omega$}_{\rm{F}}
	+ \nabla \cdot \mbox{\boldmath${\cal{R}}$}
	+ \nu \nabla^2 {\bf{U}},
	\label{eq:mean_U_eq}
\end{equation}
and the solenoidal condition
\begin{equation}
	\nabla \cdot {\bf{U}} = 0,
	\label{eq:mean_U_sol_cond}
\end{equation}
where $P$ is the mean pressure per unit density including the effect of the centrifugal force, and $\nu$ is the kinematic viscosity. Here, the Reynolds stress $\mbox{\boldmath${\cal{R}}$}$ is the sole effect of turbulence in the mean velocity equation. It is deﬁned as
\begin{equation}
	\mbox{\boldmath${\cal{R}}$}
	\equiv \langle {{\bf{u}}' {\bf{u}}'} \rangle
	= \{ {\langle {u'^i u'^j} \rangle} \},
	\label{eq:rey_strss_def}
\end{equation}
where the superindices $i$ and $j$ denote the Cartesian components, and $\{ \}$ denotes all the components of the tensor.

In the framework of the two-scale direct-interaction approximation (TSDIA) or the multiple-scale DIA (MSDIA), which is a renormalized perturbation ﬁeld theory for inhomogeneous turbulence, the turbulent fluxes such as the Reynolds stress are expressed in terms of the propagators (i.e., the correlation and response functions).\cite{yos1984,yok2020} 
\NY{In this formulation, we introduce fast and slow variables, $(\mbox{\boldmath$\xi$};\tau)$ and $({\bf{X}};T)$, respectively. They are defined as $\mbox{\boldmath$\xi$} = {\bf{x}}$, ${\bf{X}} = \delta {\bf{x}}$, $\tau = t$, and $T = \delta t$, with the scale parameter $\delta$. If $\delta$ is small, $({\bf{X}};T)$ start changing only when the original variables $({\bf{x}}; t)$ vary considerably. In this sense, the variables $({\bf{X}};T)$ are used to express large-scale and slow variations. A field quantity $f$ is then divided into the mean and fluctuating parts, $F$ and $f'$, as $f({\bf{x}};t) = F({\bf{X}};T) + f'(\mbox{\boldmath$\xi$},{\bf{X}};\tau,T)$. As a consequence of the introduction of two scales, the spatial and time derivatives are expressed as $\nabla = \nabla_{\mbox{\tiny\boldmath$\xi$}} + \delta \nabla_{\bf{X}}$ and $\partial / \partial t = \partial / \partial \tau + \delta \partial / \partial T$. The space and time derivatives of the slow variables are then of order $\delta$. In this multiple-scale formulation, the effects of large-scale variations of $f'$ in space and time enter through higher-order terms in $\delta$. This makes it possible to treat turbulence inhomogeneity, anisotoropies, and non-equilibrium properties.}

\NY{We assume that the fluctuating field is homogeneous with respect to the fast spatial variable $\mbox{\boldmath$\xi$}$, and we Fourier transform it as $f'(\mbox{\boldmath$\xi$},{\bf{X}}; \tau,T) = \int f'({\bf{k}},{\bf{X}}; \tau,T) \exp[- i {\bf{k}} \cdot (\mbox{\boldmath$\xi$} - {\bf{U}} \tau)]\ d{\bf{k}}$. Here, the Fourier transform is computed in the frame co-moving with the mean velocity ${\bf{U}}$. On the basis of the two-scale formulation, the Fourier transform of a fluctuating field quantity $f'$ is expanded with scale parameter $\delta$ as $f'({\bf{k}},{\bf{X}}; \tau,T) = \sum_{n=0}^\infty \delta^n f'_n({\bf{k}},{\bf{X}}; \tau,T)$.} 
\NY{For the velocity fluctuation ${\bf{u}}'$, the lowest-order field ${\bf{u}}'_0({\bf{k}},{\bf{X}};\tau,T)$ is assumed to be homogeneous and isotropic, and its correlation and response functions are written as\cite{yok1993}
\begin{equation}
	\frac{\langle { u'_0{}^i({\bf{k}},{\bf{X}};\tau,T) u'_0{}^j({\bf{k}}',{\bf{X}};\tau',T) } \rangle}
		{\delta_D({\bf{k}} + {\bf{k}}')}
	= D^{ij}({\bf{k}}) Q({k,{\bf{X}};\tau,\tau',T})
	+ \frac{i}{2} \frac{k^\ell}{k^2} \epsilon^{ij\ell} H(k,{\bf{X}};\tau,\tau',T),
	\label{eq:hit_correl_stat}
\end{equation}
and
\begin{equation}
	G^{ij}({\bf{k}},{\bf{X}}; \tau,\tau',T) = D^{ij}({\bf{k}}) G(k,{\bf{X}}; \tau,\tau',T),
	\label{eq:hit_response_stat}
\end{equation}
where $Q(k,{\bf{X}};\tau,\tau',T)$ and $H(k,{\bf{X}};\tau,\tau',T)$ are the spectral density functions of the turbulent energy and helicity, respectively, $G^{ij}({\bf{k}},{\bf{X}}; \tau,\tau',T)$ is the response function associated with the ${\bf{u}}'_0$ response to an infinitesimal perturbation to the velocity field, $G(k,{\bf{X}}; \tau,\tau',T)$ is the isotropic response (Green's) function, $D^{ij}({\bf{k}})$ is the usual solenoidal projector, and $\delta_D$ denotes the Dirac's delta distribution. The second, or $H$-related, term in Eq.~(\ref{eq:hit_correl_stat}) represents the contribution of the non-mirror-symmetric components of the turbulence field.}

\NY{In this formulation, the Reynolds stress tensor is calculated as
\begin{eqnarray}
	&{}& \langle {u'{}^i({\bf{x}};t) u'{}^j({\bf{x}};t)} \rangle =
	    \int {d{k}}
		\frac{\langle {u'{}^i({\bf{k}};t) u'{}^j({\bf{k}}';t)} \rangle}{\delta_D({\bf{k}} + {\bf{k}}')} =
		\int {d{k}}  \left\{
		\frac{\langle {u'_0{}^i({\bf{k}};t) u'_0{}^j({\bf{k}}';t)} \rangle}{\delta_D({\bf{k}} + {\bf{k}}')} + \right.
	\nonumber\\
	&{}& \left.	+ \delta \left[ {
		\frac{\langle {u'_0{}^i({\bf{k}};t) u'_1{}^j({\bf{k}}';t)} \rangle}{\delta_D({\bf{k}} + {\bf{k}}')}
		+ \frac{\langle {u'_1{}^i({\bf{k}};t) u'_0{}^j({\bf{k}}';t)} \rangle}{\delta_D({\bf{k}} + {\bf{k}}')}
	} \right]
	+ \mathcal{O}(\delta^2)
	 \right\} .
	\label{eq:rey_strss_cal}
\end{eqnarray}}

With the aid of the TSDIA formulation, the Reynolds stress in non-mirror-symmetric flows can be obtained from the Navier--Stokes equation as\cite{yok1993}
\begin{equation}
	\mbox{\boldmath${\cal{R}}$}_{\rm{D}}
	= - \nu_{\rm{T}} \mbox{\boldmath${\cal{S}}$}
	+ \left[ {
	\mbox{\boldmath$\Gamma$} \mbox{\boldmath$\Omega$}_\ast
	+ (\mbox{\boldmath$\Gamma$} \mbox{\boldmath$\Omega$}_\ast)^\dagger
	} \right]_{\rm{D}},
	\label{eq:rey_strss_exp}
\end{equation}
where the suffix ${\rm{D}}$ denotes the deviatoric or traceless part of a tensor as
\begin{equation}
	{\cal{A}}^{ij}_{\rm{D}}
	= {\cal{A}}^{ij} - \frac{1}{3} \delta^{ij} {\cal{A}}^{\ell\ell},
	\label{eq:dev_tens__def}
\end{equation}
$\mbox{\boldmath${\cal{S}}$} = \{ {S^{ij}} \}$ is the mean-velocity strain rate deﬁned as
\begin{equation}
	{\cal{S}}^{ij}
	= \frac{\partial U^j}{\partial x^i}
	+ \frac{\partial U^i}{\partial x^j}
	- \frac{2}{3} \delta^{ij} \nabla \cdot {\bf{U}}
	\NY{ = \frac{\partial U^j}{\partial x^i}
	+ \frac{\partial U^i}{\partial x^j}},
	\label{eq:mean_vel_strain}
\end{equation}
where the last equality results from the fact that the mean flow is incompressible, and $\mbox{\boldmath$\Omega$}_\ast$ is the mean absolute vorticity deﬁned by
\begin{equation}
	\mbox{\boldmath$\Omega$}_\ast
	= \mbox{\boldmath$\Omega$}
	+ 2 \mbox{\boldmath$\omega$}_{\rm{F}},
	\label{eq:mean_abs_vorticity_def}
\end{equation}
with $\mbox{\boldmath$\Omega$} = \nabla \times {\bf{U}}$ the mean relative vorticity. In Eq.~(\ref{eq:rey_strss_exp}), $\nu_{\rm{T}}$ and $\mbox{\boldmath$\Gamma$}$ are transport coefficients for $\mbox{\boldmath${\cal{S}}$}$ and $\mbox{\boldmath$\Omega$}_\ast$, respectively, and $\dagger$ denotes the transposed tensors. In the TSDIA analysis, the eddy viscosity $\nu_{\rm{T}}$ can be expressed in terms of the response function $G$ and of the spectral energy density $Q$ as\cite{yok1993} 
\begin{equation}
	\nu_{\rm{T}}
	= \frac{7}{15} \int d{\bf{k}} \int_{-\infty}^{\tau}\!\!\! d\tau'\ 
		G(k,{\bf{X}};\tau,\tau',T) Q(k,{\bf{X}};\tau,\tau',T).
	\label{eq:nuT_anal_exp}
\end{equation}
In the simplest case where the time integral of $G(k,{\bf{X}}; \tau,\tau',T)$ is performed separately, it becomes the mixing-length expression for the turbulent viscosity
\begin{equation}
	\nu_{\rm{T}} = \tau_{\rm{T}} K = u \ell
	\label{eq:nuT_MLT}
\end{equation}
[$\tau_{\rm{T}}$: turbulence timescale, $u$: turbulence characteristic velocity, $\ell (= u \tau)$: turbulence length scale]. This means that Eq.~(\ref{eq:nuT_anal_exp}) is a generic expression that includes the mixing-length expression as a particular case. On the other hand, the transport coefficient $\mbox{\boldmath$\Gamma$}$ for the mean absolute vorticity $\mbox{\boldmath$\Omega$}_\ast$ is expressed in terms of the spectral helicity density $H$ as
\begin{equation}
	\mbox{\boldmath$\Gamma$}
	= \frac{1}{30} \int k^{-2} d{\bf{k}} \int_{-\infty}^{\tau} d\tau'\ 
	G(k,{\bf{X}};\tau,\tau',T) \nabla H(k,{\bf{X}};\tau,\tau',T).
	\label{eq:Gamma_anal_exp}
\end{equation}
While the turbulent energy through the eddy viscosity in Eq.~(\ref{eq:nuT_anal_exp}) couples with the mean velocity strain $\mbox{\boldmath${\cal{S}}$}$ (the symmetric part of the mean velocity shear) in Eq.~(\ref{eq:rey_strss_exp}), the turbulent helicity gradient in Eq.~(\ref{eq:Gamma_anal_exp}) couples with the mean absolute vorticity $\mbox{\boldmath$\Omega$}_\ast$ (the anti-symmetric part of the mean velocity shear) in Eq.~(\ref{eq:rey_strss_exp}).

Equations~(\ref{eq:rey_strss_exp}) and (\ref{eq:Gamma_anal_exp}) show that the coupling coefficient for the mean absolute vorticity $\mbox{\boldmath$\Omega$}_\ast$ is proportional not to the turbulent helicity itself, but to its gradient. This is natural if we consider the parity or sign-change property of a ﬁeld quantity under inversion (parity is even for axial vectors and pure scalars, odd for polar vectors and pseudo-scalars). The Reynolds stress deﬁned by the correlation of velocity (odd) and velocity (odd) has even parity. Then, the turbulent helicity itself $H$ (odd) cannot be the coupling coefficient for the mean absolute vorticity (even). However, the helicity gradient $\nabla H$ (even) can enter as the coupling term to the absolute vorticity:
\begin{equation}
	\langle {
		\underbrace{{\bf{u}}'}_{\text{odd}}
		\underbrace{{\bf{u}}'}_{\text{odd}}
	} \rangle
	:= - \underbrace{\tau K}_{\text{even}}
		\underbrace{\mbox{\boldmath${\cal{S}}$}}_{\text{even}}
	+ \underbrace{\tau \ell^2}_{\text{even}} 
		(\underbrace{\nabla}_{\text{odd}}
		\underbrace{H}_{\text{odd}})
		\underbrace{\mbox{\boldmath$\Omega$}_\ast}_{\text{even}},
	\label{eq:parity_rey_strss}
\end{equation}
where, for the sake of simplicity in the notation, the suffix ${\rm{D}}$ for the deviatoric part was suppressed, and $:=$ denotes ``schematically represents.''

\subsection{\label{sec:II.C}Turbulence modeling}
\subsubsection{\label{sec:II.C.1}Choice of turbulence statistical quantities}

The analytical expression of the Reynolds stress in Eq.~(\ref{eq:rey_strss_exp}) with Eqs.~(\ref{eq:nuT_anal_exp}) and (\ref{eq:Gamma_anal_exp}) depends on the spectral and time integrals of the spectral and response functions, and is very heavy for practical use. In order to apply such an expression to practical ﬂow phenomena, we construct a turbulence model on the basis of analytical results. In place of spectral functions $Q(k,{\bf{X}}; \tau,\tau',T)$ and $H(k,{\bf{X}}; \tau,\tau',T)$, we adopt one-point statistical quantities: The turbulent kinetic energy per mass:
\begin{equation}
	K = \langle {{\bf{u}}'^2} \rangle/2,
	\label{eq:one-point_K_eq}
\end{equation}
and the turbulent kinetic helicity as defined before:
\begin{equation}
	H = \langle {{\bf{u}'} \cdot \mbox{\boldmath$\omega$}}' \rangle.
	\label{eq:one-point_H_def}
\end{equation}
The response function $G(k,{\bf{X}};\tau,\tau',T)$ represents the weight of how much past states affect the present state, and provides timescale information of turbulence. In place of the response function, we introduce the dissipation rate $\varepsilon$ of the turbulent kinetic energy $K$ as
\begin{equation}
	\varepsilon_K
	= \nu \left\langle {
		\frac{\partial u^j}{\partial x^i}
		\frac{\partial u^j}{\partial x^i}
	} \right\rangle
	(\equiv \varepsilon).
	\label{eq:epsK_def}
\end{equation}
This quantity represents how much the turbulent kinetic energy $K$ decays in time. Then we evaluate the turbulence timescale as
\begin{equation}
	\tau_K
	= \frac{K}{\varepsilon}
	(\equiv \tau).
	\label{eq:turb_timescale_Keps}
\end{equation}
In terms of these turbulent statistical quantities $K$, $H$ and $\varepsilon$, the transport coefficients in Eq.~(\ref{eq:rey_strss_exp}), $\nu_{\rm{T}}$ as given by Eq.~(\ref{eq:nuT_anal_exp}) and $\mbox{\boldmath$\Gamma$}$ as given by Eq.~(\ref{eq:Gamma_anal_exp}), can be expressed as
\begin{equation}
	\nu_{\rm{T}}
	= C_\nu \tau K
	= C_\nu \frac{K^2}{\varepsilon},
	\label{eq:nuT_model_Keps}
\end{equation}
\begin{equation}
	\mbox{\boldmath$\Gamma$}
	= C_\eta \tau L^2 \nabla H
	= C_\eta \frac{K^4}{\varepsilon^3} \nabla H
	\NY{\equiv \eta_{\rm{T}} \nabla H},
	\label{eq:Gamma_model_KepsH}
\end{equation}
\NY{
where $C_\nu$ and $C_\eta$ are model constants, and $\eta_{\rm{T}}$ is the part of $\mbox{\boldmath$\Gamma$}$ except for the gradient of the turbulent helicity, $\nabla H$, defined by  
\begin{equation}
	\eta_{\rm{T}} = C_\eta \frac{K^4}{\varepsilon^3}.
	\label{eq:etaT_def}
\end{equation}
}

From these expressions, a model for the Reynolds stress based on Eq.~(\ref{eq:rey_strss_exp}) is given by
\begin{eqnarray}
	\left\langle {
		u'^i u'^j
	} \right\rangle
	&=& \frac{2}{3} \delta^{ij} K
	- \nu_{\rm{T}} \left( {
		\frac{\partial U^j}{\partial x^i}
		+ \frac{\partial U^i}{\partial x^j}
		- \frac{2}{3} \delta^{ij} \nabla \cdot {\bf{U}}
	} \right)
	\nonumber\\
	&+& \eta_{\rm{T}} \left( {
		\frac{\partial H}{\partial x^i} \Omega^j_\ast
		+ \frac{\partial H}{\partial x^j} \Omega^i_\ast
		- \frac{2}{3} \delta^{ij} \left( {
			\mbox{\boldmath$\Omega$}_\ast \cdot \nabla
		} \right) H
	} \right).
	\label{eq:rey_strss_model_exp}
\end{eqnarray}

\subsubsection{\label{sec:II.C.2}Transport equations of statistical quantities}

In order to construct a self-consistent turbulence model, we need to know the spatio-temporal distributions of the turbulent statistical quantities $K$, $H$, and $\varepsilon$. This closure is done by considering the evolution equations of $K$, $H$, and $\varepsilon$. The equations of $K$ and $H$ are written as
\begin{equation}
	\frac{DF}{Dt}
	= \left( {
		\frac{\partial}{\partial t}
		+ {\bf{U}} \cdot \nabla
	} \right) F
	= P_F
	- \varepsilon_F
	+ \nabla \cdot {\bf{T}}_F
	\label{eq:F_transport_eq}
\end{equation}
with $F = (K,H)$. Here, $P_F$, $\varepsilon_F$, and ${\bf{T}}_F$ are the production, dissipation, and transport rates of $F$. They are given by
\begin{equation}
	P_K 
	= - {\cal{R}}^{ij} \frac{\partial U^j}{\partial x^i},
	\label{eq:P_K_def}
\end{equation}
\begin{equation}
	\varepsilon_K
	= \nu \left\langle {
		\frac{\partial u'^j}{\partial x^i}
		\frac{\partial u'^j}{\partial x^i}
	} \right\rangle
	(\equiv \varepsilon),
	\label{eq:epsK_eps_def}
\end{equation}
\begin{equation}
	{\bf{T}}_K
	= \left\langle {
		-p'{\bf{u}}'
		- \frac{1}{2} {\bf{u}}'^2 {\bf{u}}'
		+ \nu \nabla \left( {\frac{1}{2} {\bf{u}}'^2} \right)
	} \right\rangle
	\approx \frac{\nu_{\rm{T}}}{\sigma_K} \nabla K,
	\label{eq:TK_def_model}
\end{equation}
\begin{equation}
	P_H
	= - {\cal{R}}^{ij} \frac{\partial \Omega^j}{\partial x^i}
	+ \Omega_\ast^j \frac{\partial {\cal{R}}^{ij}}{\partial x^i},
	\label{eq:PH_def}
\end{equation}
\begin{equation}
	\varepsilon_H
	= 2\nu \left\langle {
		\frac{\partial u'^j}{\partial x^i}
		\frac{\partial \omega^j}{\partial x^i}
	} \right\rangle
	\approx C_H \frac{\varepsilon}{K} H,
	\label{eq:epsH_def_model}
\end{equation}
\begin{equation}
	{\bf{T}}_H
	= \left\langle {
		-p'\mbox{\boldmath$\omega$}'
		+ \frac{1}{2} {\bf{u}}'^2 \mbox{\boldmath$\omega$}'
		- ({\bf{u}}' \cdot \mbox{\boldmath$\omega$}') {\bf{u}}'
		+ \nu \nabla ({\bf{u}}' \cdot \mbox{\boldmath$\omega$}')
	} \right\rangle
	\approx \frac{\nu_{\rm{T}}}{\sigma_H} \nabla H,
	\label{eq:TH_def_model}
\end{equation}
where $C_H$ is the model constant for the helicity dissipation rate, and $\sigma_K$ and $\sigma_H$ are respectively the turbulent Prandtl numbers for the turbulent energy and helicity diffusion.

In addition to Eq.~(\ref{eq:F_transport_eq}) for $K$ and $H$, we consider the equation for the turbulent energy dissipation $\varepsilon$ as
\begin{equation}
		\frac{D\varepsilon}{Dt}
	= \left( {
		\frac{\partial}{\partial t}
		+ {\bf{U}} \cdot \nabla
	} \right) \varepsilon
	\approx C_{\varepsilon 1} \frac{\varepsilon}{K} P_K
	- C_{\varepsilon 2} \frac{\varepsilon}{K} \varepsilon_K
	+ \nabla \cdot \left( {
		\frac{\nu_{\rm{T}}}{\sigma_\varepsilon} 
		\nabla \varepsilon
	} \right),
	\label{eq:epsK_eq_model}
\end{equation}
where $C_{\varepsilon 1}$, $C_{\varepsilon 2}$, and $\sigma_\varepsilon$ are model constants, which are optimised as\NY{\cite{pop2000,rod2000}}
\begin{equation}
	C_{\varepsilon 1} = 1.4,\;
	C_{\varepsilon 2} = 1.9,\;
	\sigma_\varepsilon = 1.2.
	\label{eq:Ceps1_Ceps2_sigmaeps_consts}
\end{equation}
The following should be noted about Eq.~(\ref{eq:epsK_eq_model}) for $\varepsilon$. Unlike the transport equations for $K$ and $H$, which are connected to the conservative properties of the total energy $\int_V {\bf{u}}^2 dV$ and of the total helicity $\int_V {\bf{u}} \cdot \mbox{\boldmath$\omega$} dV$, the transport equation for $\varepsilon$ is on a less firm mathematical basis. It may seem that the equation for $\varepsilon$ can be heuristically obtained from an analogy with the $K$ equation. However, theoretically this is not the case. 
\NY{Equation}~(\ref{eq:epsK_eq_model}) for $\varepsilon$ is obtained instead from the fundamental \AP{(Navier--Stokes)} equation by assuming the turbulent energy obeys Kolmogorov scaling as
\begin{equation}
	\frac{E(k)}{\varepsilon}
	= \sigma_{E0} \varepsilon^{-1/3} k^{-5/3},
	\label{eq:Kolmogorov_scaling}
\end{equation}
($\sigma_{E0}$: Kolmogorov constant), and the vanishing of the non-equilibrium effect represented by $D\varepsilon / Dt$ and $DL/Dt$. For a detailed analysis of this relation the reader is referred to \citet{yos1987} and \citet{oka1996}.

We can also construct the transport equation of the helicity dissipation rate $\varepsilon_H$ as\cite{yok2016a}
\begin{eqnarray}
	\frac{D\varepsilon_H}{Dt}
	&=& \left( {
		\frac{\partial}{\partial t} + {\bf{U}} \cdot \nabla
	} \right) \varepsilon_H
	\nonumber\\
	&\approx& C_{\varepsilon H1} \frac{\varepsilon_H}{K} P_K
	- C_{\varepsilon H2} \frac{\varepsilon_H}{K} \varepsilon
	+ C_{\varepsilon H3} \frac{\varepsilon_H}{K} P_H
	- C_{\varepsilon H4} \frac{\varepsilon_H}{K} \varepsilon_H
	+ \nabla \cdot \left( {
		\frac{\nu_{\rm{T}}}{\sigma_{\varepsilon H}} \nabla \varepsilon_H
	} \right),
	\label{eq:epsH_eq_model}
\end{eqnarray}
where $C_{\varepsilon H1}$, $C_{\varepsilon H2}$, $C_{\varepsilon H3}$, $C_{\varepsilon H4}$, and $\sigma_{\varepsilon H}$ are model constants evaluated as
\begin{equation}
	C_{\varepsilon H1} = 0.36,\;
	C_{\varepsilon H2} = 0.49,\;
	C_{\varepsilon H3} = C_{\varepsilon H4} = 1.1,\;
	\sigma_{\varepsilon H} = \mathcal{O}(1).
	\label{eq:CepsH1_CepsH2_CepsH3_CepsH4_sigmaepsH}
\end{equation}
Equation~(\ref{eq:epsH_eq_model}) is derived by assuming the spectral dependence of helicity as
\begin{equation}
	\frac{H(k)}{\varepsilon_H}
	= \sigma_{H0} \varepsilon^{-1/3} k^{-5/3},
	\label{eq:helicity_scaling}
\end{equation}
($\sigma_{H0}$: Kolmogorov constant for helicity) and the vanishing of the non-equilibrium effect represented by $D\varepsilon / Dt$, $D\varepsilon_H /Dt$, and $DL /Dt$. Equation~(\ref{eq:helicity_scaling}) means that the helicity spectrum depends on the energy dissipation rate $\varepsilon$ as well as the helicity dissipation rate $\varepsilon_H$. However, in the present work, for the sake of model simplicity, we will not solve Eq.~(\ref{eq:epsH_eq_model}), but just adopt the algebraic model for $\varepsilon_H$ given in Eq.~(\ref{eq:epsH_def_model}).

\subsubsection{\label{sec:II.C.3}Helicity turbulence ($K-\varepsilon-H$) model}

In summary, the Reynolds-averaged helicity turbulence model, which may be called the $K-\varepsilon-H$ model, consists of the mean momentum Eq.~(\ref{eq:mean_U_eq}), the solenoidal condition in Eq.~(\ref{eq:mean_U_sol_cond}), the Reynolds stress expression in Eq.~(\ref{eq:rey_strss_model_exp}), and the transport coefficients $\nu_{\rm{T}}$ in Eq.~(\ref{eq:nuT_model_Keps}) and $\eta_{\rm{T}}$ in Eq.~(\ref{eq:Gamma_model_KepsH}). In addition, the transport Eq.~(\ref{eq:F_transport_eq}) of the turbulent energy $K$, the turbulent helicity $H$ with transport Eqs.~(\ref{eq:P_K_def})-(\ref{eq:TH_def_model}), and the energy dissipation rate $\varepsilon$ with Eq.~(\ref{eq:epsK_eq_model}) should be simultaneously solved.

\subsection{\label{sec:II.D}Validations of the helicity effect }

\begin{figure}
\includegraphics[width=0.6\textwidth]{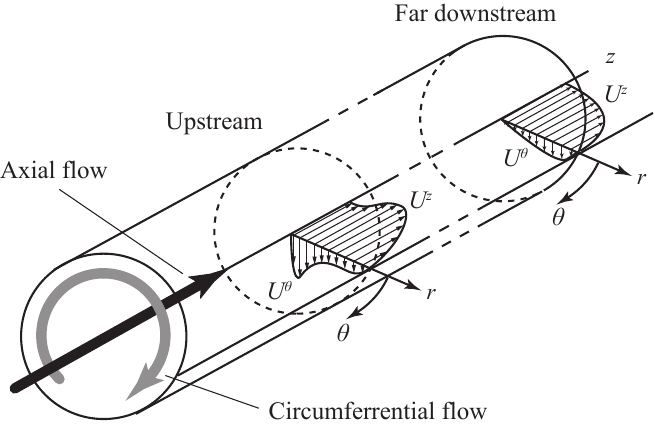}
\caption{\label{fig:swirling_flow}Configuration of a turbulent swirling ﬂow in a straight pipe. A dented axial mean velocity profile in the central axis region upstream is slowly relaxed to the usual ﬂat velocity profile far downstream.}
\end{figure}

The helicity effect represented by the second or $\mbox{\boldmath$\Omega$}_\ast$-related term in Eq.~(\ref{eq:rey_strss_exp}) has been validated numerically in several studies.
The first one was at the level of the Reynolds-averaged turbulence ($K-\varepsilon-H$) model considering the turbulent swirling ﬂow in a straight pipe.\cite{yok1993} In this flow, in addition to the axial ﬂow, a circumferential ﬂow is injected by vanes or fans at the inlet. The resulting mean axial velocity shows a dent in its radial profile in the central axis region.\cite{kit1991,ste1995} This is in marked contrast with the usual turbulent ﬂow without the swirling or circumferential velocity, where, due to the strong momentum mixing by turbulence, the mean axial velocity shows a very ﬂat radial profile in the whole region of the ﬂow except for the near wall region (see Figure~\ref{fig:swirling_flow}). It was shown experimentally that the dent profile of the mean axial velocity lasts until the mean circumferential velocity becomes a trivial rigid-rotation profile.\cite{ste1995}

With the standard $K-\varepsilon$ turbulence model with the eddy-viscosity representation of the Reynolds stress, such a dent profile of the mean axial velocity caused by circumferencial flow at the inlet cannot be reproduced in the downstream region. The eddy-viscosity effect $\nu_{\rm{T}} \mbox{\boldmath${\cal{S}}$}$ in turbulence is so strong that any non-trivial inhomogeneous mean velocity profile cannot be sustained, and is very rapidly smeared out in the downstream region even if a very strong dent profile is imposed at the inlet. In order to alleviate this situation in the turbulent swirling ﬂow, the model constant associated to the eddy viscosity, $C_\nu$, is often set much smaller than the usual optimised value $C_\nu = 0.09$, down to $0.01$. This is an {\it ad hoc} solution that is not based on the physics of the turbulent swirling ﬂow.

On the contrary, with the $K-\varepsilon-H$ model, the dent profile of the mean axial velocity is successfully reproduced in the downstream region. The non-trivial inhomogeneous axial velocity is sustained because of the balance between the eddy-viscosity and inhomogeneous helicity effects. The inhomogeneous helicity effect [the third term in Eq.~(\ref{eq:rey_strss_model_exp})] effectively suppresses the eddy-viscosity effect [the second term in Eq.~(\ref{eq:rey_strss_model_exp})] and preserves the dent profile of the mean axial velocity until the downstream region. This shows that the simple eddy-viscosity representation of the Reynolds stress is not sufficient to capture the turbulent swirling ﬂow, and that helicity effect should be implemented into the Reynolds-stress expression. At this level of turbulence model arguments, the relevance of the inhomogeneous helicity effect was then well validated.

A second validation was performed at the level of DNSs featuring large-scale ﬂow generation by inhomogeneous helicity.\cite{yok2016b} A triply periodic box is set-up with two ($x$ and $y$) directions being homogeneous, and the other one ($z$) being the direction of inhomogeneity. An inhomogeneous turbulent helicity is externally injected by forcing helicity with a sinusoidal profile with respect to $z$ during the whole simulation period. In addition, uniform rotation is imposed in the $y$ direction as
\begin{equation}
	\mbox{\boldmath$\omega$}_{\rm{F}}
	= \left( {
	\omega_{\rm{F}}^x, \omega_{\rm{F}}^y, \omega_{\rm{F}}^z
	} \right)
	= \left( {
  	0, \omega_{\rm{F}}, 0
	} \right).
	\label{eq:YB_omega_F_setup}
\end{equation}
As for the initial conditions, there is no initial large-scale or mean flows. Under these conditions, DNSs show that a large-scale velocity is generated in the rotation (or $y$) direction as time 
\AP{evolves}. The large-scale ﬂow increases from zero at the early or developing stage from $0 \le t \le 400$. \NY{The time $t$ is normalized here by the correlation time of turbulence defined by $\tau_{\rm{turb}} = 1/(u_{\rm{rms}} k_{\rm{f}})$ ($u_{\rm{rms}}$: root mean square velocity, $k_{\rm{f}}$: forcing wave number)}. Afterwards, the ﬂow reaches a steady state with a well defined mean large-scale circulation.

The origin of this large-scale flow can be explained with the helicity turbulence model. At the early stage, when no large-scale flows are present, the $z$-$y$ component of the Reynolds stress in Eq.~(\ref{eq:rey_strss_model_exp}) can be expressed as
\begin{equation}
	\left\langle {u'^z u'^y} \right\rangle
	= \eta_{\rm{T}} \frac{\partial H}{\partial z} \omega_{\rm{F}}^y.
	\label{eq:YB_rey_strss_early}
\end{equation}
In Figure~2, we plot the spatial distribution of the Reynolds stress $\langle {u'^z u'^y} \rangle$ and the helicity effect $(\partial H/\partial z) \omega_{\rm{F}}^y$ in the DNSs.

\begin{figure}
\includegraphics[width=0.7\textwidth]{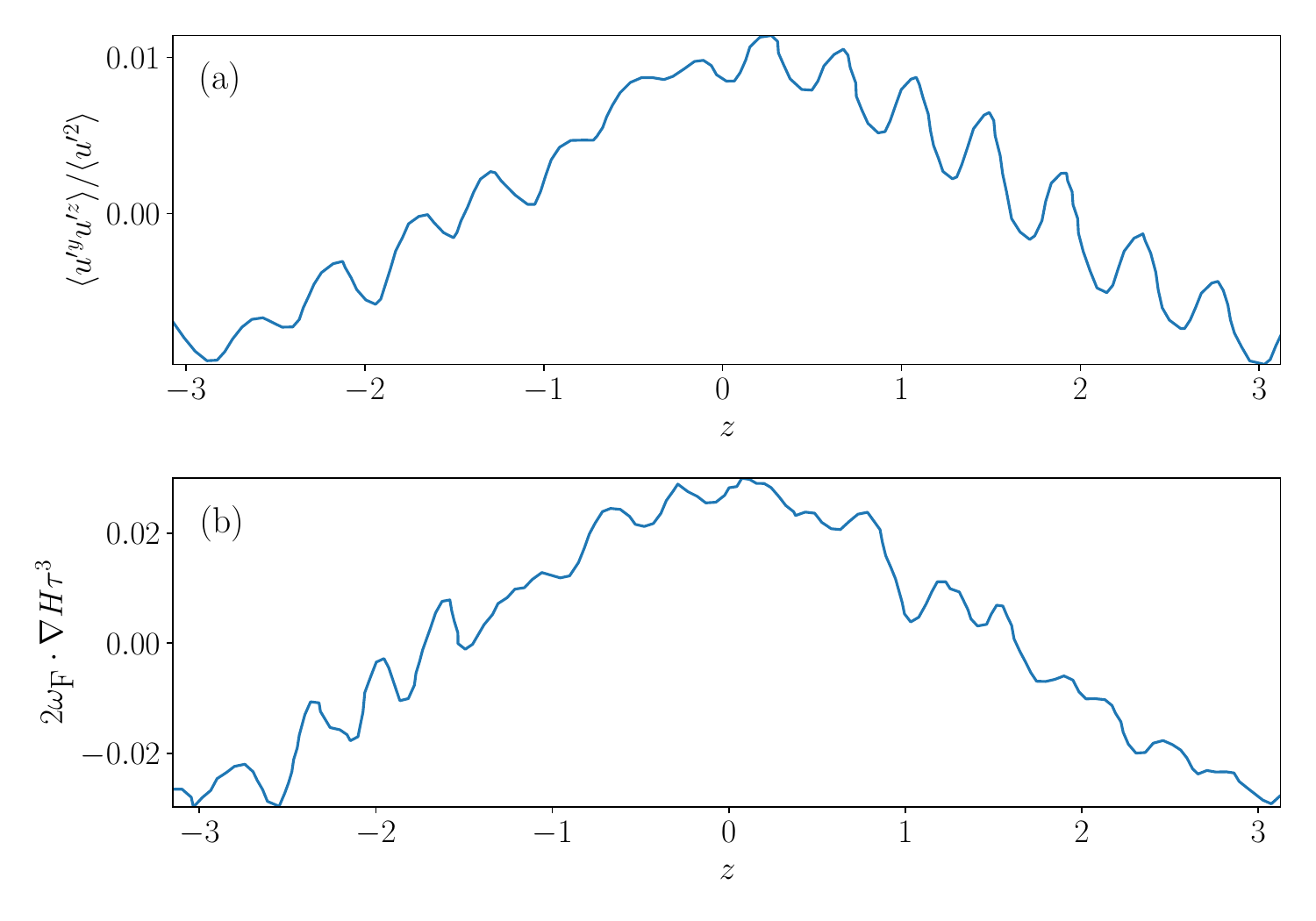}
\caption{\label{fig:rey_strss_balance_early}Spatial distributions of the Reynolds stress (a) and the inhomogeneous-helicity effect (b). The Reynolds stress $\langle {u'^y u'^z} \rangle$ is scaled by the turbulent energy. The turbulence timescale $\tau$ is deﬁned by $\tau = 1/(u_{\rm{rms}} k_{\rm{F}})$ with $u_{\rm{rms}}$ being the root mean square velocity, and $k_{\rm{F}}$ the forcing wavenumber. Redrawn from the data of \citet{yok2016b}}
\end{figure}

\begin{figure}
\includegraphics[width=0.7\textwidth]{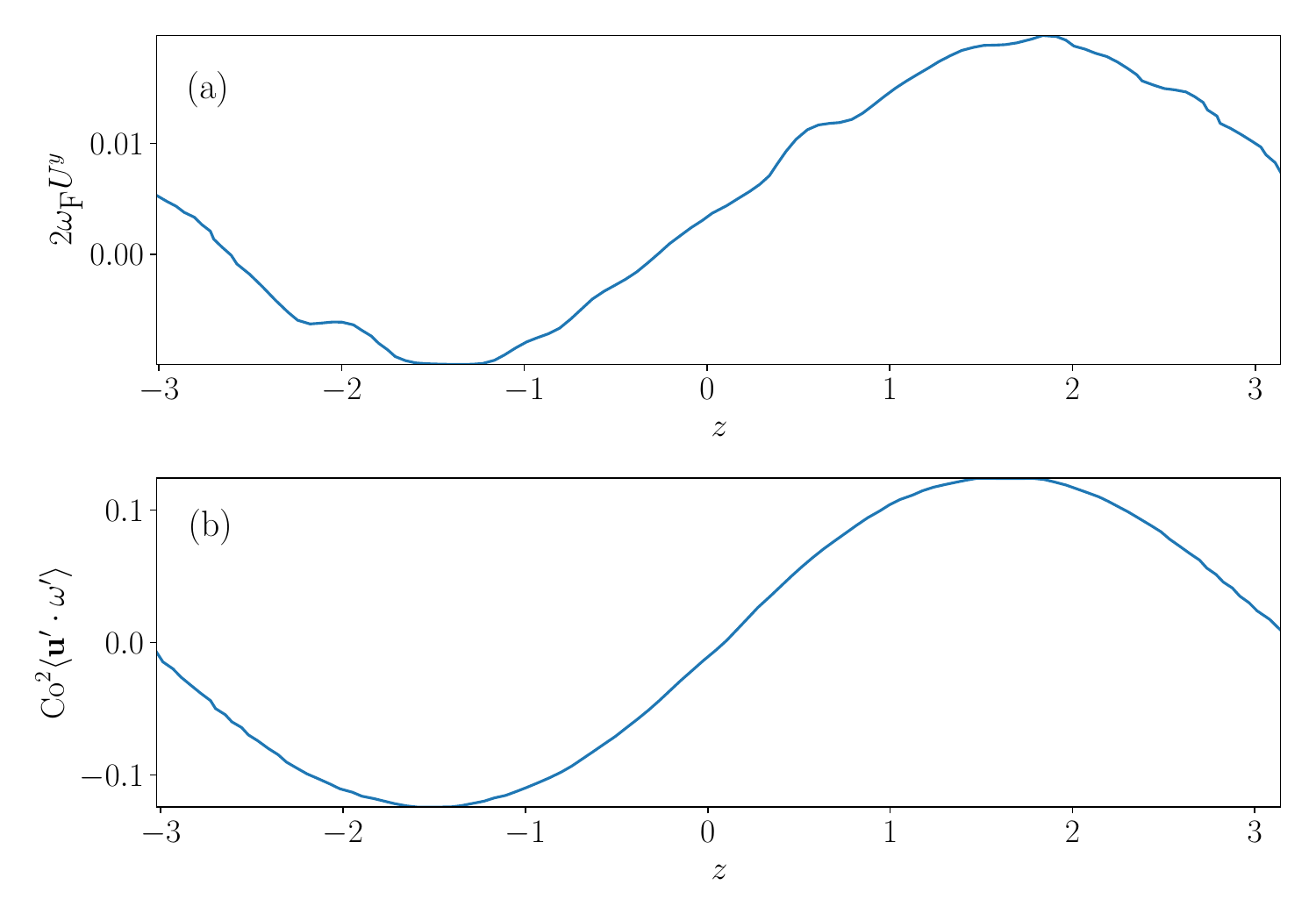}
\caption{\label{fig:rey_strss_bal_developed}Spatial distribution of the induced large-scale velocity $U^y$ (a) and the turbulent helicity $\langle {{\bf{u}}' \cdot \mbox{\boldmath$\omega$}'} \rangle$ injected by external forcing (b). The Coriolis number ${\rm{Co}}$ is deﬁned by ${\rm{Co}} = \omega_{\rm{F}} \tau$. Redrawn from the data of \citet{yok2016b}}
\end{figure}

At the later, or developed stage of evolution, the system reaches a stationary state, in which the eddy-viscosity and inhomogeneous-helicity effects balance with each other as
\begin{equation}
	0 \simeq - \nu_{\rm{T}} {\cal{S}}^{zy}
	+ \eta_{\rm{T}} \frac{\partial H}{\partial z} 2 \omega_{\rm{F}}^y.
	\label{eq:YB_rey_strss_balance_developed}
\end{equation}
Here, use has been made of the fact that, in this case, the system rotation is much larger than the relative vorticity ($\Omega^y \ll 2 \omega_{\rm{F}}^y$). Equation~(\ref{eq:YB_rey_strss_balance_developed}) is integrated with respect to $z$ as
\begin{equation}
	U^y \simeq \frac{\eta_{\rm{T}}}{\nu_{\rm{T}}} 2 \omega_{\rm{F}} H.
	\label{eq:YB_U-H_rel}
\end{equation}
Note that this results in the development of a large-scale flow $U^y$. Figure~\ref{fig:rey_strss_bal_developed} shows the induced mean velocity $U^y$ and the imposed turbulent helicity $H$. We can see a very high correlation between them. This indicates that, in agreement with the helicity turbulence model, at the developed stage of the flow evolution the eddy-viscosity effect is balanced by the inhomogeneous helicity effect.

This result shows that the balance between the eddy-viscosity and the inhomogeneous-helicity effect given by Eq.~(\ref{eq:YB_rey_strss_balance_developed}) is an important component to properly capture the evolution of inhomogeneous turbulent flows. Again, the inhomogeneous-helicity effect plays a crucial role expressing the Reynolds stress in this setup. Without the inhomogeneous-helicity effect, there is no effect that can counter-balance the eddy viscosity. These DNSs highlight the relevance of the inhomogeneous-helicity effect in the turbulent momentum transport in the presence of mean absolute vorticity.

Another validation was given by a numerical experiment of the helicity effect in the large-scale ﬂow generation in the context of the Reynolds-stress budget.\cite{ina2017} In this study, a triply periodic rectangular box was considered. The $y$ direction is the one of inhomogeneity, and the $z$ and $x$ are the directions of homogeneity. The turbulent helicity as well as the turbulent energy is locally injected into the region near the $y = 0$ plane by external forcing. In addition to the localized turbulent helicity injection, rotation is also imposed in the $x$ direction [$\mbox{\boldmath$\omega$}_{\rm{F}} = (\omega_{\rm{F}}^x, 0, 0)$]. A large-scale ﬂow is generated in the rotation (or $x$) direction. Such a large-scale ﬂow generation is observed only in the case in which both inhomogeneous helicity and rotation are present. Neither inhomogeneous-helicity injection nor system rotation by themselves are sufficient for large-scale ﬂow generation. By considering the transport or budget equation of the Reynolds stress ${\cal{R}}^{xy}$, it was shown that the Coriolis effect ${\cal{R}}^{xz} \omega_{\rm{F}}^x$ affects the difference between the pressure-diffusion effect $- (\partial/\partial y) \langle {p' u'^x} \rangle$ and the pressure-strain effect $2 \langle {p' s'^{xy}} \rangle$. This is natural since the system rotation $\mbox{\boldmath$\omega$}_{\rm{F}}$ effect can affect the fluctuating velocity evolution through the Coriolis force and the fluctuating pressure term. This study validated that the inhomogeneous-helicity effect is also relevant to explain large-scale ﬂow generation through the Reynolds stress.

Overall, all these numerical simulations conﬁrm that the inhomogeneous-helicity effect plays an essential role in the momentum transport through the Reynolds stress in the presence of system rotation and/or of large-scale vortical motions.

\section{\label{sec:III}Helicity SGS models }
\subsection{\label{sec:III.A}SGS model with helicity}

In the framework of LESs, we adopt a filter and divide a ﬁeld quantity $f$ into the GS and SGS components as
\begin{equation}
	f = \overline{f} + f'',
	\label{GS_SGS_decomp}
\end{equation}
where $\overline{f}$ is the filtered or GS component given by
\begin{equation}
	\overline{f}({\bf{x}};t)
	= \int d{\bf{y}}\ G(\Delta,{\bf{x}}-{\bf{y}}) f({\bf{y}},t),
	\label{eq:SGS_filtering}
\end{equation}
where now $G$ is a kernel of the spatial filter. The filtered ﬁeld $\overline{f}$ represents the components of $f$ at all scales larger than $\Delta$. On the other hand, $f'' = f - \overline{f}$ represents the small scales (SGS component) of the ﬁeld.

The GS velocity equations for the incompressible fluid are written as
\begin{equation}
	\left( {
		\frac{\partial}{\partial t} 
		+ \overline{\bf{u}} \cdot \nabla
	} \right) \overline{\bf{u}}
	= - \NY{\nabla} \overline{p}
	- \nabla \cdot \mbox{\boldmath$\tau$}_{\rm{SGS}}
	+ \nu \nabla^2 \cdot \overline{\bf{u}},
	\label{eq:GS_vel_eq}
\end{equation}
with the solenoidal condition
\begin{equation}
	\nabla \cdot \overline{\bf{u}} = 0,
	\label{eq:GS_sol_cond}
\end{equation}
where $\overline{p}$ is the GS pressure, and $\mbox{\boldmath$\tau$}_{\rm{SGS}} = \{ {\tau_{\rm{SGS}}^{ij}} \}$ is the SGS stress deﬁned by
\begin{equation}
	\tau_{\rm{SGS}}^{ij}
	= \overline{u^i u^j}
	- \overline{u}^i \overline{u}^j.
	\label{eq:SGS_stres_def}
\end{equation}
In the standard SGS model, the SGS stress is modeled as
\begin{equation}
	\tau_{\rm{SGS}}^{ij} 
	= \frac{2}{3} \delta^{ij} K_{\rm{S}}
	- \nu_{\rm{S}} \overline{s}^{ij},
	\label{eq:standard_SGS_strss_model}
\end{equation}
where $K_{\rm{S}} = (\overline{{\bf{u}}^2} - \overline{\bf{u}}^2)/2$ is the SGS energy, and $\overline{\bf{s}} = \{ {s^{ij}} \}$ is the GS velocity strain deﬁned by
\begin{equation}
	\overline{s}^{ij}
	= \frac{\partial \overline{u}^j}{\partial x^i}
	+ \frac{\partial \overline{u}^i}{\partial x^j}
	- \frac{1}{3} \delta^{ij} \nabla \cdot \overline{\bf{u}}
	\NY{= \frac{\partial \overline{u}^j}{\partial x^i}
	+ \frac{\partial \overline{u}^i}{\partial x^j}},
	\label{eq:GS_strain_rate_def}
\end{equation}
where again the last equality holds as the flow is incompressible. In the LES context, in Eq.~(\ref{eq:standard_SGS_strss_model}) the transport coefficient $\nu_{\rm{S}}$ is called the SGS viscosity. As already mentioned, in the Smagorinsky model, $\nu_{\rm{S}}$ is expressed in terms of the filter width $\Delta$ and the magnitude of the GS strain rate as\NY{\cite{sma1963,sag2006}}
\begin{equation}
	\nu_{\rm{S}} = (C_{\rm{S}} \Delta)^2 \overline{s},
	\label{eq:smag_model}
\end{equation}
where $\overline{s}$ is the magnitude of the \NY{GS} strain rate deﬁned by
\begin{equation}
	\overline{s} = \sqrt{(\overline{s}^{ij})^2/2},
	\label{eq:strain_rate_amplitude}
\end{equation}
and $C_{\rm{S}}$ is the Smagorinsky model constant.

On the basis of the helicity effect analyzed in Sec.~\ref{sec:II.C} in the framework of Reynolds averaging, we now construct some types of SGS turbulence models in which small or SGS turbulent structure effects are incorporated through the SGS helicity. Hereafter, we generally call them the helicity SGS (or H-SGS) models.
	
In the H-SGS models, the SGS stress is modeled in terms of the GS quantities as
\begin{equation}
	\tau_{\rm{SGS}}^{ij}
	= \frac{2}{3} K_{\rm{S}} \delta^{ij}
	- \nu_{\rm{S}} \overline{s}^{ij}
	+ \eta_{\rm{S}} \left[ { 
		\frac{\partial H_{\rm{S}}}{\partial x^i} \overline{\omega}_\ast^j
		+ \frac{\partial H_{\rm{S}}}{\partial x^j} \overline{\omega}_\ast^i
		- \frac{2}{3} \delta^{ij} 
		(\overline{\mbox{\boldmath$\omega$}}_\ast \cdot \nabla)
		H_{\rm{S}}
	} \right],
	\label{eq:SGS_strss_exp}
\end{equation}
where $\overline{\mbox{\boldmath$\omega$}}_\ast$ is the GS absolute vorticity defined by the GS relative vorticity $\overline{\mbox{\boldmath$\omega$}} = \nabla \times \overline{\bf{u}}$ and the system rotation angular velocity $\mbox{\boldmath$\omega$}_{\rm{F}}$ as
\begin{equation}
	\overline{\mbox{\boldmath$\omega$}}_\ast
	= \overline{\mbox{\boldmath$\omega$}}
	+ 2 \mbox{\boldmath$\omega$}_{\rm{F}}
	= \nabla \times \overline{\bf{u}}
	+ 2 \mbox{\boldmath$\omega$}_{\rm{F}}.
	\label{eq:GS_absolute_vort}
\end{equation}

In Eq.~(\ref{eq:SGS_strss_exp}) we have two transport coefficients, as we had before in the helical RANS case. The first one, $\nu_{\rm{S}}$, is the SGS viscosity, and $\eta_{\rm{S}}$ is the coefficient related to helicity terms. These transport coefficients are determined by the statistical properties of the SGS fields; for them will be given later. Also, in Eq.~(\ref{eq:SGS_strss_exp}) $K_{\rm{S}}$ is the SGS energy defined as before, and $H_{\rm{S}}$ is the SGS helicity; for completeness both are given by
\begin{equation}
	K_{\rm{S}}
	= (\overline{{\bf{u}}^2} - \overline{\bf{u}}^2)/2,
	\label{eq:SGS_K_def}
\end{equation}
\begin{equation}
	H_{\rm{S}}
	= \overline{{\bf{u}}\cdot \mbox{\boldmath$\omega$}} 
	- \overline{\bf{u}} \cdot \overline{\mbox{\boldmath$\omega$}}.
	\label{eq:SGS_H_def}
\end{equation}
\DR{In order to incorporate the SGS stress tensort $\mbox{\boldmath$\tau$}_{\rm{SGS}} = \{ {\tau_{\rm{SGS}}^{ij}} \}$ given by Eq.~(\ref{eq:SGS_strss_exp}) into solutions of Eq.~(\ref{eq:GS_vel_eq}), we must evaluate the SGS energy $K_{\rm{S}}$ and the SGS helicity $H_{\rm{S}}$, as well as the transport coefficients $\nu_{\rm{S}}$ and $\eta_{\rm{S}}$ in a self-consistent manner. Depending on how we evaluate these SGS statistical quantities and transport coefficients, we construct different types of helicity SGS models.} 
\AP{We explore} them one by one 
\AP{in the following sub-sections.}

\subsection{\label{sec:III.B}Two-equation helicity SGS model}

We want to simplify this set of equations to construct models with two, one, or zero transport equations for the SGS quantities. 
\NY{In the two- and one-equation models, in addition to solving the GS momentum Eq.~(\ref{eq:GS_vel_eq}) with the SGS stress given by Eq.~(\ref{eq:SGS_strss_exp}), the transport equations of the SGS statistical quantities $K_{\rm{S}}$ and/or $H_{\rm{S}}$ are solved. Apart from the SGS helicity transport equation, a one-equation model based on the SGS energy transport equation was first presented by \citet{sch1975} and applied to plane channels and annuli. In the presence of several production mechanisms of SGS energy, including buoyancy and velocity shear, the one-equation type SGS models are expected to work well, specifically in the meteorological field. However, it was also pointed out that even solving the $K_{\rm{S}}$ transport equation with fixed model constants could not lead to results consistent with observations in mixing-layer and channel flows.\cite{oka1996} This indicates that the improvement of the SGS-viscosity representation can be of essential importance.}

In the particular case of the two-equation H-SGS model, we solve two transport equations for the SGS energy $K_{\rm{S}}$ and the SGS helicity $H _{\rm{S}}$. In Eq.~(\ref{eq:SGS_strss_exp}), the SGS viscosity $\nu_{\rm{S}}$ and the transport coefficient of the absolute vorticity $\eta_{\rm{S}}$, apart from the SGS helicity gradient $\nabla H_{\rm{S}}$, are expressed in terms of $\Delta$ and $K_{\rm{S}}$ as
\begin{equation}
	\nu_{\rm{S}}
	= \nu_{\rm{S}} ( {\Delta, K_{\rm{S}}} )
	= C_{\nu{\rm{S}}} \Delta K_{\rm{S}}^{1/2},
	\label{eq:nuS_in_Del_KS}
\end{equation}
\begin{equation}
	\eta_{\rm{S}}= \eta_{\rm{S}} ( {\Delta, K_{\rm{S}}} )
	= C_{\eta{\rm{S}}} \Delta^2 \tau_{\rm{S}}
	= C_{\eta{\rm{S}}} \Delta^2 (\Delta / K_{\rm{S}}^{1/2})
	= C_{\eta{\rm{S}}} \Delta^3 K_{\rm{S}}^{-1/2},
	\label{eq:etaS_in_Del_KS}
\end{equation}
where \NY{$\tau_{\rm{S}} = \Delta / K_{\rm{S}}^{1/2}$ is the timescale of the SGS turbulent component,} and $C_{\nu{\rm{S}}}$ and $C_{\eta {\rm{S}}}$ are model constants.

As already mentioned, in addition to the GS velocity equation, we solve the SGS energy and the SGS helicity transport equations. The SGS energy equation is written as
\begin{equation}
	\left( {
		\frac{\partial}{\partial t} 
		+ \overline{\bf{u}} \cdot \nabla
	} \right) K_{\rm{S}}
	= P_{\rm{KS}}
	- \varepsilon_{\rm{KS}}
	+ T_{\rm{KS}},
	\label{eq:SGS_KS_eq_in_two}
\end{equation}
where $P_{\rm{KS}}$ is the SGS energy production rate, $\varepsilon_{\rm{KS}}$ is its dissipation rate, and $T_{\rm{KS}}$ is the SGS energy transport rate. They are defined or modeled as
\begin{equation}
	P_{\rm{KS}}
	= - {\tau_{\rm{SGS}}^{ab}} \frac{\partial \overline{u}^b}{\partial x^a},
	\label{eq:SGS_PKS_def_in_two}
\end{equation}
\begin{subequations}\label{eq:SGS_epsKS_def_two}
\begin{eqnarray}
	\varepsilon_{\rm{KS}}
	&=& \nu \left( {
		\overline {
		\frac{\partial u^j}{\partial x^i}
		\frac{\partial u^j}{\partial x^i}
		}
	- \frac{\partial \overline{u}^j}{\partial x^i}
		\frac{\partial \overline{u}^j}{\partial x^i}
	} \right)\\
	\label{eq:SGS_epsKS_def_in_two}
	&\approx& C_{\varepsilon{\rm{S}}} \frac{K_{\rm{S}}^{3/2}}{\Delta},
	\label{eq:SGS_epsKS_approx_in_two}
\end{eqnarray}
\end{subequations}
\begin{equation}
	T_{\rm{KS}}
	\approx \nabla \cdot \left( {
		\frac{\nu_{\rm{S}}}{\sigma_{\rm{S}}}
		\nabla K_{\rm{S}}
	} \right),
	\label{eq:SGS_TKS_def_two}
\end{equation}
where $C_{\varepsilon {\rm{S}}}$ and $\sigma_{\rm{S}}$ are model constants. 

On the other hand, the SGS helicity transport equation is written as
\begin{equation}
	\left( {
		\frac{\partial}{\partial t} 
		+ \overline{\bf{u}} \cdot \nabla
	} \right) H_{\rm{S}}
	= P_{\rm{HS}}
	- \varepsilon_{\rm{HS}}
	+ T_{\rm{HS}},
	\label{eq:SGS_HS_eq_in_two}
\end{equation}
where $P_{\rm{HS}}$ is the SGS helicity production rate, $\varepsilon_{\rm{HS}}$ is its dissipation rate, and $T_{\rm{HS}}$ is the SGS helicity transport rate. They are defined or modeled as
\begin{equation}
	P_{\rm{HS}}
	= - {\tau_{\rm{SGS}}^{ij}} \frac{\partial \overline{\omega}^j_\ast}{\partial x^i}
	+ \frac{\partial \tau_{\rm{SGS}}^{ij}}{\partial x^i} \overline{\omega}^j_\ast,
	\label{eq:SGS_PHS_def_in_two}
\end{equation}
\begin{subequations}\label{SGS_epsHS_def_two}
\begin{eqnarray}
	\varepsilon_{\rm{HS}}
	&= &2\nu \left( {
		\overline {
		\frac{\partial u^j}{\partial x^i}
		\frac{\partial \omega^j}{\partial x^i}
	}
 	- \frac{\partial \overline{u}^j}{\partial x^i}
		\frac{\partial \overline{\omega}^j}{\partial x^i}
	} \right)
	\label{eq:SGS_epsHS_def_in_two}\\
	&\approx& C_{\varepsilon \rm{HS}} \frac{H_{\rm{S}}}{K_{\rm{S}}/\varepsilon_{\rm{KS}}}
	= C_{\varepsilon{\rm{HS}}}
		\frac{\varepsilon_{\rm{KS}} H_{\rm{S}}}{K_{\rm{S}}},
	\label{eq:SGS_epsHS_approx_in_two}
\end{eqnarray}
\end{subequations}
\begin{equation}
	T_{\rm{HS}}
	\approx \nabla \cdot \left( {
		\frac{\nu_{\rm{S}}}{\sigma_{\rm{HS}}} \nabla H_{\rm{S}}
	} \right),
	\label{eq:SGS_THS_def_in_two}
\end{equation}
where $C_{\varepsilon{\rm{HS}}}$ and $\sigma_{\rm{HS}}$ are model constants.

\NY{It is worth noting that the production rate in the equation for the GS helicity has exactly the same two terms as in Eq.~(\ref{eq:SGS_PHS_def_in_two}) but with opposite signs. In this sense, $P_{\rm{HS}}$ represents the cascade of helicity between GS and SGS.} In other words, the production of SGS helicity of one sign results from the destruction of GS helicity with the same sign at larger scales. It is also worth noting that a part of the production rate $P_{\rm{HS}}$ in Eq.~(\ref{eq:SGS_PHS_def_in_two}) plays an important role in helicity generation due to inhomogeneities along the direction of rotation. Substituting Eq.~(\ref{eq:SGS_strss_exp}) into the second term of Eq.~(\ref{eq:SGS_PHS_def_in_two}), we have a contribution of the SGS energy inhomogeneity to the SGS helicity generation given by
\begin{equation}
	\frac{\partial \tau_{\rm{SGS}}^{ij}}{\partial x^i}
		\overline{\omega}_\ast^j
	\simeq \overline{\omega}_\ast^j \frac{\partial}{\partial x^i} \left( {
		\frac{2}{3} \delta^{ij} K_{\rm{S}}
	} \right)
	= \frac{2}{3} \left( {
		\overline{\mbox{\boldmath$\omega$}}_\ast
		\cdot \nabla
	} \right) K_{\rm{S}},
	\label{eq:SGS_KS_prod_mech}
\end{equation}
which suggests that SGS helicity can be generated by SGS energy gradients $\nabla K_{\rm{S}}$ along the mean absolute vorticity $\mbox{\boldmath$\omega$}_\ast$ direction.

\subsection{\label{sec:III.C}One-equation helicity model}

If we have a proper evaluation either of $K_{\rm{S}}$ or $H_{\rm{S}}$, we do not need to solve both of the transport Eqs.~(\ref{eq:SGS_KS_eq_in_two}) and (\ref{eq:SGS_HS_eq_in_two}). Compared with the SGS helicity $H_{\rm{S}}$, the behavior and properties of the SGS energy $K_{\rm{S}}$ is better understood. If we assume local equilibrium of $K_{\rm{S}}$ production, $P_{\rm{KS}}$, and dissipation $\varepsilon_{\rm{KS}}$ in Eq.~(\ref{eq:SGS_epsKS_def_two}),
\begin{equation}
	P_{\rm{KS}} = \varepsilon_{\rm{KS}},
	\label{eq:SGS_KS_local_equil}
\end{equation}
then we have
\begin{equation}
	- {\tau_{\rm{SGS}}^{ij}} \frac{\partial \overline{u}^j}{\partial x^i}
	\simeq \varepsilon_{\rm{KS}}.
	\label{eq:SGS_PKS_epsS_local_equil}
\end{equation}
Substituting $\mbox{\boldmath$\tau$}_{\rm{SGS}}$ [Eq.~(\ref{eq:standard_SGS_strss_model})] and $\varepsilon_{\rm{KS}}$ [Eq.~(\ref{eq:SGS_epsKS_approx_in_two})] into Eq.~(\ref{eq:SGS_PKS_epsS_local_equil}), we have
\begin{equation}
	\nu_{\rm{S}} \overline{s}^2
	= C_{\varepsilon{\rm{S}}} 
	\frac{K_{\rm{S}}^{3/2}}{\Delta}.
	\label{eq:SGS_KS_local_equil_model}
\end{equation}
Eliminating $K_{\rm{S}}$ by Eq.~(\ref{eq:nuS_in_Del_KS}) and solving with respect to $\nu_{\rm{S}}$, we have
\begin{equation}
	\nu_{\rm{S}} 
	= \left( {
		\frac{C_{\nu{\rm{S}}}^3}{C_{\varepsilon{\rm{S}}}}
	} \right)^{1/2} \Delta^2 \overline{s}.
	\label{eq:SGS_nuS_in_Del_s_one}
\end{equation}
This is equivalent to the Smagorinsky model [Eq.~(\ref{eq:smag_model})] for the SGS viscosity with
\begin{equation}
	C_{\rm{S}} 
	= \left( {
		\frac{C_{\nu{\rm{S}}}^3}{C_{\varepsilon{\rm{S}}}}
	} \right)^{1/4}.
	\label{eq:CS_CnuS_Ceps}
\end{equation}

In the one-equation model we derive here, $K_{\rm{S}}$, its dissipation rate $\varepsilon_{\rm{KS}}$, and the helicity-related coefficient $\eta_{\rm{S}}$ are expressed in terms of the filter width $\Delta$ and the magnitude of the GS velocity strain rate $\overline{s}$ as
\begin{equation}
	K_{\rm{S}}
	= K_{\rm{S}}( {\Delta, \overline{s}} )
	= \left( {\frac{\nu_{\rm{S}}}{C_{\nu{\rm{S}}} \Delta}} \right)^2
	= C_{\rm{KS}} \Delta^2 \overline{s}^2,
	\label{eq:SGS_KS_in_Del_s_one}
\end{equation}
\begin{equation}
	\varepsilon_{\rm{KS}}
	= \varepsilon_{\rm{KS}}( {\Delta, \overline{s}} )
	\simeq - \tau_{\rm{SGS}}^{ij} 
		\frac{\partial \overline{u}^j}{\partial x^i}
	\simeq \nu_{\rm{S}} \overline{s}^2
	= (C_{\rm{S}} \Delta)^2 \overline{s}^3,
	\label{eq:SGS_epsS_in_Del_s_one}
\end{equation}
\begin{equation}
	\eta_{\rm{S}}
	= \eta_{\rm{S}}( {\Delta, \overline{s}} )
	= C_{\eta{\rm{S}}} \Delta^3 K_{\rm{S}}^{-1/2}
	= C_{\rm{HS}} \Delta^2 / \overline{s},
	\label{eq:etaS_in_Del_s_one}
\end{equation}
with $C_{\rm{KS}} = (C_{\varepsilon{\rm{S}}} C_{\nu{\rm{S}}})^{-1/2}$ and $C_{\rm{HS}} = C_{\eta{\rm{S}}} C_{\varepsilon{\rm{S}}}^{1/2}/C_{\nu{\rm{S}}}^{1/2}$. From Eqs.~(\ref{eq:SGS_KS_in_Del_s_one}) and (\ref{eq:SGS_epsS_in_Del_s_one}), the turbulence timescale is expressed as
\begin{equation}
	\tau_{\rm{S}}
	= \frac{K_{\rm{S}}}{\varepsilon_{\rm{KS}}}
	= \frac{C_{\rm{KS}}}{C_{\rm{S}}^2} \overline{s}^{-1}.
	\label{eq:SGS_tauS_in_s_one}
\end{equation}

In the one-equation model, in addition to the GS velocity given by Eq.~(\ref{eq:GS_vel_eq}), the transport equation of $H_{\rm{S}}$ [Eq.~(\ref{eq:SGS_HS_eq_in_two})] should be simultaneously solved. In the $H_{\rm{S}}$ equation, the dissipation rate $\varepsilon_{\rm{HS}}$ is given by the SGS helicity $H_{\rm{S}}$ divided by the SGS timescale $\tau_{\rm{S}}$ as
\begin{equation}
	\varepsilon_{\rm{HS}}
	= \varepsilon_{\rm{HS}} ( {\Delta,\overline{s},H_{\rm{S}}} )
	= C_{\varepsilon {\rm{H}}} \frac{H_{\rm{S}}}{\tau_{\rm{S}}}
	= \frac{C_{\varepsilon {\rm{H}}} C_{\rm{S}}^2}{C_{\rm{KS}}} H_{\rm{S}} \overline{s},
	\label{eq:SGS_epsHS_inHS_s_one}
\end{equation}
and the transport rate $T_{\rm{HS}}$ is given from Eqs.~(\ref{eq:SGS_THS_def_in_two}) and (\ref{eq:SGS_KS_prod_mech}) as
\begin{eqnarray}
	T_{\rm{HS}}
	&=& T_{\rm{HS}} ( {\Delta,\overline{s}, \overline{\mbox{\boldmath$\omega$}}_\ast, H_{\rm{S}}} )
	\nonumber\\
	&=& \nabla \cdot \left( {
		\mbox{\boldmath$\tau$}_{\rm{SGS}}
		\overline{\mbox{\boldmath$\omega$}_\ast}
	} \right)
	+ \nabla \cdot \left( {
		\frac{\nu_{\rm{S}}}{\sigma_{\rm{HS}}} \nabla H_{\rm{S}}
	} \right)
	\simeq \frac{2}{3} \overline{\mbox{\boldmath$\omega$}_\ast} \cdot \nabla K_{\rm{S}}
	+ \nabla \cdot \left( {
		\frac{\nu_{\rm{S}}}{\sigma_{\rm{HS}}} \nabla H_{\rm{S}}
	} \right)
	\nonumber\\
	&=& \frac{2}{3} C_{\rm{KS}} \Delta^2
		\overline{\mbox{\boldmath$\omega$}_\ast} \cdot \nabla \overline{s}^2
	+ \nabla \cdot \left( {
		\frac{\nu_{\rm{S}}}{\sigma_{\rm{HS}}} \nabla H_{\rm{S}}
	} \right).
	\label{eq:SGS_THS_in_Del_s_HS_one}
\end{eqnarray}

\subsection{\label{sec:III.D}Zero-equation helicity model}

The Smagorinsky model is one of the simplest SGS models in LESs. In this model, no transport equations are solved other than the GS momentum equation. In other words, the Smagorinsky model is a zero-equation SGS model. In order to construct a SGS model at the same level as the Smagorinsky model, we have to express the SGS helicity in terms of $\Delta$ and the GS quantities. In this class of modeling, the SGS statistical quantities and the SGS transport coefficient should be expressed in terms of the filter width $\Delta$ and the GS field quantities: the GS velocity strain rate $\overline{\mbox{\boldmath$s$}}$, its magnitude $\overline{s}$, and the GS (absolute) vorticity $\overline{\mbox{\boldmath$\omega$}}_\ast$. For the SGS energy $K_{\rm{S}}$ and SGS viscosity $\nu_{\rm{S}}$, and the absolute vorticity-related coefficient $\eta_{\rm{S}}$, we use $K_{\rm{S}}$ [Eq.~(\ref{eq:SGS_KS_in_Del_s_one})], $\nu_{\rm{S}}$ [Eq.~(\ref{eq:smag_model})], and $\eta_{\rm{S}}$ [Eq.~(\ref{eq:etaS_in_Del_s_one})], respectively. For the SGS helicity, we further assume local equilibrium in the SGS helicity evolution, namely, that the production rate $P_{\rm{HS}}$ [Eq.~(\ref{eq:SGS_PHS_def_in_two})] and its dissipation rate $\varepsilon_{\rm{HS}}$ [Eq.~(\ref{SGS_epsHS_def_two})] balance with each other as
\begin{equation}
	P_{\rm{HS}}
	\simeq \varepsilon_{\rm{HS}}.
	\label{SGS_HS_local_equil_zero}
\end{equation}
From Eqs.~(\ref{eq:SGS_PHS_def_in_two}) and (\ref{eq:SGS_epsHS_inHS_s_one}), we then have
\begin{equation}
	- \tau_{\rm{SGS}}^{ij} 
	\frac{\partial \overline{\omega}_\ast^j}{\partial x^i}
	+ \overline{\omega}_\ast^i
	\frac{\partial \tau_{\rm{SGS}}^{ij}}{\partial x^j}
	\simeq \frac{C_{\rm{\varepsilon H}} C_{\rm{S}}^2}{C_{\rm{KS}}} \overline{s} H_{\rm{S}}.
	\label{eq:SGS_PHS_epsH_balance_zero}
\end{equation}
With the expression of the SGS stress tensor $\mbox{\boldmath$\tau$}_{\rm{SGS}}$ in Eq.~(\ref{eq:standard_SGS_strss_model}), we have
\begin{eqnarray}
	H_{\rm{S}}
	&\simeq& \frac{C_{\rm{KS}}}{C_{\rm{\varepsilon H}} C_{\rm{S}}^2} \frac{1}{\overline{s}} 
	\left( {
		- \tau_{\rm{SGS}}^{ij} 
		\frac{\partial \overline{\omega}_\ast^j}{\partial x^i}
		+ \overline{\omega}_\ast^i
	\frac{\partial \tau_{\rm{SGS}}^{ij}}{\partial x^j}
	} \right)
	\nonumber\\
	&\simeq& \frac{C_{\rm{KS}}}{C_{\rm{\varepsilon H}} C_{\rm{S}}^2} \frac{1}{\overline{s}} 
	\left[ {
	\nu_{\rm{S}} \overline{s}^{ij} 
	\frac{\partial \overline{\omega}_\ast^j}{\partial x^i}
	+ \overline{\omega}_\ast^i
	\frac{\partial}{\partial x^j}\nu_{\rm{S}} \overline{s}^{ij}
	+ \frac{2}{3} \left( {
		\overline{\mbox{\boldmath$\omega$}}_\ast \cdot \nabla
	} \right) K_{\rm{S}}
	} \right].
	\label{eq:SGS_HS_eval_zero}
\end{eqnarray}
Using $\nu_{\rm{S}} ( {\Delta,\overline{s}} )$ [Eq.~(\ref{eq:smag_model})] and $K_{\rm{S}} ( {\Delta,\overline{s}} )$ [Eq.~(\ref{eq:SGS_KS_in_Del_s_one})], we have an estimation of $H_{\rm{S}}$ as
\begin{eqnarray}
	H_{\rm{S}}
	&\simeq& \frac{C_{\rm{KS}}}{C_{\rm{\varepsilon H}} C_{\rm{S}}^2} \frac{1}{\overline{s}} 
	\left[ {
		(C_{\rm{S}} \Delta)^2 \left( {
			\overline{s}\ \overline{s}^{ij} 
			\frac{\partial \overline{\omega}_\ast^j}{\partial x^i}
			+ \overline{\omega}_\ast^i
			\frac{\partial}{\partial x^j} \overline{s}\ \overline{s}^{ij}
		} \right)
	+ \frac{2}{3} C_{\rm{KS}} \Delta^2 \left( {
			\overline{\mbox{\boldmath$\omega$}}_\ast \cdot \nabla
		} \right) \overline{s}^2
	} \right]
	\nonumber\\
	&=& \frac{1}{C_{\rm{\varepsilon H}}} \frac{\Delta^2}{\overline{s}} 
	\left[ {
		C_{\rm{KS}} \left( {
		\overline{s}\ \overline{s}^{ij} 
		\frac{\partial \overline{\omega}_\ast^j}{\partial x^i}
	+ \overline{\omega}_\ast^i
		\frac{\partial}{\partial x^j} \overline{s}\ \overline{s}^{ij}
	} \right)
	+ \frac{2}{3} \left( {
		\frac{C_{\rm{KS}}}{C_{\rm{S}}}
	} \right)^2 \left( {
		\overline{\mbox{\boldmath$\omega$}}_\ast \cdot \nabla
	} \right) \overline{s}^2
	} \right].
	\label{eq:SGS_HS_eval_Del_s_zero}
\end{eqnarray}

In summary, the zero-equation helicity SGS model is constituted of the GS velocity Eq.~(\ref{eq:GS_vel_eq}) with the SGS stress $\mbox{\boldmath$\tau$}_{\rm{SGS}}$ [Eq.~(\ref{eq:SGS_strss_exp})]. The transport coefficients $\nu_{\rm{S}}$ and $\eta_{\rm{S}}$, and the SGS quantities $K_{\rm{S}}$ and $H_{\rm{S}}$ are given by
\begin{equation}
	\nu_{\rm{S}} 
	= \nu_{\rm{S}} ( {\Delta,\overline{s}} )
	= (C_{\rm{S}} \Delta)^2 \overline{s},
	\label{eq:SGS_nuS_zero}
\end{equation}
\begin{equation}
	\eta_{\rm{S}}
	= \eta_{\rm{S}} ( {\Delta, \overline{s}} )
	= C_{\rm{HS}} \Delta^2 / \overline{s},
	\label{eq:SGS_etaS_zero}
\end{equation}
\begin{equation}
	K_{\rm{S}}
	= K_{\rm{S}} ( {\Delta, \overline{s}} )
	= \left( {\frac{\nu_{\rm{S}}}{C_{\nu{\rm{S}}} \Delta}} \right)^2
	= C_{\rm{KS}} \Delta^2 \overline{s}^2,
	\label{eq:SGS_KS_zero}
\end{equation}
\begin{eqnarray}
	H_{\rm{S}}
	&=& H_{\rm{S}} ( {
		\Delta,\overline{s},\overline{\mbox{\boldmath$s$}},\overline{\mbox{\boldmath$\omega$}}_\ast
		} )
	\nonumber\\
	&=& \frac{1}{C_{\rm{\varepsilon H}}} \frac{\Delta^2}{\overline{s}} 
	\left[ {
		C_{\rm{KS}} \left( {
		\overline{s}\ \overline{s}^{ij} 
		\frac{\partial \overline{\omega}_\ast^j}{\partial x^i}
	+ \overline{\omega}_\ast^i
		\frac{\partial}{\partial x^j} \overline{s}\ \overline{s}^{ij}
		} \right)
	+ \frac{2}{3} \left( {
		\frac{C_{\rm{KS}}}{C_{\rm{S}}}
		} \right)^2 \left( {
		\overline{\mbox{\boldmath$\omega$}}_\ast \cdot \nabla
		} \right) \overline{s}^2
	} \right],
	\label{eq:SGS_HS_eval_zero_sum}
\end{eqnarray}
where $C_{\rm{S}}$, $C_{\rm{KS}}$, $C_{\rm{HS}}$, and $C_{\varepsilon{\rm{H}}}$ are model constants, whose values are given by comparison of the model with DNSs, and should be fixed for all the possible applications of the model. Instead of adjustable parameters, they are expected to be universal constants.

\section{\label{sec:IV}Numerical set-up and basic statistics of direct numerical simulations}

The validity of the helicity effect in the SGS models can be examined with the aid of direct numerical simulations (DNSs). To this end, we compare the results of the model calculations with those of the DNSs, filtered to separate GS and SGS components.

We perform DNSs in a triply periodic box with the Geophysical High-Order Suit for Turbulence (GHOST) code.\cite{min2011,ros2020} GHOST is an accurate and highly scalable pseudo-spectral code, which uses a hybrid parallelisation method combining MPI and Open MP, and with support for GPUs. The code solves a variety of partial differential equations encountered in studies of turbulence, including rotating and/or stratiﬁed turbulence, MHD, and Hall-MHD turbulence, particle diffusion in turbulence, quantum turbulence, etc. In order to validate turbulence models, we implement several types of filters into GHOST, and decompose a ﬁeld quantity $f$ into the filtered or GS component $\overline{f}$, and unfiltered or SGS component $f''$ as in Eq.~(\ref{GS_SGS_decomp}). The implemented filters include the box or top-hat filter, the Gaussian filter, the spectral or sharp cut filter, and the Lorentzian filter. In the following we present results using the Gaussian filter deﬁned as
\begin{equation}
	G(x-\xi) 
	= \left( {
		\frac{\gamma}{\pi \Delta^2}
	} \right)
	\exp\left( {
		-\frac{\gamma |x - \xi|^2}{\Delta^2}
	} \right),
	\label{eq:Gaussian_fil_config}
\end{equation}
\begin{equation}
	\widehat{G}(k)
	= \exp \left( {- \frac{\Delta^2 k^2}{4\gamma}} \right),
	\label{eq:Gaussian_fil_wave}
\end{equation}
where $\gamma=6$. Note that for the Gaussian filter, both the convolution kernel $G(x-\xi)$ and the transfer function $\widehat{G}(k)$ are written in the same exponential form.

\begin{figure}
\includegraphics[width=0.6\textwidth]{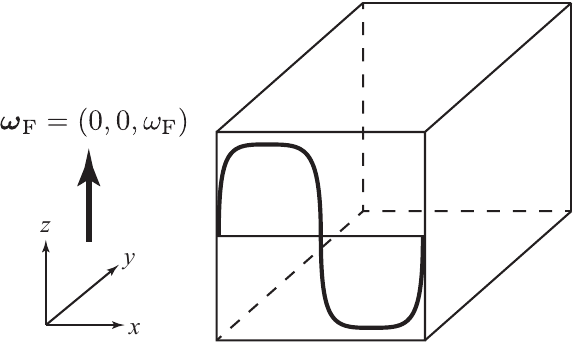}
\caption{\label{fig:dns_setup}Schematic representation of the simulation domain with the Cartesian coordinates, the mean inhomogeneous helicity profile $H(x)$ indicated by the thick black line, and the direction of rotation $\mbox{\boldmath$\omega$}_{\rm{F}}$ indicated by the thick arrow.}
\end{figure}

A mechanical forcing is applied in the simulations with correlation at the forcing wave number $k_{\rm{F}} = 5$, using the method to inject helicity in the flow described in \citet{pou1978}. In particular, here we generate two random forcing functions. One of them, ${\bf f}^{(1)}$, has positive helicity while the other, ${\bf f}^{(2)}$, has the same absolute mean helicity but with negative sign. Both are homogeneous in space. Then, a forcing with a large-scale hyperbolic tangent modulation of the helicity in the $x$ direction is generated as (see also Fig.~\ref{fig:dns_setup})
\begin{equation}
    {\bf f} = \mathcal{P} \left\{ \Theta(x) {\bf f}^{(1)} + \left[1-\Theta(x)\right] {\bf f}^{(2)} \right\} ,
	\label{eq:DNS_forcing}
\end{equation}
where $\mathcal{P}$ is the projector that ensures the total forcing remains solenoidal, and $\Theta(x) = [\tanh(\alpha \sin x)+1]/2$ where $\alpha$ controls the slope of the hyperbolic tangent modulation.

\begin{figure}
\includegraphics[width=0.9\textwidth]{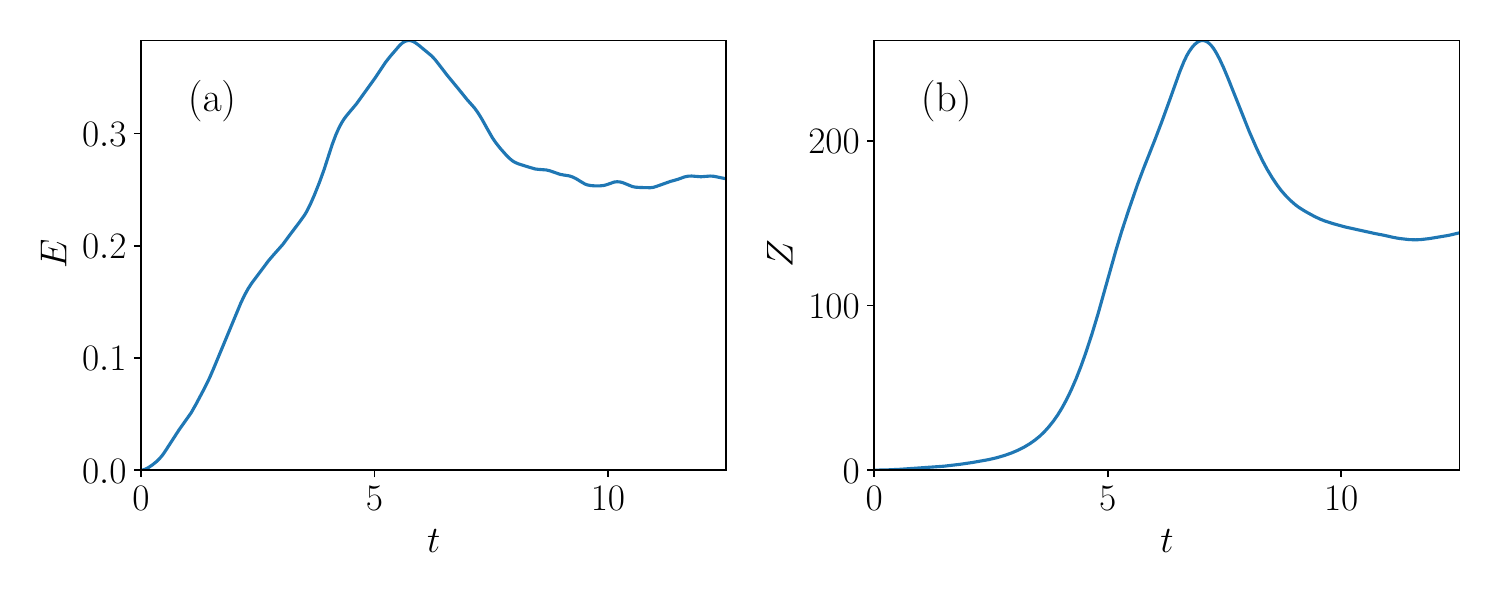}
\caption{\label{fig:kin_en_enstrophy}Time evolutions of (a) the kinetic energy $E$, and (b) the enstrophy $Z$, in the simulation without rotation. After the flow reaches the turbulent steady state, we keep integrating the flow and analyze GS and SGS quantities.}
\end{figure}

We perform simulations using this forcing without and with rotation, using a grid with $1024^3$ points. The case without rotation corresponds, in a way, to homogeneous isotropic turbulence, although the turbulence helicity is injected with the large-scale inhomogeneity previously described. In the simulation with rotation, the net angular velocity is set as
\begin{equation}
	\mbox{\boldmath$\omega$}_{\rm{F}}
	= \left( {
		\omega_{\rm{F}}^{(x)},
		\omega_{\rm{F}}^{(y)},
		\omega_{\rm{F}}^{(z)}
	} \right)
	= \left( {
		0,
		0,
		\omega_{\rm{F}}
	} \right),
	\label{DNS_omegaF_setup}
\end{equation}
with $\omega_{\rm{F}} = 8.0$. As for the SGS filtering, the validity of helicity SGS model may depend on the filter width $\Delta = 2\pi/k_{\rm{C}}$. To examine such a dependence, we adopt two filter widths in each simulation. The wavenumber of the first filter considered is $k_{\rm{C}} = 7$, and that of the second filter is $k_{\rm{C}} = 14$ (compare these values with $k_\textrm{F}=5$).

Figure \ref{fig:kin_en_enstrophy} shows the temporal evolution of the mean turbulent energy $E = V^{-1} \int_V {\bf{u}}^2/2 \, dV$ and of the mean enstrophy $Z = V^{-1} \int_V \mbox{\boldmath$\omega$}^2 \, dV$ in the simulation without rotation. As a result of the forcing, the turbulent energy increases up to some maximum value about $t \approx 6$. The enstrophy follows the energy evolution and reaches its maximum value around $t \approx 7$. Then the system starts relaxing and reaches a turbulent steady state around $t \approx 9$. For the analysis that follows, we use data stemming from the integration of the flow after this steady state is reached, by continuing the simulation for ten large-scale turnover times.

\begin{figure}
\includegraphics[width=0.9\textwidth]{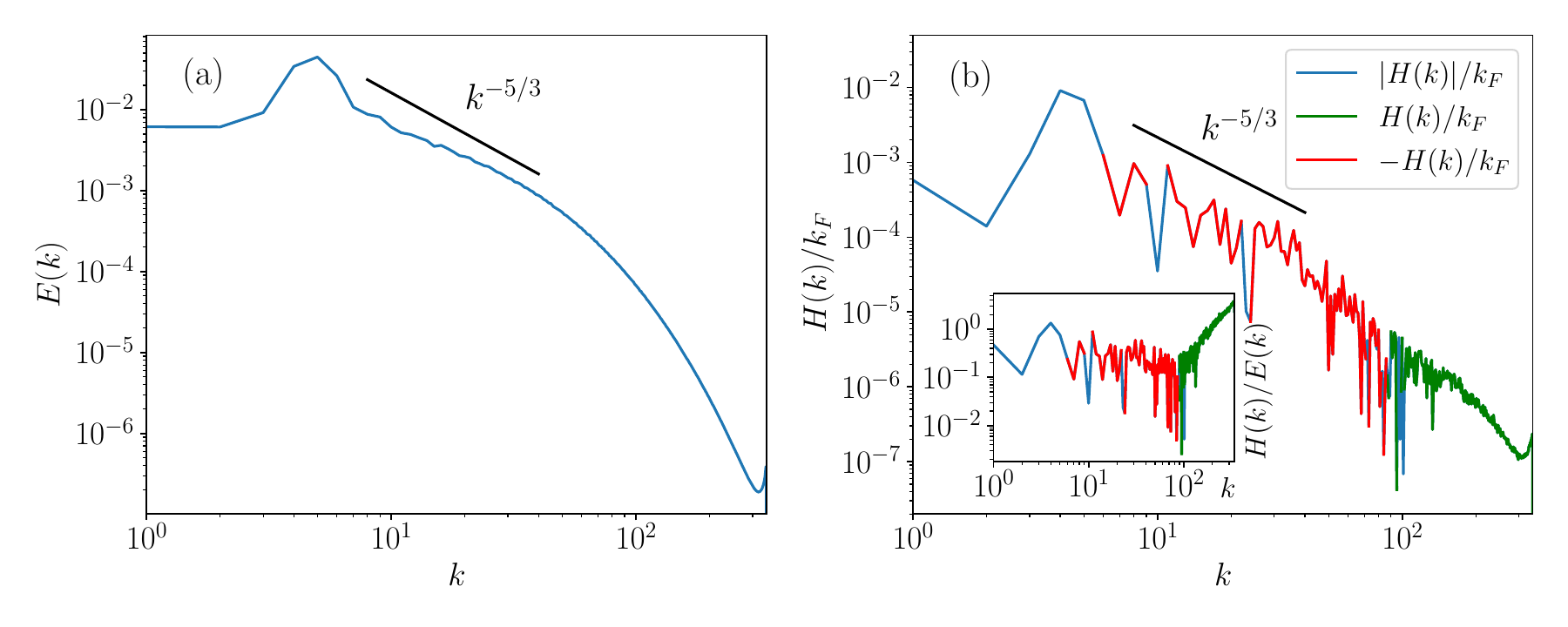}
\caption{\label{fig:dns_en_hel_spectra}Spectra of (a) kinetic energy, $E(k)$, and (b) helicity, $H(k)$, normalized by the forcing wavenumber $k_{\rm{F}} (= 5)$. As for the helicity spectrum, absolute values are shown, with positive and negative values denoted by green and red colors (see the labels in the inset). The inset in (b) shows the relative helicity spectrum $|H(k)|/[k E(k)]$ compensated by $k^{-1}$, i.e., $|H(k)|/E(k)$.}
\end{figure}

In Fig.~\ref{fig:dns_en_hel_spectra}, the spectra of the turbulent kinetic energy and of the turbulent helicity in the turbulent steady state of the simulation without rotation are shown. Note the peak at the forcing wave number $k_\textrm{F}=5$, followed by an inertial range and a dissipative range. A straight line indicating Kolmogorov scaling $k^{-5/3}$ is shown as a reference. The normalized helicity spectrum scaled by the forcing wavenumber $k_{\rm{F}}$, $|H(k)|/k_{\rm{F}}$, is compatible with this scaling $k^{-5/3}$, but it changes signs in several wave numbers. As a result, the figure shows the absolute value of the spectrum, and positive and negative values in different colors. The inset in panel (b) also shows the relative helicity spectrum $|H(k)|/[k E(k)]$ compensated by $k^{-1}$, i.e., $|H(k)|/E(k)$. Note that as the relative helicity is expected to scale as $\sim k^{-1}$, the compensated spectrum is expected to be approximately flat in the inertial range.

\begin{figure}
\includegraphics[width=0.9\textwidth]{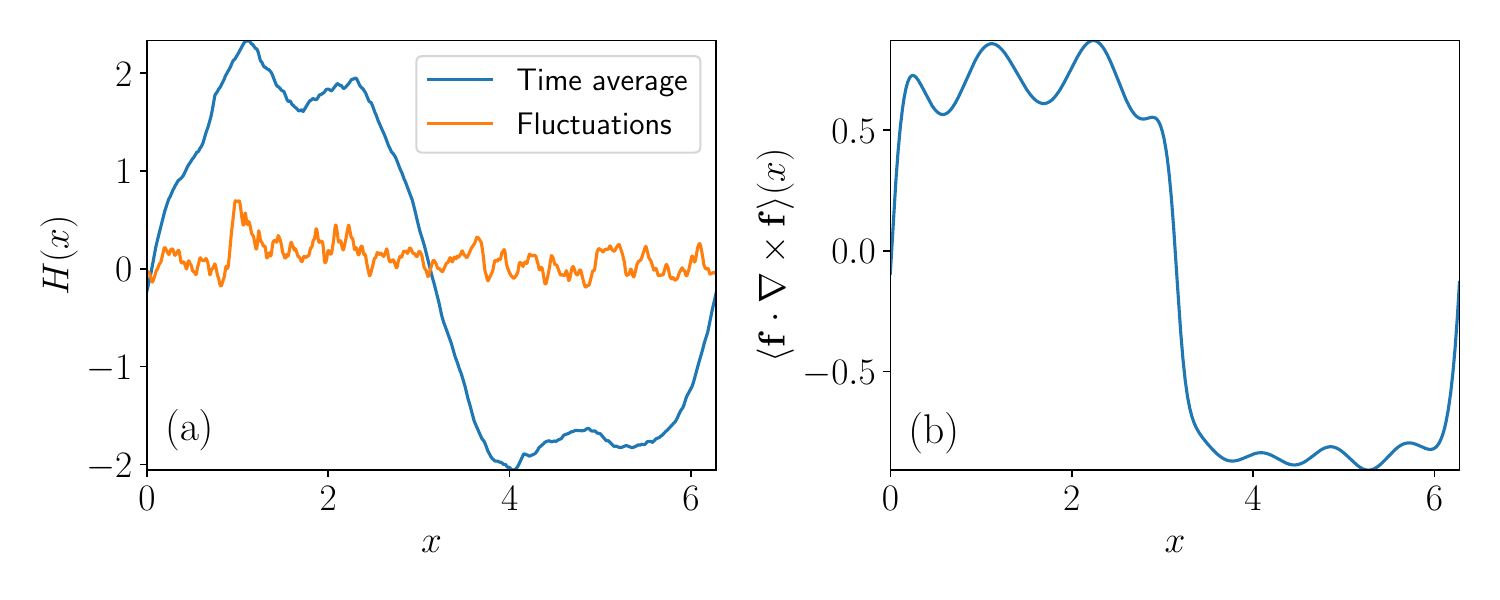}
\caption{\label{fig:helicity_in_forcing}Spatial profile of (a) helicity and (b) helicity in the forcing. (a) The spatial distribution of the time-averaged helicity is colored in blue, and that of the fluctuations are colored in orange.}
\end{figure}

Finally, Fig.~\ref{fig:helicity_in_forcing} shows the spatial proﬁle of helicity $H$ in the flow, as well as the helicity of the forcing ${\bf{f}} \cdot \nabla \times {\bf{f}}$, in both cases averaged over the $y$ and $z$ direction and as a function of $x$. For the flow, the spatial distribution $H(x)$ is time averaged, and we also show the instantaneous fluctuations around this average. The time-averaged helicity in Fig.~\ref{fig:helicity_in_forcing}(a) reflects the profile of the forcing shown in Fig.~\ref{fig:helicity_in_forcing}(b). With this analysis we confirm that the forcing acts as expected, generating a mostly isotropic flow with energy containing wave number around $k=5$, but with a large-scale inhomogeneity of the helicity in the $x$ direction.

\section{\label{sec:V}Numerical results and model validation}

The SGS stress model in Eq.~(\ref{eq:SGS_strss_exp}) suggests that in the presence of inhomogeneous SGS helicity, the SGS stresses $\mbox{\boldmath$\tau$}_{\rm{SGS}}$ depend not only on the GS strain rate $\overline{\mbox{\boldmath$s$}}$ but also on the GS absolute vorticity $\overline{\mbox{\boldmath$\omega$}}_\ast$. The transport coefficient for $\overline{\mbox{\boldmath$s$}}$ is the SGS viscosity $\nu_{\rm{S}}$, while the transport coefficient for $\overline{\mbox{\boldmath$\omega$}}_\ast$, $\eta_S$, is scaled by the gradient of the SGS helicity $\nabla H_{\rm{S}}$.

\begin{figure}
\includegraphics[width=0.8\textwidth]{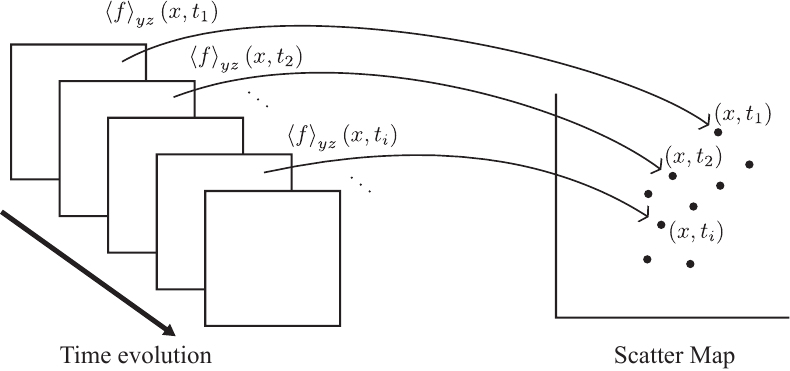}
\caption{\label{fig:scatter_map}Schematically depicted concept of a scatter map. We compute a statistical quantity $f$ averaged over the directions of homogeneity ($y$ and $z$) as $\langle {f} \rangle_{yz}(x, t)$. The quantity is a function of $x$ and $t$. We then compare the model expression for the turbulent correlation with the DNS results.}
\end{figure}

To validate the helicity SGS models, we construct scatter plots of several SGS quantities, as a function of the location in the direction of inhomogeneity ($x$) and in time ($t$), to see the correlations between the values obtained by the DNS and those given by the model (estimated from the DNS by separating the fields in GS and SGS components as explained before), as schematically drawn in Fig.~\ref{fig:scatter_map}. Then, we can compare the agreement of the helicity SGS model with the ground truth DNS data, as well as with the estimations using the classical \AP{(non-helical)} Smagorinsky model, to asses possible improvements when taking into account the role of helicity in the SGS stress tensor.

To further understand the model properties, in the following analysis and corresponding figures, we distinguish data points depending on the magnitude of SGS helicity gradient, $|\nabla H_{\rm{S}}|$. In most figures, colors will represent the pointwise level of SGS helicity gradient with high (yellow and green) through low (blue and dark blue) magnitudes of SGS helicity gradient $|\nabla H_{\rm{S}}|$. This will allow us to identifiy whether the correlation depends on the local spatial inhomogeneity of the SGS helicity.

As was described in Sec.~\ref{sec:IV}, we performed simulations with and without net rotation, the latter corresponding to a state similar to homogeneous isotropic turbulence (except for the large-scale helicity inhomogeneity), and the former to rotating turbulence.

\subsection{\label{sec:V.A}Subgrid-scale (SGS) quantities and their models}
\subsubsection{\label{sec:V.A.1}Timescales of SGS motions}

\begin{figure}
\includegraphics[width=1.0\textwidth]{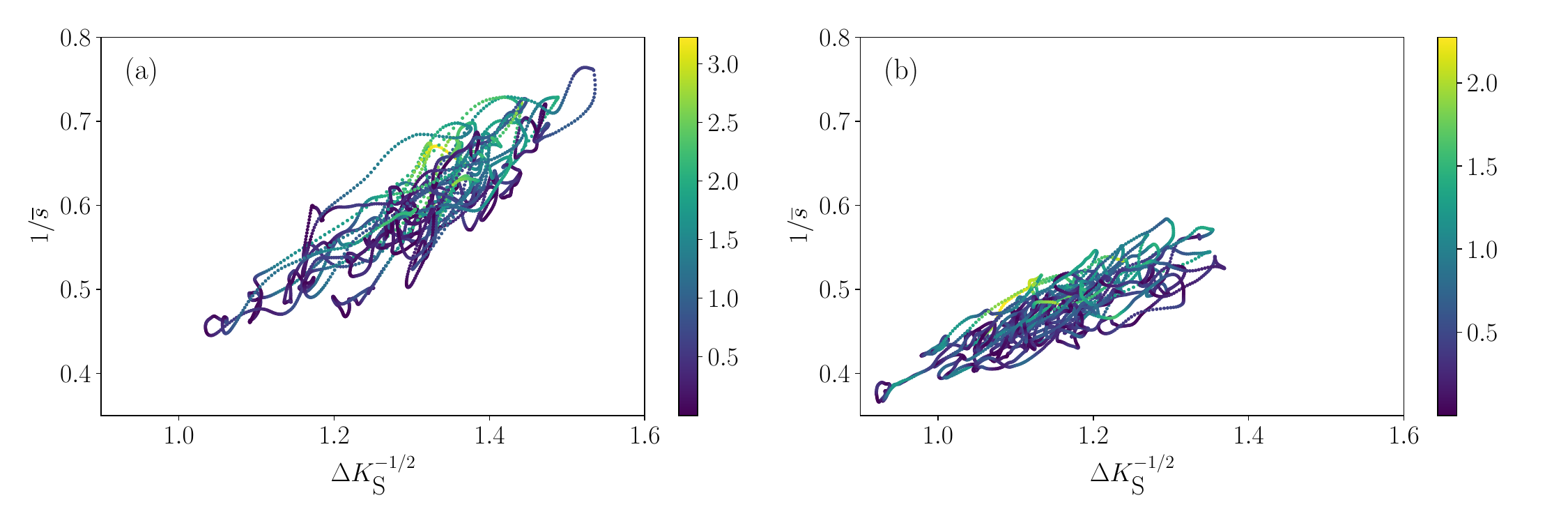}
\caption{\label{fig:timescale_hit}Timescales obtained from the magnitude of the strain rate $\overline{s}$ [Eq.~(\ref{eq:strain_rate_amplitude})], and the filter width $\Delta$ times the SGS energy $K_{\rm{S}}^{-1/2}$, in simulations without rotation. (a) Large scale filtering with filter width $k_{\rm{C}} = 2 \pi/\Delta = 7$, and (b) smaller scale filtering at $k_{\rm{C}} = 14$. The colors in this figure and in the following figures indicate the pointwise magnitude of $|\grad H_S|$ (as indicated by the colorbars).}
\end{figure}

\paragraph{Without rotation:} Figure \ref{fig:timescale_hit} shows the scatter plots of the SGS timescales. The timescale of the large-scale strain is given by $\overline{s}^{-1}$. On the other hand, the timescale of SGS motions can be expressed as the maximum lengthscale of the unresolved (SGS) motions, $\Delta$, divided by the characteristic velocity of the SGS component $K_{\rm{S}}^{1/2}$. Therefore Fig.~\ref{fig:timescale_hit} shows the correlation between $\overline{s}^{-1}$ and $\Delta K_{\rm{S}}^{-1/2}$ computed using the procedure shown in Fig.~\ref{fig:scatter_map}. In this figure and in the following, the ranges of the axes are kept the same in panels (a) and (b), to allow a direct comparison between cases with different filter widths.

A good correlation is seen between $\Delta K_{\rm{S}}^{-1/2}$ and $\overline{s}^{-1}$. Their relation can be expressed as
\begin{equation}
	\frac{1}{\overline{s}}
	= C_\tau \frac{\Delta}{K_{\rm{S}}^{1/2}},
	\label{eq:timescale_Del_KS}
\end{equation}
where the proportionality coefficient $C_\tau$ is
\begin{equation}
	C_\tau \approx 0.38.
	\label{eq:Ctau}
\end{equation}
The correlations are good irrespective of the magnitude of the SGS gradient of SGS helicity, $|\nabla H_{\rm{S}}|$. These results suggest that the timescale of SGS motion is well represented by the reciprocal of the magnitude of GS velocity strain $\overline{s}$.

\begin{figure}
\includegraphics[width=1.0\textwidth]{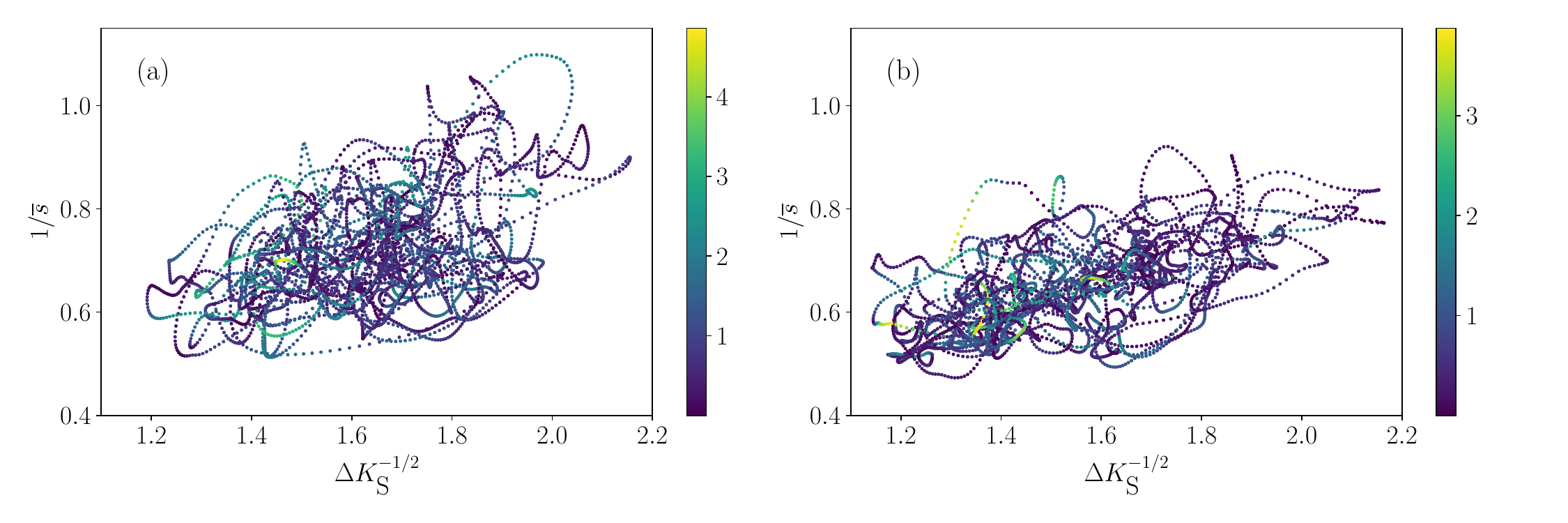}
\caption{\label{fig:timescale_rot}Timescales evaluated from the magnitude of strain rate $\overline{s}$ [Eq.~(\ref{eq:strain_rate_amplitude})], and the filter width $\Delta$ times the SGS energy $K_{\rm{S}}^{-1/2}$ in the simulation with rotation ($\omega_{\rm{F}} = 8$). (a) Large scale filtering with filter width $k_{\rm{C}} = 2\pi/\Delta = 7$, and (b) small scale filtering at $k_{\rm{C}} = 14$.}
\end{figure}

\paragraph{With rotation:} In Fig.~\ref{fig:timescale_rot} we present the scatter plots of the same SGS timescales, deﬁned by $\Delta K_{\rm{S}}^{-1/2}$ and by $\overline{s}^{-1}$. 
\NY{Some correlation is seen between $\Delta K_{\rm{S}}^{-1/2}$ and $\overline{s}^{-1}$, for both the larger scale filtering ($k_{\rm{C}} = 7$) and the smaller scale filtering ($k_{\rm{C}} = 14$) cases [Fig.~\ref{fig:timescale_rot}(a) and (b)]. However, the correlations are smaller than those seen without rotation [Fig.~\ref{fig:timescale_hit}(a) and (b)]. In general, rotation is expected to decrease fluctuations. As a result, the SGS timescale, which is the reciprocal of the fluctuating velocity, may scatter more wildly reflecting this property of rotating turbulence.} Of course, the correlation could be improved by increasing resolution and moving $k_C$ to larger wave numbers such that the Zeman scale, $\sqrt{\epsilon/\omega_F^3}$, becomes resolved. At this scale the flow starts to become isotropic, and therefore results should asymptotically approach those seen in the case without rotation. Indeed, the case with $k_{\rm{C}} = 14$ seems to have less dispersion, or in other words, \NY{the correlation between $\Delta/K^{1/2}$ and $\overline{s}^{-1}$ is higher in the smaller filtering case ($k_{\rm{C}} = 14$) with rotation. Therefore, in the present simulations with rotation, we can conclude that the model is capturing eddies at some scales that lead to fluctuations. We should also note that no prominent dependence on the magnitude of the SGS helicity gradient, $\nabla H_{\rm{S}}$, represented by the colors, is seen in the scatter plots.}

\subsubsection{\label{sec:V.A.2}SGS energy}

\begin{figure}
\includegraphics[width=1.0\textwidth]{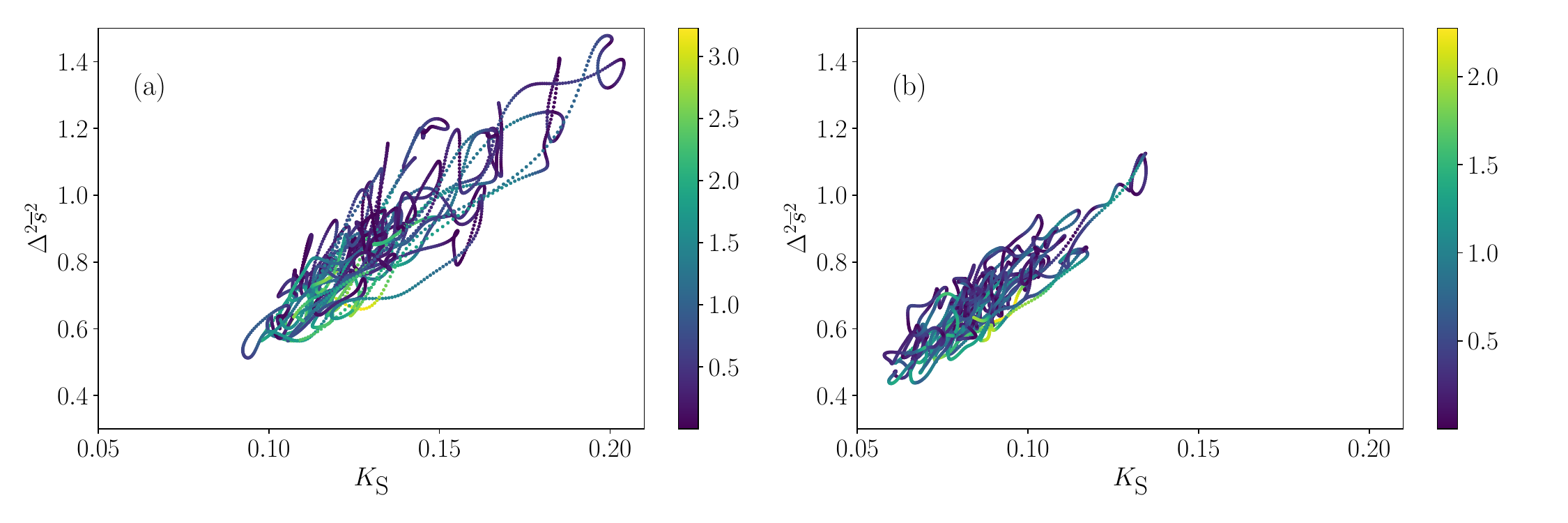}
\caption{\label{fig:KS_s_Del_rel_hit}SGS energy $K_{\rm{S}}$ and magnitude of the squared GS strain rate $\overline{s}^2$ times the squared filter width $\Delta^2$ in the run without rotation. (a) Large scale filtering with $k_{\rm{C}} = 2\pi/\Delta = 7$, and (b) smaller scale filtering at $k_{\rm{C}} = 14$.}
\end{figure}

\paragraph{Without rotation:} 
As is the case with the Smagorinsky model, expressing the SGS energy $K_{\rm{S}}$ in terms of the filter width $\Delta$ and the magnitude of the GS strain rate $\overline{s}$, as in Eq.~(\ref{eq:SGS_KS_in_Del_s_one}), greatly contributes to the simplification of SGS modeling. Figure~\ref{fig:KS_s_Del_rel_hit} shows the scatter plot of the SGS energy $K_{\rm{S}}$ and of $(\Delta \overline{s})^2$.

We see a good correlation between them with
\begin{equation}
	K_{\rm{S}} = C_{\rm{KS}} (\Delta \overline{s})^2,
	\label{eq:KS_Del_s_correl}
\end{equation}
where the proportionality coefficient $C_{\rm{KS}}$ is
\begin{equation}
	C_{\rm{KS}} \approx 0.13
	\label{eq:C_KS_value}
\end{equation}
for $k_{\rm{C}} = 14$. The correlation does not depend on the magnitude of the SGS helicity gradient $\nabla H_{\rm{S}}$. This suggests that the SGS energy can be well expressed by $(\Delta \overline{s})^2$.

\begin{figure}
\includegraphics[width=1.0\textwidth]{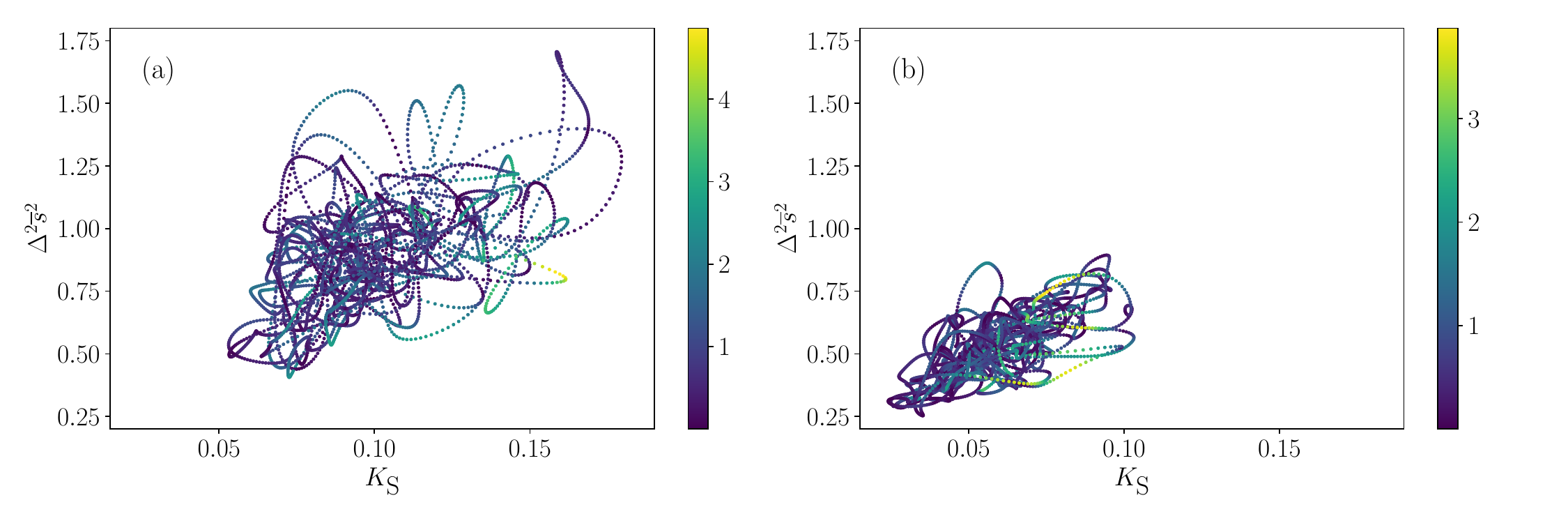}
\caption{\label{fig:KS_s_Del_rel_rot}SGS energy $K_{\rm{S}}$ against the squared magnitude of the GS strain rate $\overline{s}^2$ multiplied by the squared filter width $\Delta^2$ in the run with rotation ($\omega_{\rm{F}} = 8$). (a) Large scale filtering with $k_{\rm{C}} = 2\pi/\Delta = 7$, and (b) small scale filtering at $k_{\rm{C}} = 14$.}
\end{figure}

\paragraph{With rotation:} Figure~\ref{fig:KS_s_Del_rel_rot} shows the scatter plots of the SGS energy $K_{\rm{S}}$ and $(\overline{s})^2$  in the simulation with rotation, in the large scale filtering case, and the smaller scale filtering case. A reasonable correlation is seen between $K_{\rm{S}}$ and $(\Delta \overline{s})^2$ for both large and small scale ﬁltering cases [Fig.~\ref{fig:KS_s_Del_rel_rot}(a) and (b)]. However, the data points display more dispersion than in the simulation without rotation [Fig.~\ref{fig:KS_s_Del_rel_hit}(a) and (b)]. Again, we believe this is associated with an increased sensitivity to fluctuations due to the decrease in turbulent intensity in the rotating case. No prominent dependence on the magnitude of the SGS helicity gradient (color dependence) is seen in the scatter plots.

\subsubsection{\label{sec:V.A.3}SGS energy dissipation $\varepsilon_{\rm{KS}}$ and its model}

The energy dissipation rate $\varepsilon_{\rm{KS}}$ is deﬁned by Eq.~(54a). For homogeneous turbulence, it can be rewritten as
\begin{equation}
	\varepsilon_{\rm{KS}}
	= \nu \left( { \overline{
		\frac{\partial u^j}{\partial x^i} 
		\frac{\partial u^j}{\partial x^i}} 
	- \frac{\partial \overline{u}^j}{\partial x^i}
		\frac{\partial \overline{u}^j}{\partial x^i}
	} \right)
	= \nu \left( {
		\overline{\mbox{\boldmath$\omega$}^2}
		- \overline{\mbox{\boldmath$\omega$}}^2
	} \right)
	\NY{\equiv \nu \mbox{\boldmath$\omega$}_{\rm{SGS}}^2}.
	\label{eq:epsKS_def_re}
\end{equation}
This should be compared with the model expression of $\varepsilon_{\rm{KS}}$ expressed in terms of the characteristic SGS velocity $v \sim K_{\rm{S}}^{1/2}$ and the largest scale of the SGS, $\Delta$, as
\begin{equation}
	\varepsilon_{\rm{KS}}
	= C_{\varepsilon {\rm{S}}} \frac{K_{\rm{S}}^{3/2}}{\Delta},
	\label{eq:epsKS_KS_Del_rel}
\end{equation}
where $C_{\rm{\varepsilon S}}$ is the model constant.

\begin{figure}
\includegraphics[width=1.0\textwidth]{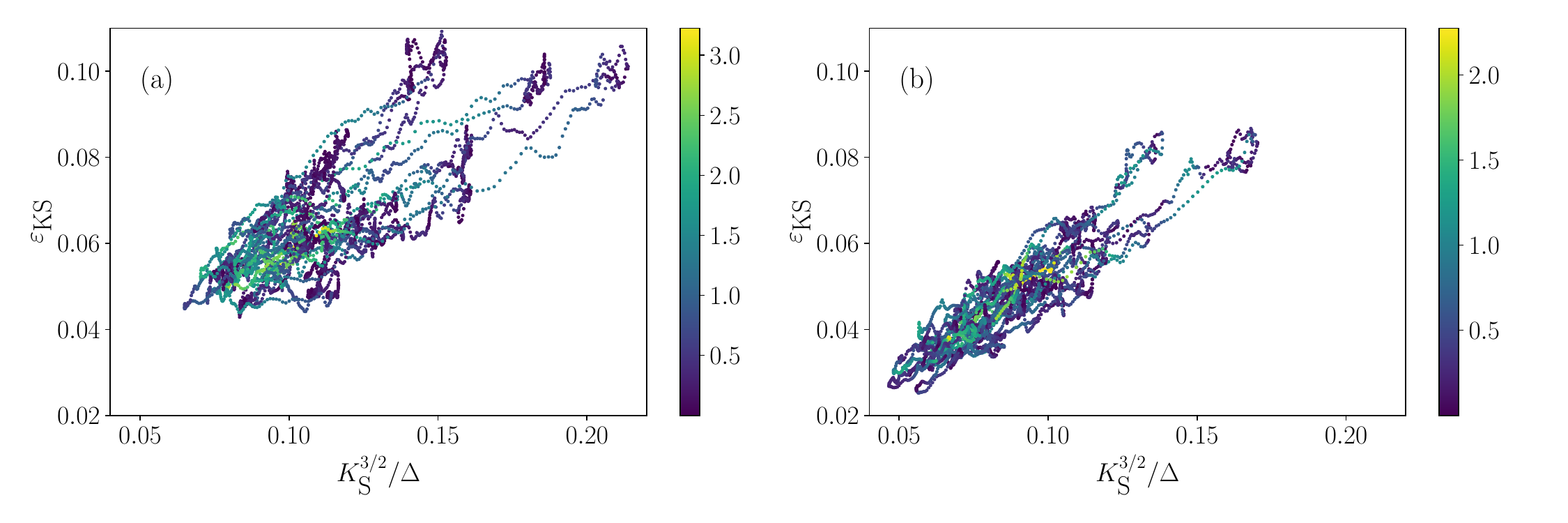}
\caption{\label{fig:epsK_KS_Del_rel_hit}SGS dissipation rate $\varepsilon_{\rm{KS}}$ as a function of $K_\textrm{S}^{3/2}/\Delta$ in the simulation without rotation, for (a) large scale filtering ($k_{\rm{C}}= 7$), and (b) smaller scale filtering ($k_{\rm{C}} = 14$).}
\end{figure}

\paragraph{Without rotation:} In Fig.~\ref{fig:epsK_KS_Del_rel_hit} we present the scatter plots of the SGS dissipation rate $\varepsilon_{\rm{KS}}$, comparing the model expression, $K_{\rm{S}}^{3/2} / \Delta$ [Eq.~(\ref{eq:epsKS_KS_Del_rel})] and the DNS values from Eq.~(\ref{eq:epsKS_def_re}). These scatter plots indicate that there is a strong correlation between $\nu (\overline{\mbox{\boldmath$\omega$}^2} - \overline{\mbox{\boldmath$\omega$}}^2)$ and its model $K_{\rm{S}}^{3/2} / \Delta$. The evaluated proportionality coefficient $C_{\varepsilon {\rm{S}}}$ is
\begin{equation}
	C_{\varepsilon {\rm{S}}} \approx 0.47\;\;
	\mbox{for}\;\;
	k_{\rm{C}} = 7,\;\;
	C_{\varepsilon {\rm{S}}}\approx 0.48\;\;
	\mbox{for}\;\;
	k_{\rm{C}} = 14.
	\label{eq:CepsS_model_const}
\end{equation}
The shapes of the scatter plots for the large and small scale filtering cases are similar. However, the SGS energy dissipation $\varepsilon_{\rm{KS}}$ is more concentrated in the region of smaller values in the case with the \AP{filter} at smaller scales ($k_{\rm{C}} = 14$) than in the case with the filter at larger scales ($k_{\rm{C}} = 7$). This just represents the smaller amount of SGS energy that needs to be modeled when using a filter that resolves more of the inertial range. Also, from the color distribution, we see no strong dependence on the magnitude of the SGS helicity gradient, except for the fact that excursions of the dissipation to large values seem to be associated with regions of smaller SGS helicity gradients.

\begin{figure}
\includegraphics[width=1.0\textwidth]{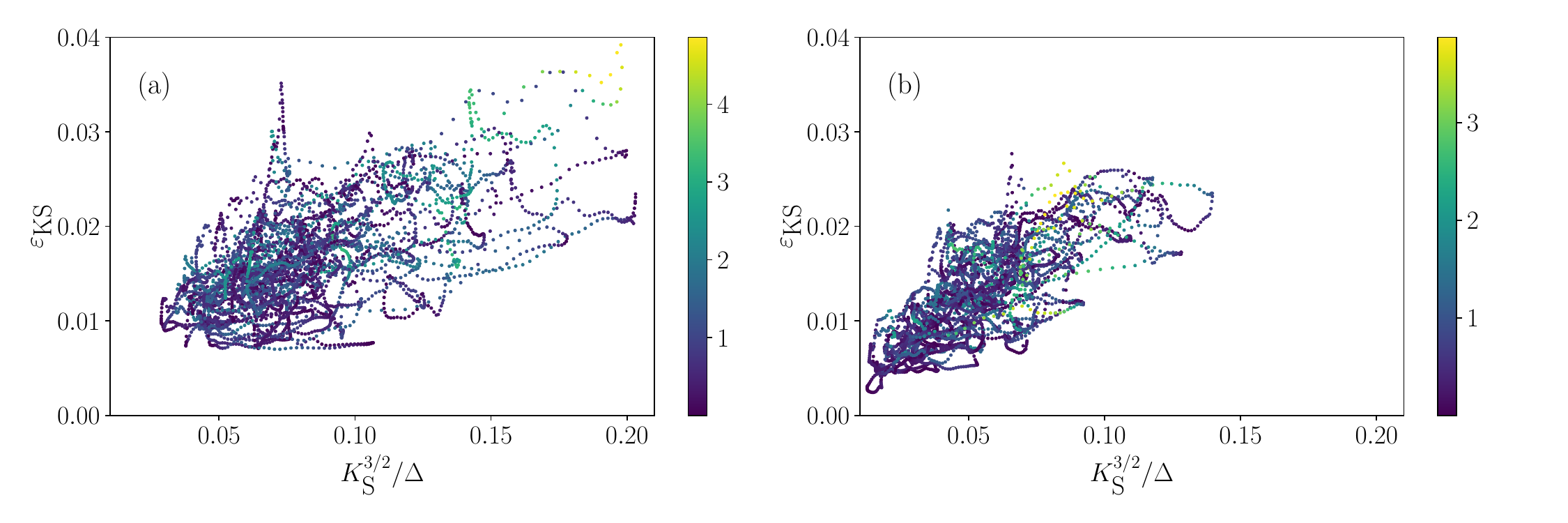}
\caption{\label{fig:epsK_KS_Del_rel_rot}SGS energy dissipation rate $\varepsilon_{\rm{KS}}$ as a function of $K_\textrm{S}^{3/2}/\Delta$ in the simulation with rotation $\omega_{\rm{F}} = 8$ for (a) large scale filtering ($k_{\rm{C}}= 7$), and (b) smaller scale filtering ($k_{\rm{C}} = 14$).}
\end{figure}

\paragraph{With rotation:} Figure~\ref{fig:epsK_KS_Del_rel_rot} shows that, with rotation ($\omega_{\rm{F}} = 8$), there are also good correlations between the DNS value of the SGS dissipation rate, $\nu (\overline{\mbox{\boldmath$\omega$}^2} - \overline{\mbox{\boldmath$\omega$}}^2) \equiv \nu \overline{\mbox{\boldmath$\omega$}}_{\rm{SGS}}^2$, and its model expression, $K_{\rm{S}}^{3/2} /\Delta$, both when using a large scale filter ($k_{\rm{C}} = 7$) and with the smaller scale filter ($k_{\rm{C}} = 14$). As in the non-rotating case, the SGS dissipation rate takes smaller values as $k_C$ increases. However, unlike the non-rotating case, excursions to large values of the SGS energy dissipation rate now seem to correspond to larger values of the SGS helicity gradient.

\subsubsection{\label{sec:V.A.4}Helicity dissipation $\varepsilon_{\rm{HS}}$ and its model}

The dissipation rate of the SGS helicity is deﬁned by Eq.~(\ref{eq:SGS_epsHS_def_in_two}) as
\begin{equation}
	\varepsilon_{\rm{HS}}
	= 2 \nu \left( {
	\overline{
		\frac{\partial u^j}{\partial x^i}
		\frac{\partial \omega^j}{\partial x^i}
 		}
	- \frac{\partial \overline{u}^j}{\partial x^i}
		\frac{\partial \overline{\omega}^j}{\partial x^i}
	} \right),
	\label{eq:SGS_epsHS_def}
\end{equation}
and its model [Eq.~(\ref{eq:SGS_epsHS_approx_in_two})] can be expressed by the SGS helicity $H_{\rm{S}}$ divided by the SGS turbulence timescale $K_{\rm{S}} / \varepsilon_{\rm{KS}}$ as
\begin{equation}
	\varepsilon_{\rm{HS}}
	= C_{\varepsilon {\rm{HS}}}
	\frac{\varepsilon_{\rm{KS}} H_{\rm{S}}}{K_{\rm{S}}},
	\label{eq:SGS_epsHS_model}
\end{equation}
where $C_{\varepsilon {\rm{HS}}}$ is the model constant. This is the simplest possible algebraic model of the SGS helicity dissipation rate $\varepsilon_{\rm{HS}}$.

\begin{figure}
\includegraphics[width=1.0\textwidth]{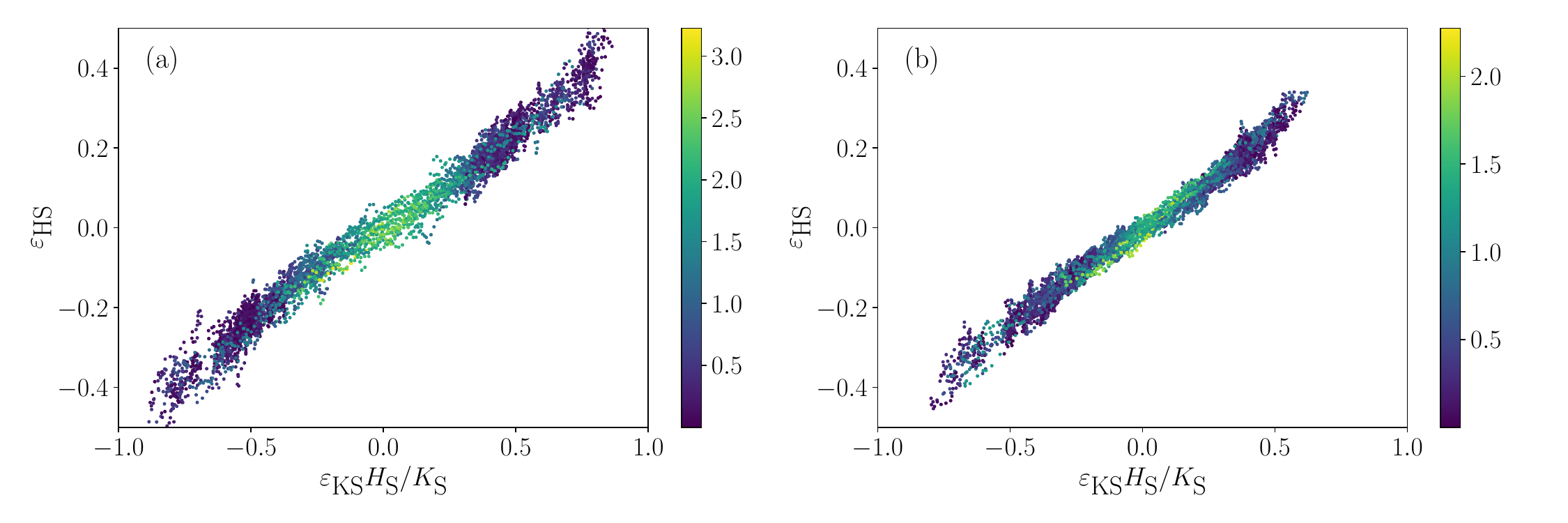}
\caption{\label{fig:epsH_KS_epsK_HS_rel_hit}SGS helicity dissipation rate and its algebraic model in the simulation without rotation for (a) the large scale filter ($k_{\rm{C}} = 7$), and (b) for the smaller scale filter ($k_{\rm{C}} = 14$).}
\end{figure}

\paragraph{Without rotation:} In Fig.~\ref{fig:epsH_KS_epsK_HS_rel_hit} we present scatter plots of $\varepsilon_{\rm{HS}}$ in the simulation without rotation, comparing the DNS values [Eq.~(\ref{eq:SGS_epsHS_def})] and its model [Eq.~(\ref{eq:SGS_epsHS_model})] for the large scale filter ($k_{\rm{C}} = 7$) and the smaller scale filter ($k_{\rm{C}} = 14$). The correlation between the algebraic model $H_{\rm{S}}/(K_{\rm{S}}/\varepsilon_{\rm{KS}})$ \NY{[Eq.~(\ref{eq:SGS_epsHS_model})]} and the SGS helicity dissipation \NY{$\varepsilon_{\rm{HS}}$ [Eq.~(\ref{eq:SGS_epsHS_def})]} is very good. This shows that the SGS helicity dissipation rate can be well represented by the algebraic model. The proportionality coefficient is
\begin{equation}
	C_{\varepsilon{\rm{HS}}} \approx 0.5\;\;
	\mbox{for}\;\;
	k_{\rm{C}} = 7,\;\;
	C_{\varepsilon{\rm{HS}}} \approx 0.48\;\;
	\mbox{for}\;\;
	k_{\rm{C}} = 14.
	\label{eq:C_epsHS_values}
\end{equation}
As the color map shows, strong SGS helicity gradients (yellow and green colors) are localized in the region of small $\varepsilon_{\rm{HS}}$ values. This is because the SGS helicity values are small where the magnitude of $H_{\rm{S}}$ gradients is large. 
\AP{Moreover,} the shapes and ranges of the scatter plots for the large scale and small scale filters are very similar. This means that the SGS helicity dissipation rate does not depend strongly on the filtering scale, further confirming that it can be well represented by the simple algebraic model, i.e., by the SGS helicity divided by the SGS timescale $K_{\rm{S}} / \varepsilon_{\rm{KS}}$.

\begin{figure}
\includegraphics[width=1.0\textwidth]{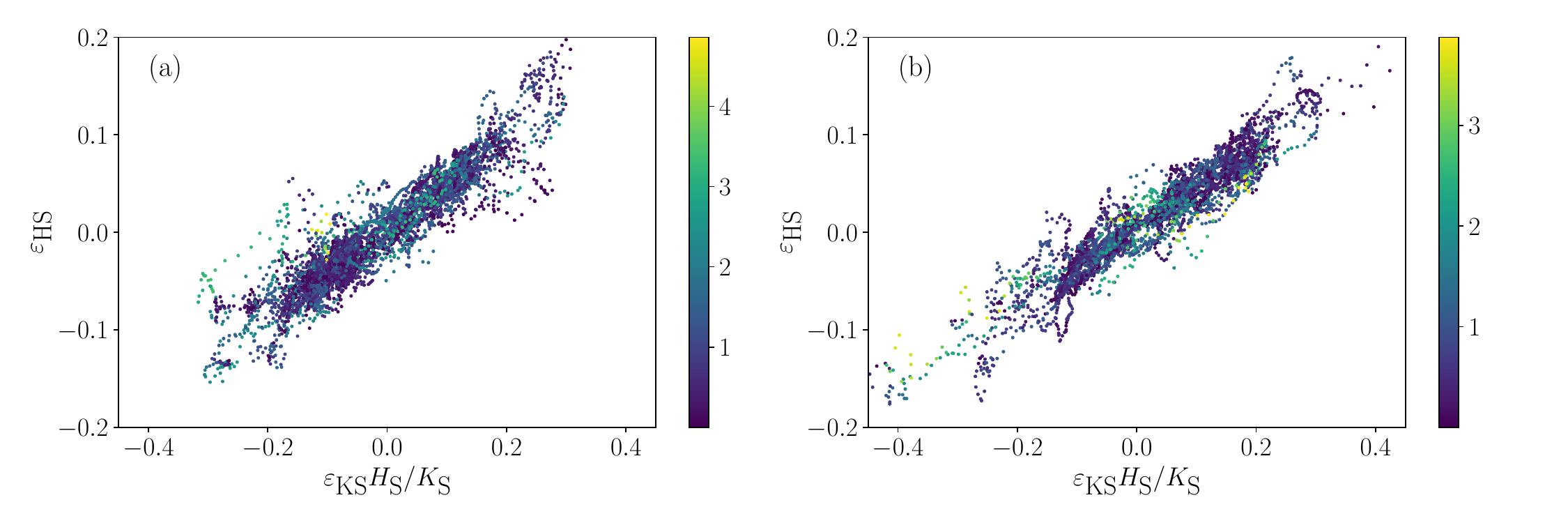}
\caption{\label{fig:epsH_KS_epsK_HS_rel_rot}SGS helicity dissipation rate $\varepsilon_{\rm{HS}}$ and its algebraic model in the simulation with rotation ($\omega_{\rm{F}} = 8$) for (a) the large scale filter ($k_{\rm{C}} = 7$), and (b) the small scale filter ($k_{\rm{C}} = 14$).}
\end{figure}

\paragraph{With rotation:} The scatter plots of $\varepsilon_{\rm{H}}$ with rotation ($\omega_{\rm{F}} = 8$) are presented in Fig.~\ref{fig:epsH_KS_epsK_HS_rel_rot}. The correlations between the algebraic model $H_{\rm{S}} (K_{\rm{S}}/ \varepsilon_{\rm{KS}})$ and the DNS values $\nu (\overline{\mbox{\boldmath$\omega$}^2} - \overline{\mbox{\boldmath$\omega$}}^2)$ are also good with rotation. These results show that the dissipation rate of the SGS helicity, $\varepsilon_{\rm{HS}}$, without and with rotation can be well modeled by the proposed model.

\subsection{\label{sec:V.B}SGS Stresses: diagonal component}

We now construct scatter plots of each component of the SGS stress, calculated using the DNSs and the models' expressions, as a function of the location in the inhomogeneous direction ($x$) and of time ($t$).

\subsubsection{\label{sec:V.B.1}Smagorinsky model without rotation}

\begin{figure}
\includegraphics[width=1.0\textwidth]{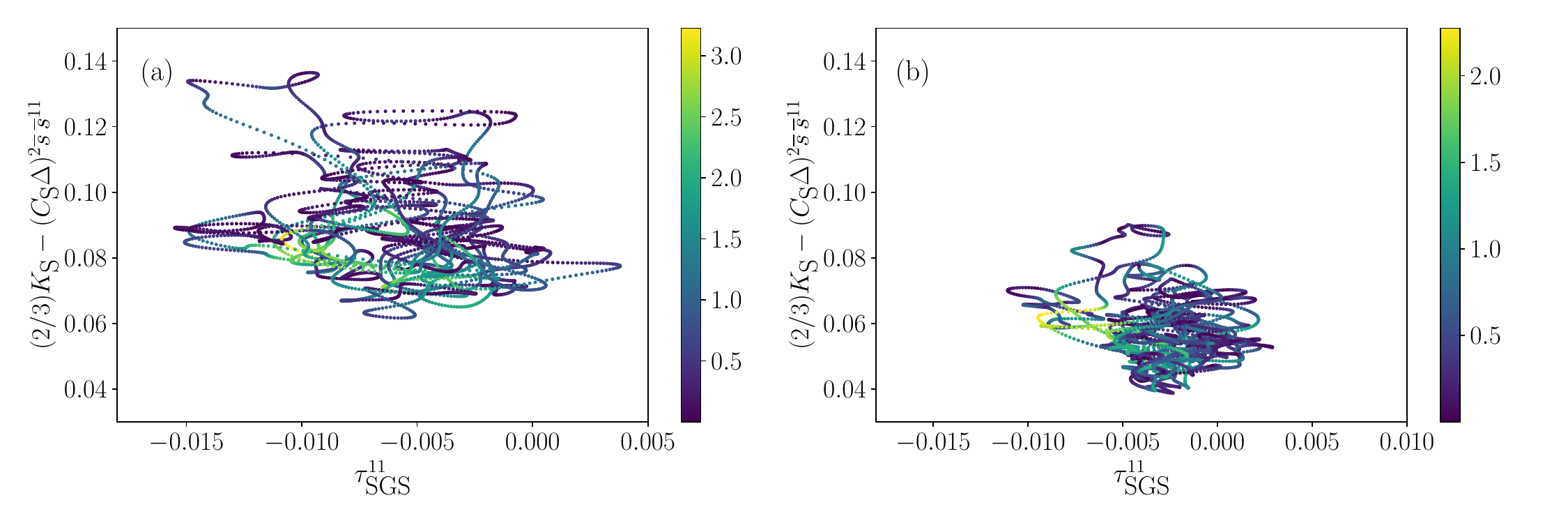}
\caption{\label{fig:s11_smag_hit}The diagonal component of the SGS stress $\tau_{\rm{SGS}}^{11}$ and the Smagorinsky model $(2/3)K_{\rm{S}} - (C_{\rm{S}} \Delta)^2 \overline{s}\ \overline{s}^{11}$ for (a) large scale filtering ($k_{\rm{C}} = 7$), and (b) small scale filtering ($k_{\rm{C}} = 14$), in both cases in the simulation without rotation.}
\end{figure}

The Smagorinsky model consists of the SGS stress $\mbox{\boldmath$\tau$}_{\rm{SGS}}$ [Eq.~(\ref{eq:standard_SGS_strss_model})] with the Smagorinsky model expression for the SGS viscosity $\nu_{\rm{S}}$ [Eq.~(\ref{eq:smag_model})]. It is explicitly written as
\begin{equation}
	\tau_{\rm{SGS}}^{ij}
	= \frac{2}{3} \delta^{ij} K_{\rm{S}}
	- (C_{\rm{S}} \Delta)^2  \overline{s}\ \overline{s}^{ij}.
	\label{eq:DNS_SGS_stress_Smag_model}
\end{equation}
Figure~\ref{fig:s11_smag_hit} shows the correlation between these two quantities for a diagonal component. In particular, the component $\tau_{\rm{SGS}}^{11}$ and $(2/3)K_{\rm{S}} - (C_{\rm{S}} \Delta)^2 \overline{s}\ \overline{s}^{11}$ are considered for (a) the large scale filter with $k_{\rm{C}} = 7$, and (b) the smaller scale filter with $k_{\rm{C}} = 14$.

Note there is very little correlation between the DNS results of $\tau_{\rm{SGS}}^{11}$ and those of the Smagorinsky model, irrespective of the filtering scales. In the larger scale filtering case ($k_{\rm{C}} = 7$), the magnitude of the SGS stress is larger than the counterpart in the small-scale filtering case ($k_{\rm{C}} = 14$), as the total amount of energy represented by the SGS components is smaller than that in the large scale filtering case.

\begin{figure}
\includegraphics[width=1.0\textwidth]{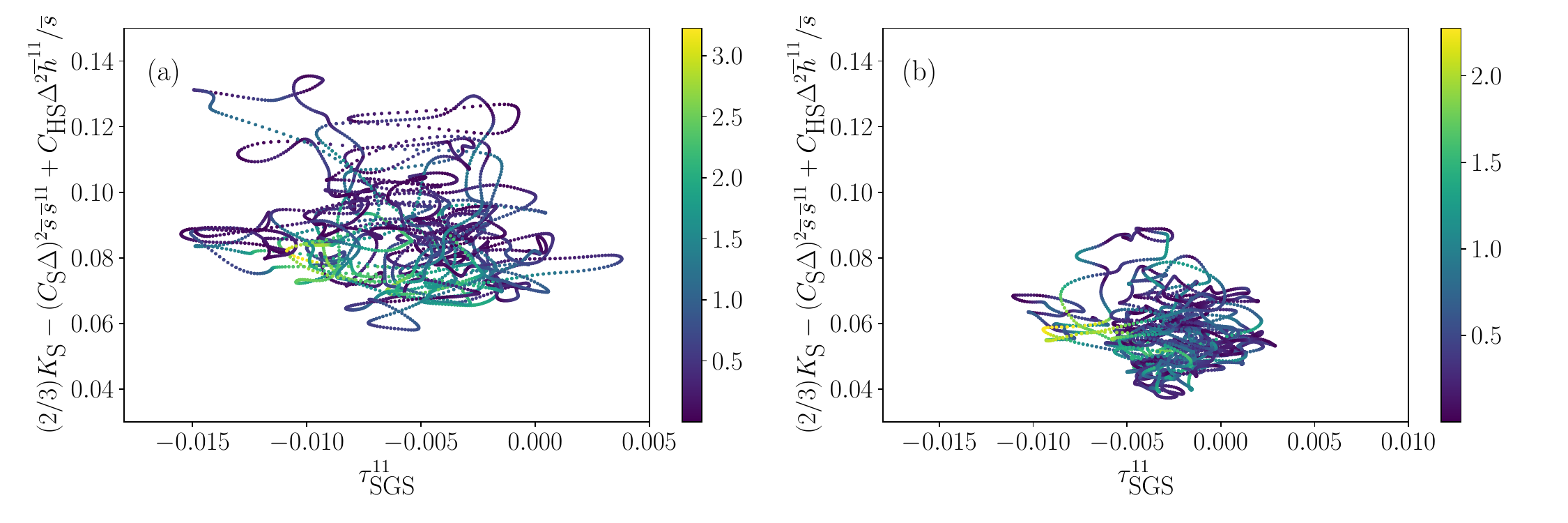}
\caption{\label{fig:s11_hel_hit}The diagonal component of the SGS stress $\tau_{\rm{SGS}}^{11}$ and the helicity SGS model $(2/3)K - (C_{\rm{S}} \Delta)^2 \overline{s}\ \overline{s}^{11} + C_{\eta {\rm{S}}} \Delta^2 \overline{s}^{-1}  (\nabla H_{\rm{S}}) \overline{\omega}$ [$(\nabla H_{\rm{S}})^1 \overline{\omega}_\ast^1$ is denoted by $\overline{h}^{11}$], for (a) the large scale filtering case ($k_{\rm{C}} = 7$), and (b) the small scale filtering case ($k_{\rm{C}} = 14$).}
\end{figure}

\subsubsection{\label{sec:V.B.2}Helicity SGS model without rotation}

In Fig.~\ref{fig:s11_hel_hit}, we show the results using the helicity SGS model, where the helicity-related contributions, the third (or $\eta_{\rm{S}}$-related) term in Eq.~(\ref{eq:SGS_strss_exp}) is included into the model expression in addition to the SGS viscosity model, the second (or $\nu_{\rm{S}}$-related) term in Eq.~(\ref{eq:SGS_strss_exp}). The scatter plots of the model with the SGS helicity included [Fig.~\ref{fig:s11_hel_hit}(a) and (b)] are very similar to the scatter plots without helicity [Fig.~\ref{fig:s11_smag_hit}(a) and (b), respectively]. This is to be expected as the helicity terms do not affect the diagonal components of the SGS stresss.

\subsubsection{\label{sec:V.B.3}Smagorinsky model with rotation}

The system rotation $\mbox{\boldmath$\omega$}_{\rm{F}}$ enters the helicity SGS model [Eq.~(\ref{eq:SGS_strss_exp})] through the GS absolute vorticity, $\overline{\mbox{\boldmath$\omega$}}_\ast = \overline{\mbox{\boldmath$\omega$}} + 2 \mbox{\boldmath$\omega$}_{\rm{F}}$. This results from the local equivalence of the GS vorticity $\overline{\mbox{\boldmath$\omega$}} = \nabla \times \overline{\bf{u}}$ and the system rotation angular velocity $2 \mbox{\boldmath$\omega$}_{\rm{F}}$.

\begin{figure}
\includegraphics[width=1.0\textwidth]{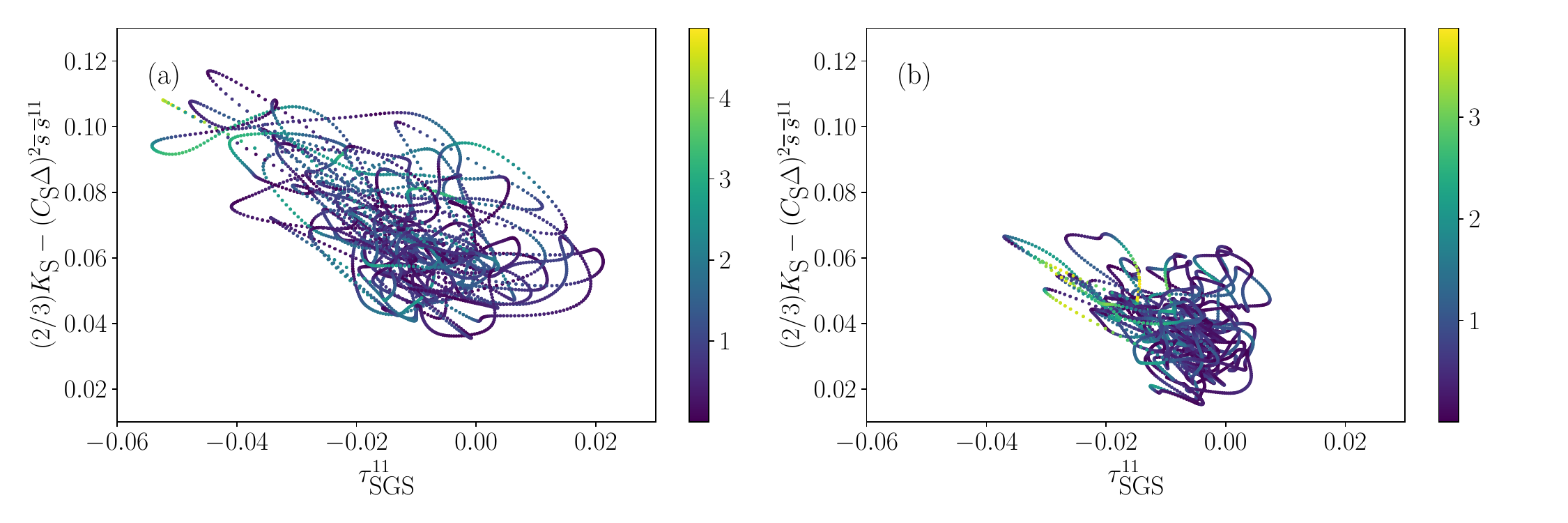}
\caption{\label{fig:s11_smag_rot}The diagonal component of the SGS stress $\tau_{\rm{SGS}}^{11}$ and its expression in the Smagorinsky model $(2/3)K - (C_{\rm{S}} \Delta)^2 \overline{s}\ \overline{s}^{11}$ in the simulation with rotation ($\omega_{\rm{F}} = 8$) for (a) the large scale filtering case ($k_{\rm{C}} = 7$), and (b) the small scale filtering case ($k_{\rm{C}} = 14$).}
\end{figure}

We construct scatter plots of each component of the SGS stress, calculated using the DNSs and the models' expressions, as a function of the location in the inhomogeneous direction ($x$) and time ($t$). Figure~\ref{fig:s11_smag_rot} illustrates the behavior of the diagonal component of SGS stresses. In particular, the correlation between $\tau_{\rm{SGS}}^{11}$ and its Smagorinsky model expression $(2/3)K_{\rm{S}} - (C_{\rm{S}} \Delta)^2 \overline{s}\ \overline{s}^{11}$ are shown for (a) the large scale filtering case ($k_{\rm{C}} = 7$), and (b) the small scale filtering case ($k_{\rm{C}} = 14$).

\begin{figure}
\includegraphics[width=1.0\textwidth]{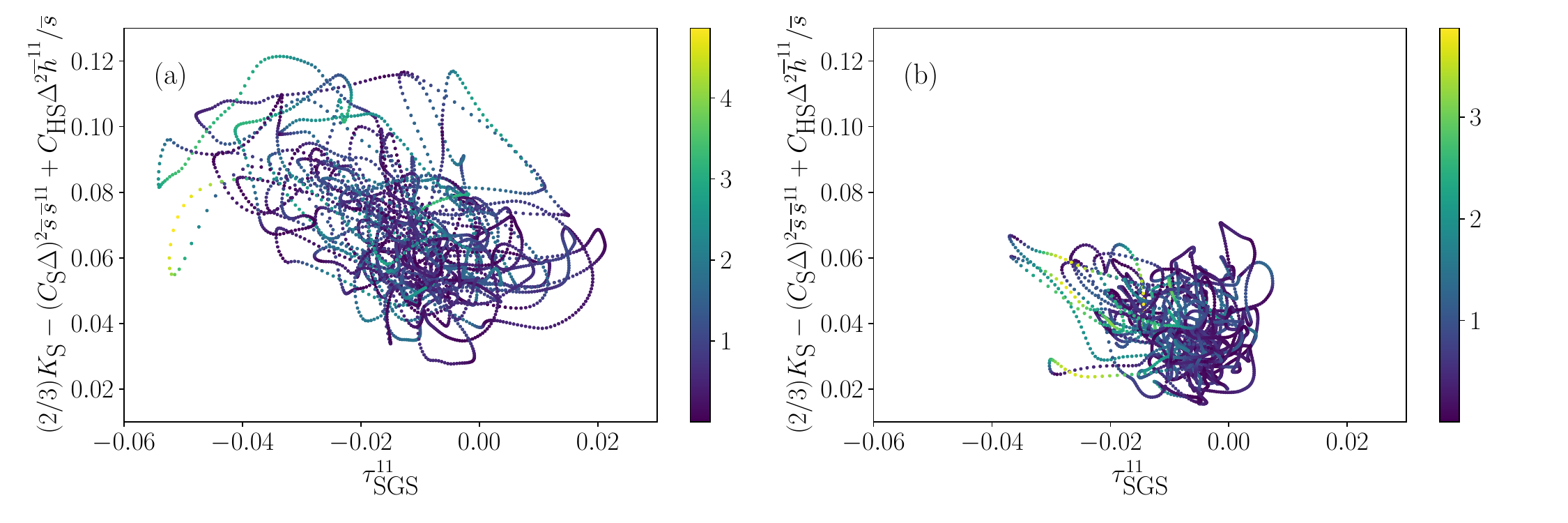}
\caption{\label{fig:s11_hel_rot}The diagonal component of SGS stress $\tau_{\rm{SGS}}^{11}$ (denoted by $R_{11}$) and the helicity SGS model $(2/3)K - (C_{\rm{S}} \Delta)^2 \overline{s}\ \overline{s}^{11} + C_{\eta {\rm{S}}} \Delta^2 \overline{s}^{-1}  (\nabla H_{\rm{S}})^1 \overline{\omega}_\ast^1$ ($(\nabla H)^1 \omega^1$ is denoted by $\overline{h}^{11}$) in the simulation with rotation ($\omega_{\rm{F}} = 8$) for (a) the large scale filtering case ($k_{\rm{C}} = 7$), and (b) the small scale filtering case ($k_{\rm{C}} = 14$).}
\end{figure}

\subsubsection{\label{sec:V.B.4}Helicity SGS model with rotation}

In Fig.~\ref{fig:s11_hel_rot}, we present the scatter plots of the diagonal components of the SGS stress, $\tau_{\rm{SGS}}^{11}$ in the DNS, and the expressions calculated by the helicity SGS model, in the case with rotation ($\omega_{\rm{F}} = 8$). As in the case without rotation, the scatter plots using the helicity SGS model [Fig.~\ref{fig:s11_hel_rot}(a) and (b)] are very similar to their counterparts using the Smagorinsky model [Fig.~\ref{fig:s11_smag_rot}(a) and (b)] both in the large and small scale filtering cases. No significant contribution from helicity, or significant improvements to the SGS model, can be expected in the diagonal components.

\subsection{\label{sec:V.C}SGS stress: off-diagonal components}

\subsubsection{\label{sec:V.C.1}Smagorinsky model without rotation}

\begin{figure}
\includegraphics[width=1.0\textwidth]{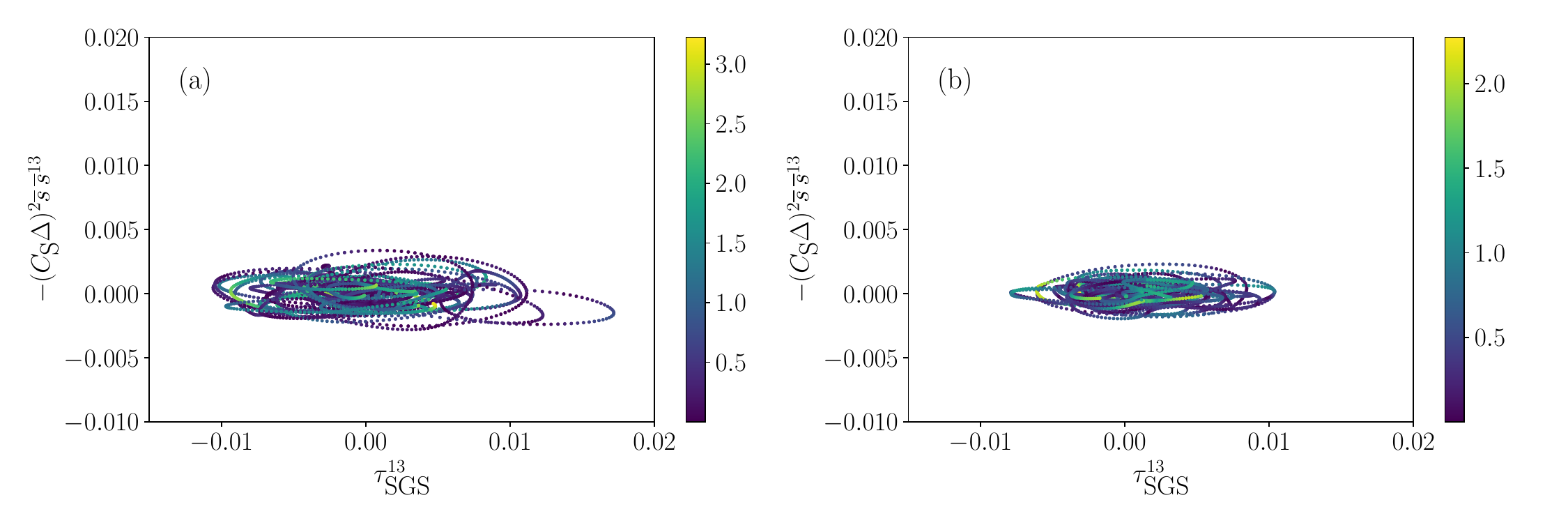}
\caption{\label{fig:s13_smag_hit}Off-diagonal component of SGS stress $\tau_{\rm{SGS}}^{13}$ and the Smagorinsky model $-(C_{\rm{S}} \Delta)^2 \overline{s}\ \overline{s}^{13}$ for (a) the large scale filtering case ($k_{\rm{C}} = 7$), and (b) the small scale filtering case ($k_{\rm{C}} = 14$), in the case without rotation.}
\end{figure}

\begin{figure}
\includegraphics[width=1.0\textwidth]{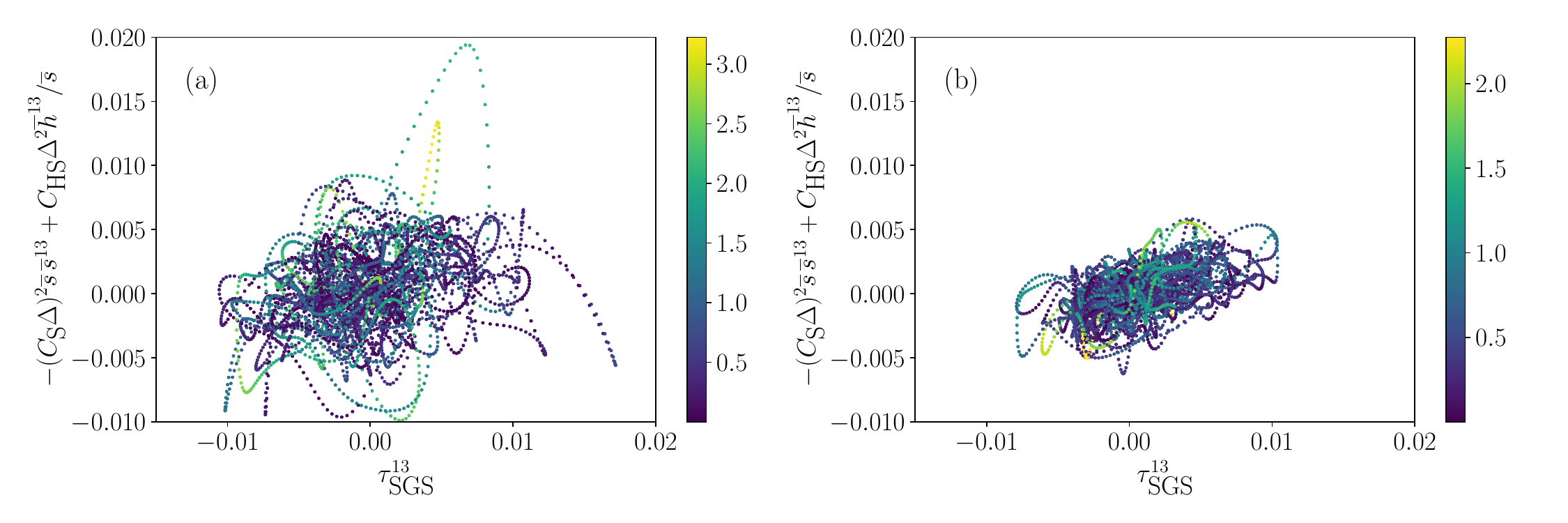}
\caption{\label{fig:s13_hel_hit}Off-diagonal component of SGS stress $\tau_{\rm{SGS}}^{13}$ and the helicity SGS model $-(C_{\rm{S}} \Delta)^2 \overline{s}\ \overline{s}^{13} + C_{\eta{\rm{S}}} \Delta^2 \overline{s}^{-1} (\nabla H_{\rm{S}})^1 \overline{\omega}_\ast^3$ [$(\nabla H_{\rm{S}})^1 \overline{\omega}_\ast^3$ is denoted by $\overline{h}^{13}$] for (a) the large scale filtering case ($k_{\rm{C}} = 7$), and (b) the small scale filtering case ($k_{\rm{C}} = 14$), in the case without rotation.}
\end{figure}

Figure \ref{fig:s13_smag_hit} shows the scatter plots of the off-diagonal component of the SGS stress, $\tau_{\rm{SGS}}^{13}$. The correlations between the values $\tau_{\rm{SGS}}^{13}$ obtained from the DNS and those of the Smagorisky model, $-(C_{\rm{S}} \Delta)^2 \overline{s}\ \overline{s}^{13}$, are presented in the scatter plots for (a) the large scale filtering case ($k_{\rm{C}} = 7$), and (b) the small scale filtering case ($k_{\rm{C}} = 14$).

We see from Fig.~\ref{fig:s13_smag_hit} that there is no correlation between the values of the DNS for $\tau_{\rm{SGS}}^{13}$ and those of the Smagorinsky model $-(C_{\rm{S}} \Delta)^2 \overline{s}\ \overline{s}^{13}$, irrespective of the filtering scale used. It is obvious that this situation will not improve by increasing the model constant, $C_{\rm{S}}$. Another point to be noted is that the shapes of the scatter plots do not show any evident correlation with the point colors. This means that the lack of correlation does not depend on the magnitude of the SGS-helicity gradients.

\subsubsection{\label{sec:V.C.2}Helicity SGS model without rotation}

In Fig.~\ref{fig:s13_hel_hit}, we present the off-diagonal components of the SGS stress, $\tau_{\rm{SGS}}^{13}$, in  scatter plots computed using the expressions from the helicity SGS model. To help with the comparison against the Smagorinsky model in Fig.~\ref{fig:s13_smag_hit}, the same range is used in all axes.

We see from these figures that the correlation between the values obtained from the DNS, and the values expected from the models' expressions, are significantly improved when using the helicity SGS model [Fig.~\ref{fig:s13_hel_hit}(a) and (b)] as compared against the Smagorinsky model [Fig.~\ref{fig:s13_smag_hit}(a) and (b)] both for large scale filtering ($k_{\rm{C}} = 7$) and small scale filtering ($k_{\rm{C}} = 14$). The correlation between the DNS and the model in the large scale filtering case ($k_{\rm{C}} = 7$) is $r = 0.54$, while the correlation in the small scale filtering case is $r = 0.90$. The performance of the helicity SGS model is better when filtering at smaller scales, which is to be expected as SGS models need some of the turbulent inertial range to be resolved to properly estimate SGS quantities. For both filtering cases, we observe that the data with large values of $\nabla H_{\rm{S}}$, which correspond to data points colored yellow and green, shows stronger correlation in the scatter plots. This fact conﬁrms that it is the third or SGS-helicity-related terms that contribute to the improvement of the model performance.

\begin{figure}
\includegraphics[width=1.0\textwidth]{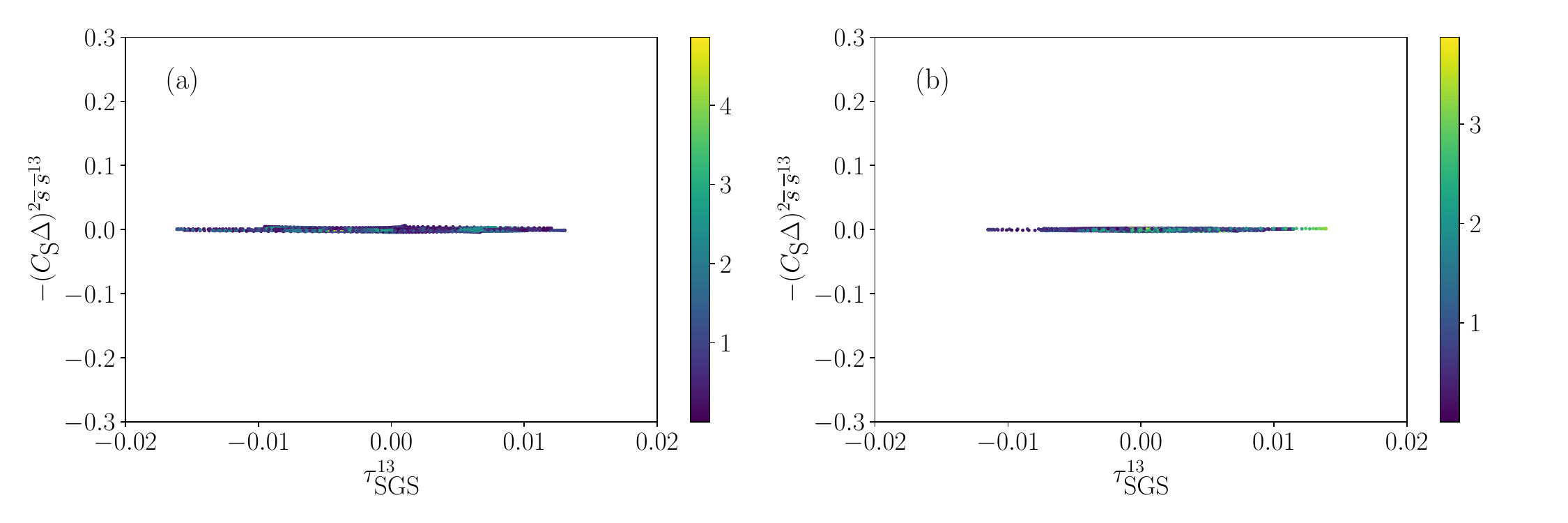}
\caption{\label{fig:s13_smag_rot}Off-diagonal component of the SGS stress $\tau_{\rm{SGS}}^{13}$ and the Smagorinsky model $-(C_{\rm{S}} \Delta)^2 \overline{s}\ \overline{s}^{13}$ in the simulation with rotation ($\omega_{\rm{F}} = 8$) for (a) the large scale filtering case ($k_{\rm{C}} = 7$), and (b) the small scale filtering case ($k_{\rm{C}} = 14$).}
\end{figure}

\begin{figure}
\includegraphics[width=1.0\textwidth]{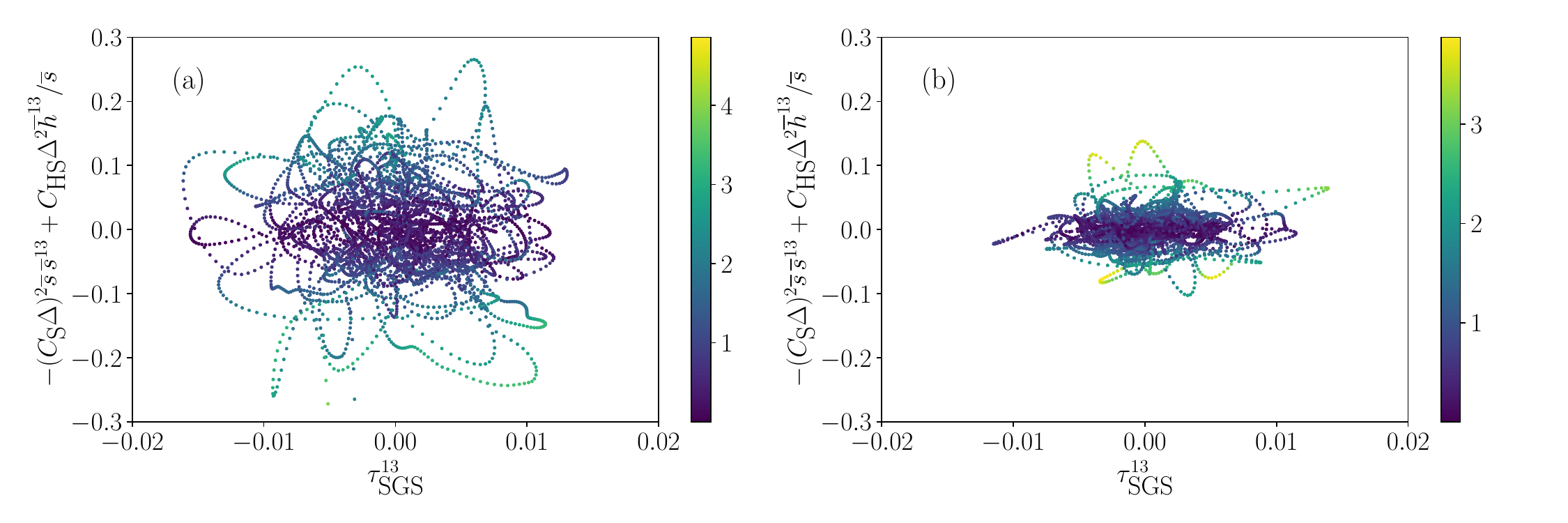}
\caption{\label{fig:s13_hel_rot}Off-diagonal component of the SGS stress $\tau_{\rm{SGS}}^{13}$ and the helicity SGS model $-(C_{\rm{S}} \Delta)^2 \overline{s}\ \overline{s}^{13} + C_{\eta{\rm{S}}} \Delta^2 \overline{s}^{-1} (\nabla H_{\rm{S}})^1 \overline{\omega}_\ast^3$ [$(\nabla H_{\rm{S}})^1 \overline{\omega}_\ast^3$ is denoted by $\overline{h}^{13}$] in the simulation with rotation ($\omega_{\rm{F}} = 8$) for (a) the large scale filtering case ($k_{\rm{C}} = 7$), and (b) the small scale filtering case ($k_{\rm{C}} = 14$).}
\end{figure}


\subsubsection{\label{sec:V.C.3}Smagorinsky model with rotation}

Figure~\ref{fig:s13_smag_rot} shows an off-diagonal component of SGS stresses in the DNS with rotation. The correlation between $\tau_{\rm{SGS}}^{13}$ and its Smagorinsky model expression $- (C_{\rm{SGS}} \Delta)^2 \overline{s}\ \overline{s}^{13}$ are displayed for (a) the large scale filtering case ($k_{\rm{C}} = 7$), and (b) the small scale filtering case ($k_{\rm{C}} = 14$). There are no correlations between the off-diagonal components of SGS stress, as illustrated by $\tau_{\rm{SGS}}^{13}$, and the values obtained by the Smagorinsky model.

\subsubsection{\label{sec:V.C.4}Helicity SGS model with rotation}

Figure~\ref{fig:s13_hel_rot} shows the scatter plots of the same off-diagonal component of the SGS stress, $\tau_{\rm{SGS}}^{13}$, and its 
\AP{corresponding} values as obtained from the helicity SGS model with rotation ($\omega_{\rm{F}} = 8$). As in the previous comparison without rotation, the same ranges in the axes are used in Figs.~\ref{fig:s13_smag_rot} and \ref{fig:s13_hel_rot}.

We see from Fig.~\ref{fig:s13_hel_rot} that some improvement in the correlation results from using the helicity SGS model, both for the large and small scale filtering cases. This makes a sharp contrast to the results with the Smagorinsky model [Fig.~\ref{fig:s13_smag_rot}(a) and (b)]. As in the non-rotating case, the improvement of model performance is more prominent in the small scale filtering case ($k_{\rm{C}} = 14$). The improvement in the correlations can be attributed to the data with strong SGS helicity gradients, colored with yellow and green. This shows that, for the improvement of model performance, the SGS helicity gradient component in Eq.~(\ref{eq:SGS_strss_exp}) plays a 
\AP{relevant} role.

\section{\label{sec:VI}Discussions}

\subsection{\label{sec:VI.A}Validation of the helicity SGS model}

In Sec.~\ref{sec:V} we presented the numerical results of a detailed comparison between the DNSs and the corresponding expressions of different SGS models, which were made with the aid of scatter plots to study the correlation between them. We should remember that even if we have good predictions of the mean or grid-scale (GS) components, detailed local correlations between the DNS results and the turbulence or subgrid-scale (SGS) model expressions cannot be generally expected. In this sense, this type of validation with one-point to one-point correlations is a very severe test for any turbulence model. However, in the numerical validation in Sec.~\ref{sec:V}, we saw that the incorporation of the inhomogeneous SGS helicity effects into the SGS model gives in general an improvement in the evaluations of the SGS stresses. In particular, the evaluation of the off-diagonal components of the SGS stress is improved by the inhomogeneous SGS helicity effect, while the standard Smagorinsky-type model cannot reproduce the behavior observed in the SGS stress. These results indicate that the inhomogeneous SGS helicity effect should be implemented into SGS turbulence models in order to improve model predictions in flows with GS absolute vorticity (relative vorticity and rotation).

In the following subsections, we brieﬂy discuss other properties of the helicity SGS model seen in the simulations. They include the filter-width dependence (Sec.~\ref{sec:VI.B}), the temporal evolution of the correlation (Sec.~\ref{sec:VI.C}), the behavior of the transport equation of SGS helicity (Sec.~\ref{sec:VI.D}), and the GS ﬂow generation in the direction parallel to rotation in regions of larger $\nabla H_{\rm{S}}$ (Sec.~\ref{sec:VI.E}).

\subsection{\label{sec:VI.B}Filter-width dependence of the correlation}

In the Reynolds-averaged model, the relative importance of helicity effect to the eddy-viscosity effect can be evaluated from Eq.~(\ref{eq:rey_strss_exp}) with Eqs.~(\ref{eq:nuT_anal_exp}) and (\ref{eq:Gamma_anal_exp}) as
\begin{equation}
	\frac{(\mbox{Helicity effect})}{(\mbox{Eddy-viscosity effect})}
	\sim \frac{|\mbox{\boldmath$\Gamma$}\mbox{\boldmath$\Omega$}_\ast|}
			{|\nu_{\rm{T}} \mbox{\boldmath${\cal{S}}$}|}
	\sim \frac{|\mbox{\boldmath$\Omega$}_\ast|}{|\mbox{\boldmath${\cal{S}}$}|}
		\frac{|H(k)|}{kQ(k)},
	\label{eq:hel_effect_eddy-visc_ratio}
\end{equation}
where $|\mbox{\boldmath$\Omega$}_\ast|$ is the magnitude of the mean absolute vorticity, $|\mbox{\boldmath${\cal{S}}$}| [= \sqrt{({\cal{S}}^{ij})^2/2}]|$ is the magnitude of mean velocity strain rate, $H(k)$ is the helicity spectral function, and $Q(k)$ is the energy spectral function. If the spectral function of helicity $H(k)$ shows the same scaling as the energy spectral function $Q(k)$, as is often observed, the relative importance of the helicity effect decreases with the wavenumber $k$. This means that the relative relevance of helicity effects is expected to be reduced in the smaller scales, provided that the ratio of the magnitudes of $|\mbox{\boldmath$\Omega$}_\ast|$ and $|\mbox{\boldmath${\cal{S}}$}|$ remains constant or also decreases with scale (see also the compensated relative helicity spectrum in the inset of Fig.~\ref{fig:dns_en_hel_spectra}).

The counterpart of the helicity effect in the SGS model may be evaluated as
\begin{equation}
	\frac{(\mbox{SGS-helicity effect})}{(\mbox{SGS-viscosity effect})}
	\sim \frac{\overline{\omega}_\ast}{\overline{s}}
		\frac{\Delta |H_{\rm{S}}|}{K_{\rm{S}}},
	\label{eq:sgs_hel_effect_eddy-visc_ratio}
\end{equation}
where $\overline{\omega}_\ast (= |\overline{\mbox{\boldmath$\omega$}}_\ast|)$ is the magnitude of GS absolute vorticity, $\overline{s} [= \sqrt{(\overline{s}^{ij})^2 /2}]$ is the magnitude of GS strain rate, and $\Delta$ is the fixed filter width. This suggests that the SGS-helicity effect relative to the SGS-viscosity effect would be decreased with the decreasing filter width $\Delta$.

In Sec.~\ref{sec:V}, we discussed the scatter plots for both the large scale filtering case ($k_{\rm{C}} = 7$) and the small scale filtering case ($k_{\rm{C}} = 14$). As Fig.~\ref{fig:s13_hel_hit}(a) and (b) representatively show, our numerical simulations indicate that the improvement in the correlation between the SGS stress and the helicity-model is better in the small scale filtering case [Fig.~\ref{fig:s13_hel_hit}(b)]. At a casual glance, this seems to contradict the evaluation of the relative importance of the SGS-helicity effect to the SGS-viscosity effect expressed by Eq.~(\ref{eq:sgs_hel_effect_eddy-visc_ratio}), where the relative importance is proportional to the filter width $\Delta$. \DR{This apparent discrepancy may be eliminated if the relative magnitude of the GS vorticity $\overline{\omega}_*$ to that of the GS strain rate $\overline{s}$ were to increase with decreasing filter width $\Delta$, and a future study will be needed to examine this dependence.}

\subsection{\label{sec:VI.C}Temporal evolution of the correlation}

In our analysis, all scatter plots were computed once the flow reached a turbulent steady state. However, we can briefly consider how the system reaches this state, and how a SGS model would perform during this transient. At early times (see Fig.~\ref{fig:kin_en_enstrophy}), the SGS energy $K_{\rm{S}}$ increases with time, while the SGS helicity does not that much (as most of the relative helicity is concentrated at large scales in the flow). We therefore expect the relative helicity, deﬁned by $\Delta |H_{\rm{S}}| /K_{\rm{S}}$, to either remain constant or decrease with time. On the other hand, the magnitudes of the GS absolute vorticity, $\overline{\omega}_\ast (= |\overline{\mbox{\boldmath$\omega$}}_\ast|)$, and of the GS velocity strain, $\overline{s} [= \sqrt{(\overline{s}^{ij})^2 /2}]$, do not change much with time. As a result, the relative importance of the helicity effect to the SGS viscosity effect could be time dependent, and helicity could play a relevant role in transient dynamics. This problem is left for future studies.

\subsection{\label{sec:VI.D}Evaluation of the SGS helicity}

The evolution of the SGS energy $K_{\rm{S}}$ is subject to Eq.~(\ref{eq:SGS_KS_eq_in_two}) with Eqs.~(\ref{eq:SGS_PKS_def_in_two})-(\ref{eq:SGS_TKS_def_two}), and the SGS helicity is subject to Eq.~(\ref{eq:SGS_HS_eq_in_two}) with Eqs.~(\ref{eq:SGS_PHS_def_in_two})-(\ref{eq:SGS_THS_def_in_two}). The production rate of SGS helicity, $P_{\rm{HS}}$ [Eq.~(\ref{eq:SGS_PHS_def_in_two})], requires some inhomogeneous GS absolute vorticity or an inhomogeneous SGS stress along the GS absolute vorticity for SGS helicity generation. As Eq.~(\ref{eq:SGS_KS_prod_mech}) suggests, the inhomogeneity of the SGS energy along rotation, $(\mbox{\boldmath$\omega$}_{\rm{F}} \cdot \nabla) K_{\rm{S}}$, is very important. However, in the present simulations, as turbulence is homogeneous in the $y$ and $z$ directions, this has no substantial contribution to $P_{\rm{ HS}}$.

In the present work, the instantaneous production of helicity (both GS and SGS helicities) $\widetilde{P}_{\rm{HS}}$ is supplied instead by the external forcing. The instantaneous hellicity production is given by
\begin{equation}
	\widetilde{P}_{\rm{HS}}
	= {\bf{f}} \cdot \nabla \times {\bf{f}},
	\label{eq:PHS_forcing}
\end{equation}
and the instantaneous dissipation rate of helicity $\widetilde{\varepsilon}_{\rm{HS}}$ is given by
\begin{equation}
	\widetilde{\varepsilon}_{\rm{HS}}
	= 2 \nu \frac{\partial u^j}{\partial x^i}
		\frac{\partial \omega^j}{\partial x^i}.
	\label{eq:epsHS_inst}
\end{equation}

\begin{figure}
\includegraphics[width=0.55\textwidth]{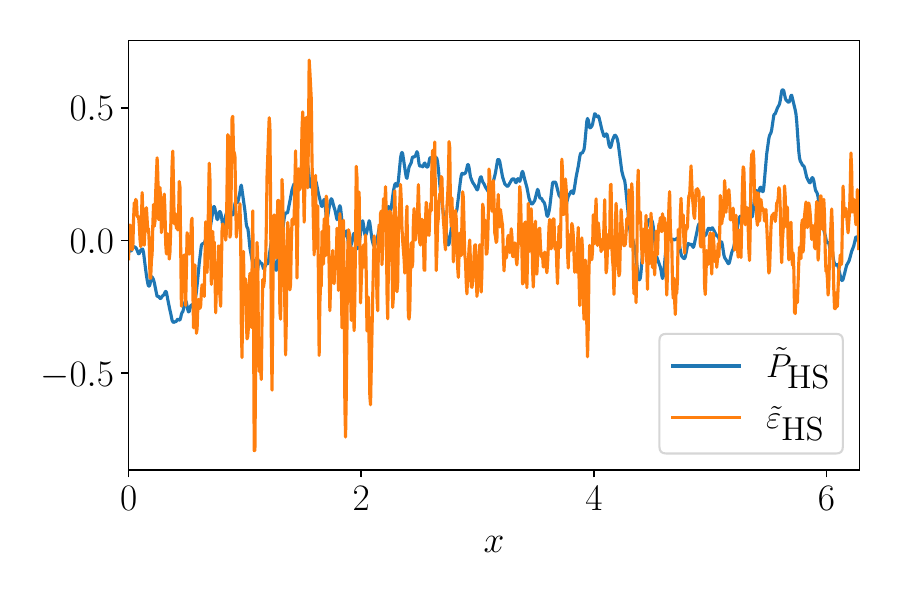}
\caption{\label{fig:inst_hel_prod_dissip}Spatial distributions of the instantaneous fluctuations of the production and dissipation rates of helicity, $\widetilde{P}_{\rm{HS}}$ [Eq.~\ref{eq:PHS_forcing}] and $\widetilde{\varepsilon}_{\rm{HS}}$ [Eq.~(\ref{eq:epsHS_inst})], respectively. The production fluctuations (drawn in blue) show slow variations in space, while the dissipation fluctuations (drawn in orange) show fast variations.}
\end{figure}

The spatial profiles (in the $x$ direction) of instantaneous fluctuations in the production and dissipation of helicity in the DNS without rotation are shown in Fig.~\ref{fig:inst_hel_prod_dissip}. Note that the r.m.s.~values of production and dissipation rates are of the same order, although production has fluctuations at larger scales (which is to be expected as production is dominated by the forcing), while dissipation has fluctuations closer to the Kolmogorov scale.

\begin{figure}
\includegraphics[width=0.55\textwidth]{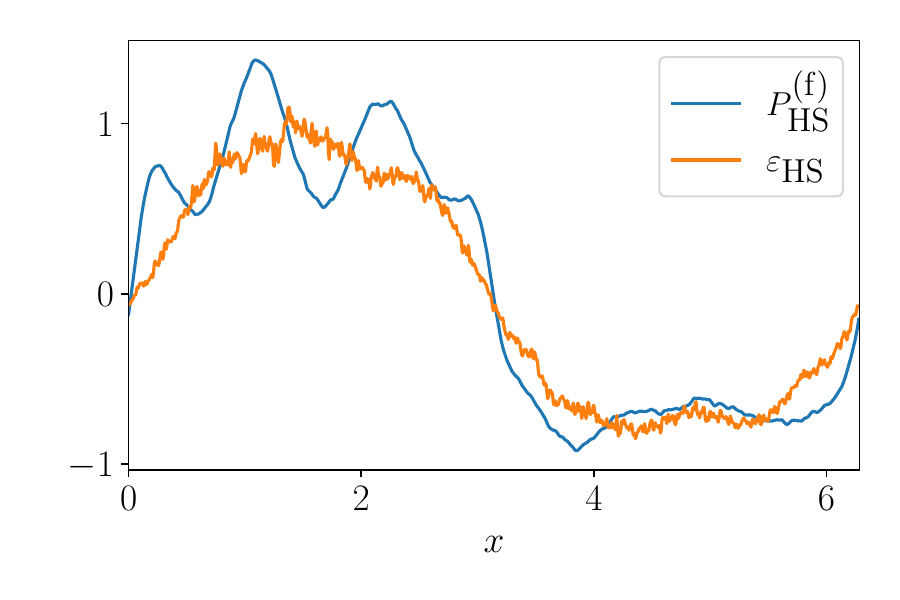}
\caption{\label{fig:mean_prod_dissip}Spatial distributions of the mean production rate of SGS helicity due to the forcing, $P_{\rm{HS}}^{({\rm{f}})}$, and dissipation rate $\varepsilon_{\rm{HS}}$. The production rate (drawn in blue) shows relatively large-scale variations, while the dissipation rate (drawn in orange) has rapid fluctuations.}
\end{figure}

The production rate of the SGS helicity $H_{\rm{S}}$ due to the forcing, $P_{\rm{HS}}^{({\rm{f}})}$, is defined by
\begin{equation}
	P_{\rm{HS}}^{({\rm{f}})}
	= \overline{{\bf{f}} \cdot \nabla \times {\bf{f}}}
	- \overline{\bf{f}} \cdot \overline{\nabla \times {\bf{f}}},
	\label{eq:P_HS_force_def}
\end{equation}
and the dissipation rate of $H_{\rm{S}}$, $\varepsilon_{\rm{HS}}$, is defined by Eq.~(\ref{eq:SGS_epsHS_def_in_two}) or (\ref{eq:SGS_epsHS_def}). The spatial profiles (in the $x$ direction) of the production and dissipation rates of SGS helicity, $P_{\rm{HS}}^{({\rm{f}})}$ and $\varepsilon_{\rm{HS}}$, are shown in Fig.~\ref{fig:mean_prod_dissip}. Note that the production rate of helicity due to the forcing, $P_{\rm{HS}}^{({\rm{f}})}$ [Eq.~(\ref{eq:P_HS_force_def})], is correlated with the dissipation rate of helicity $\varepsilon_{\rm{HS}}$ [Eq.~(\ref{eq:SGS_epsHS_def})], having the same amplitudes in each half-box (i.e., they are approximately balanced). Production has fluctuations at larger scales (which is to be expected as production is dominated by the forcing), while dissipation again has fluctuations closer to the Kolmogorov scale.

\begin{figure}
\includegraphics[width=0.55\textwidth]{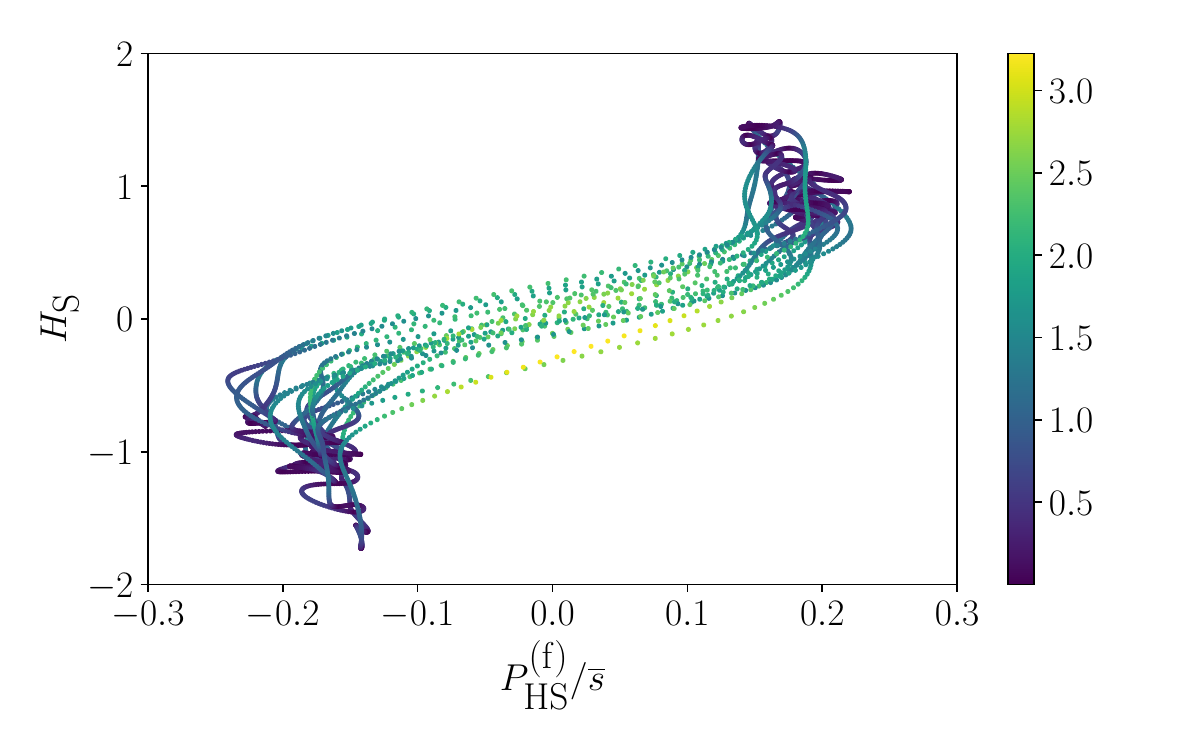}
\caption{\label{fig:SGS_HS_eval}The SGS helicity $H_{\rm{S}}$ [Eq.~(\ref{eq:SGS_H_def})] and its evaluation due to helical forcing term $P_{\rm{HS}}^{({\rm{f}})} / \overline{s} = (\overline{{\bf{f}} \cdot \nabla \times {\bf{f}}} - \overline{\bf{f}} \cdot \overline{\nabla \times {\bf{f}}}) / \overline{s}$ [see Eq.~(\ref{eq:SGS_HS_eval_forcing})].}
\end{figure}

The local equilibrium of helicity, $P_{\rm{HS}} \simeq \varepsilon_{\rm{HS}}$ [Eq.~(\ref{SGS_HS_local_equil_zero})], in this case can be written using Eqs.~(\ref{eq:P_HS_force_def}) and (\ref{eq:SGS_epsHS_inHS_s_one}) as
\begin{equation}
	P_{\rm{HS}}^{({\rm{f}})}
	\simeq \frac{C_{\varepsilon {\rm{H}}} C_{\rm{S}}^2}{C_{\rm{KS}}} H_{\rm{S}} \overline{s},
	\label{eq:SGS_local_equil_forcing}
\end{equation}
which gives an evaluation of the SGS helicity as
\begin{equation}
	H_{\rm{S}}
	\simeq \frac{C_{\rm{KS}}}{C_{\varepsilon {\rm{H}}} C_{\rm{S}}^2}
		\frac{P_{\rm{HS}}^{({\rm{f}})}}{\overline{s}}
	= \frac{C_{\rm{KS}}}{C_{\varepsilon {\rm{H}}} C_{\rm{S}}^2}
		\frac{\overline{{\bf{f}} \cdot \nabla \times {\bf{f}}}
		- \overline{{\bf{f}}} \cdot \overline{\nabla \times {\bf{f}}}}{\overline{s}}.
	\label{eq:SGS_HS_eval_forcing}
\end{equation}
Reflecting the good balancing between $P_{\rm{HS}}$ [Eq.~(\ref{eq:P_HS_force_def})] and $\varepsilon_{\rm{HS}}$ [Eq.~(\ref{eq:SGS_epsHS_def_in_two})] shown in Fig.~\ref{fig:mean_prod_dissip}, and the good correlation between $\varepsilon_{\rm{HS}}$ and its algebraic model [Eq.~(\ref{eq:SGS_epsHS_approx_in_two})], we verified that $H_{\rm{S}}$ also shows a good correlation with the r.h.s.\ of Eq.~(\ref{eq:SGS_HS_eval_forcing}) (see Fig.~\ref{fig:SGS_HS_eval}).
For the zero-equation helicity model mentioned in Sec.~\ref{sec:III.D}, we can therefore use this evaluation of $H_{\rm{S}}$ [Eq.~(\ref{eq:SGS_HS_eval_forcing})] without solving the full $H_{\rm{S}}$ transport equation [Eq.~(\ref{eq:SGS_HS_eq_in_two})].

\subsection{\label{sec:VI.E}GS ﬂow generation in the direction parallel to rotation at high $|\nabla H_{\rm{S}}|$ region}

The SGS stress [Eq.~(\ref{eq:SGS_strss_exp})] suggests that, against the SGS viscosity $\nu_{\rm{S}} \overline{{\bf{s}}} = \nu_{\rm{S}} \{\overline{s}^{ij} \}$, a large-scale or GS ﬂow can be generated due to the inhomogeneous SGS helicity effect $\eta_{\rm{S}} [\nabla H_{\rm{S}} \overline{\mbox{\boldmath$\omega$}} + (\nabla H_{\rm{S}} \overline{\mbox{\boldmath$\omega$}})^\dag]_{\rm{D}} = \eta_{\rm{S}} \{ [(\nabla H_{\rm{S}})^i \overline{\omega}_\ast^j + (\nabla H_{\rm{S}})^j \overline{\omega}_\ast^i]_{\rm{D}} \}$. In the present set-up, we have imposed rotation $\mbox{\boldmath$\omega$}_{\rm{F}}$ in the $z$ direction. If we have a non-trivial spatial distribution of $H_{\rm{S}}$ represented by $\nabla^2 H_{\rm{S}}$, we may have a large-scale or GS ﬂow in the direction of $\mbox{\boldmath$\omega$}_{\rm{F}}$ proportional $-\nabla^2 H_{\rm{S}}$.

Indeed, in the present simulations we have some velocity structure developing in the direction of the rotation $\mbox{\boldmath$\omega$}_{\rm{F}}$. However, the spatial proﬁle of $\nabla^2 H_{\rm{S}}$ in the present simulations is not as simple as \AP{in} the cases considered in \citet{yok2016b}, and therefore it is more complicated to separate this flow from the flow that would develop just as a consequence of the Taylor--Proudman theorem. We defer the reader to \citet{yok2016b} for a study of large-scale flow generation resulting from helicity inhomogeneities.

\section{\label{sec:VII}Conclusions}

The turbulent helicity in a flow represents the geometrical or structural properties of turbulence, in the sense that it is directly related to the cross-ﬂow correlation, while the turbulent energy captures only the intensity of turbulence. In the Reynolds-stress expression, the turbulent helicity is the transport coefficient coupling with the mean or large-scale vortical motion (the anti-symmetric part of the mean velocity shear). This is in marked contrast with the turbulent energy, constituting the eddy or turbulent viscosity, which couples with the mean strain rate of the velocity (the symmetric part of the mean velocity shear). The helicity effect in the Reynolds stress counter-balances the eddy viscosity, and may contribute to the suppression of the momentum mixing and to the large-scale ﬂow generation.

Following this helicity effect in the Reynolds-averaged turbulence model, we constructed a SGS model where the small-scale turbulence structure effect is incorporated into the SGS stress expression through the SGS helicity. This model is called here the helicity SGS model. For the SGS momentum flux, the Smagorinsky model has been ubiquitously adopted, where the SGS stress is expressed in terms of the SGS viscosity $\nu_{\rm{S}}$ coupled with the GS (grid-scale) velocity strain rate, where $\nu_{\rm{S}}$ is expressed in terms of the filter width $\Delta$ and the GS strain rate $\overline{s}$. In the helicity SGS model, in addition to the Smagorinsky term associated with the GS strain rate, we consider the SGS helicity term, where the gradient of the SGS helicity couples with the GS absolute vorticity (GS relative vorticity and total rotation) in the SGS stresses.

In the present work, the helicity SGS model was validated with the aid of direct numerical simulations in a triply periodic boundary box \AP{at a resolution of $1024^3$ grid points}. To explicitly test the effect of gradients of helicity in SGS modeling, large-scale spatially inhomogeneous helicity was imposed by external forcing. Using a filter, a given ﬁeld quantity was divided into the resolved or GS component and the unresolved or SGS component. In the GS component equations, the effect of the SGS components shows up as a SGS turbulent flux. For the mean momentum evolution equation, the SGS stress is the sole turbulent flux that represents the effects of the SGS component. As for the SGS statistical quantities such as the SGS energy, its dissipation rate, the SGS timescale, and the SGS stresses, the results of the SGS model were compared with the counterparts of the DNSs.

Our findings can be summarised as follows:
\begin{enumerate}

\item{SGS stresses:}
\begin{itemize}

\item  The diagonal components of the SGS stress are independent of the SGS helicity effect. This is to be expected as there is no coupling between the GS absolute vorticity and the inhomogeneous SGS helicity in the present setup.

\item The off-diagonal components of the SGS stress have much higher correlation with the DNS results in the helicity SGS model than in the Smagorinsky model. This is a clear 
\AP{difference} as in the Smagorinsky model no substantial correlation between the SGS model and DNS results could be identified.

\item The correlation improvements in the off-diagonal component of SGS stress are observed both in cases without and with rotation. This indicates that, in helical turbulence even without rotation, the inhomogeneous helicity effect should be implemented into SGS models.

\end{itemize}

\item{SGS quantities (timescale, energy, helicity, and dissipation rates). 

In Sec.~\ref{sec:III}, we constructed three levels of helicity SGS models. The simplest model is the zero-equation helicity SGS model (Sec.~\ref{sec:III.D}), where no transport equations of the SGS energy or SGS helicity are solved explicitly, in the same sense as the Smagorinsky model. In the zero-equation model, the SGS helicity is evaluated by Eq.~(\ref{eq:SGS_HS_eval_zero}) under the assumption of the local equilibrium of the SGS helicity production and its dissipation rates [Eq.~(\ref{SGS_HS_local_equil_zero})]. We should note that, in the present work, we have no GS ﬂow inhomogeneities that contribute to produce SGS helicity. Instead, the SGS helicity is injected by the external force as in \NY{Eq.~(\ref{eq:P_HS_force_def})}. For these quantities:}

\begin{itemize}

\item The timescale of SGS fluctuations can be well represented by the reciprocal of the GS strain rate, $\overline{s}^{-1}$ both in the cases without and with rotation. The correlation between them is less clear in the presence of rotation. This is because the magnitude of the fluctuations, whose reciprocal is proportional to the SGS timescale, is suppressed by rotation.

\item The SGS energy $K_{\rm{S}}$ can be well represented in terms of the ﬁlter width $\Delta$ and the GS strain rate $\overline{s}$ by $(\Delta \overline{s})^2$. The correlation between $K_{\rm{S}}$ and $(\Delta \overline{s})^2$ is stronger in the non-rotating case.

\item In this specific case, the SGS helicity and its dissipation rate are well evaluated by Eqs.~(\ref{eq:P_HS_force_def}) and (\ref{SGS_epsHS_def_two}), respectively. In this sense, the zero-equation helicity SGS model is supposed to provide a reasonable turbulence model.

\end{itemize}
\end{enumerate}

In realistic ﬂow configurations with GS ﬂow inhomogeneities, the evaluation of the SGS helicity based on Eqs.~(\ref{eq:SGS_HS_eval_zero}) or (\ref{eq:SGS_HS_eval_Del_s_zero}) should be further examined. This is left for future work.

The high correlation in the SGS stresses with the helicity SGS model indicates that a drawback of the Smagorinsky model---too much dissipative behavior---can be alleviated by incorporating the helicity effect into the SGS stress model. In this sense, we expect that one of the main drawbacks of the Smagorinsky model referred to in Sec.~\ref{sec:I} can be mitigated by the present model.

Of course, in order to fully argue the model constant adjustment problem, we have to apply the helicity SGS model not only \AP{to} isotropic turbulence but also to mixing-layer flows or/and wall-turbulence flow configurations. 
\NY{
As a follow-up of this work, we should apply the present H-SGS models to a channel turbulent flow, where strong streamwise vortical structures are ubiquitously observed. There, the inhomogeneous SGS helicity coupled with the GS vorticity should substantially reduce the SGS-viscosity effect. If the current H-SGS models properly work there, a channel flow might be treated without resorting to the adjustment of the Smagorinsky constant $C_{\rm{S}}$ from $0.18$ to $0.10$ for wall turbulence. We could then use the H-SGS models with the universal value of $C_{\rm{S}} = 0.18$ as in isotropic turbulence. Such elaborated applications of the helicity SGS models to more complicated ﬂow configurations are left for the future. In this sense, this work is certainly in progress. However, the present results indicate that the H-SGS models are promising prescriptions for alleviating some difficulties of the Smagorinsky model from the physical perspective of modeling.}


\begin{acknowledgments}
One of the authors (NY) would like to thank the Isaac Newton Institute for Mathematical Sciences, Cambridge, for support and hospitality during the programme Anti-diffusive dynamics: from sub-cellular to astrophysical scales, where work on this paper was undertaken. This work was supported by EPSRC grant EP/R014604/1. Part of this work was performed during the period NY stayed at the Nordic Institute for Theoretical Physics (NORDITA) at the program ``Stellar Convection: Modelling, Theory and Observations'' (Aug.-Sep.\ 2024) and at the Department of Physics, University of Buenos Aires (Sep.-Oct.\ 2024). Part of this work was supported by the Japan Society for the Promotion of Science (JSPS) Grants-in-Aid for Scientific Research JP23K25895.
\end{acknowledgments}


\end{document}